\documentclass[12pt,a4paper]{iopart}

\newif\ifpdf
\ifx\pdfoutput\undefined
  \pdffalse
\else
  \pdfoutput=1
  \pdftrue
\fi
\ifpdf
   \usepackage[pdftex,colorlinks=true,urlcolor=blue,pdfstartview=FitH]
              {hyperref}
\fi

\usepackage[latin1]{inputenc}        
\usepackage[dvips]{graphicx}         
\usepackage{fancyheadings}           
\usepackage{amsmath}                 
\usepackage{iopams}                  
\usepackage{exscale}                 
\begin{document} 
\eqnobysec
\def\openone{\leavevmode\hbox{\small1\kern-3.8pt\normalsize1}}
\newcommand{\rd}{{\rm d}}
\newcommand{\re}{{\rm e}}
\newcommand{\ri}{i}
\newcommand{\refg}[1]{eq.{\hskip 2pt}(\ref{#1})}
\newcommand{\refgs}[1]{eqs.{\hskip 2pt}(\ref{#1})}
\newcommand{\refabb}[1]{Abb. (\ref{#1})}
\newcommand{\reftab}[1]{Tab. (\ref{#1})}
\newcommand{\refkap}[1]{section \ref{#1}}
\newcommand{\refanh}[1]{appendix \ref{#1}}

\jl{1}
\title{Semiclassical Approximations in Phase Space \\
 with Coherent States}
\author{M. Baranger\dag, M. A. M. de Aguiar\S\dag, F. Keck\ddag,
H. J. Korsch\ddag, \\ 
and B. Schellhaaß\ddag}

\address{\dag\ Center for Theoretical Physics, Laboratory for Nuclear Science and Department of Physics, Massachusetts Institute of Technology, Cambridge, MA 02139, USA}
\address{\S\ Instituto de F\'{\i}sica 'Gleb Wataghin', 
Universidade Estadual de Campinas, 13083-970, Campinas, Brazil}
\address{\ddag\ FB Physik, Universität Kaiserslautern, 
         D-67653 Kaiserslautern, Germany}

\begin{abstract}

We present a complete derivation of the semiclassical limit of the
coherent state propagator in one dimension, starting from path
integrals in phase space.  We show that the arbitrariness in the path
integral representation, which follows from the overcompleteness of
the coherent states, results in many different semiclassical limits. We
explicitly derive two possible semiclassical formulae for the
propagator, we suggest a third one, and we discuss their
relationships. We also derive an initial value representation for the
semiclassical propagator, based on an initial gaussian wavepacket.  It
turns out to be related to, but different from, Heller's thawed
gaussian approximation.  It is very different from the Herman--Kluk
formula, which is not a correct semiclassical limit.  We point out
errors in two derivations of the latter.  Finally we show how the
semiclassical coherent state propagators lead to WKB-type quantization
rules and to approximations for the Husimi distributions of stationary
states.

\end{abstract}

\scriptsize
\tableofcontents
\normalsize


\section{Introduction}
  \label{kap1}

Semiclassical approximations in phase space using coherent states have
been discussed extensively for several decades.  This attractive
topic, a favorite of many theoretical physicists and chemists, turns
out to be very difficult.  In this contribution to its literature, we
shall attempt to sort out and clarify the web of contradictions and
inconsistencies that have characterized the recent state of the field.
We shall do so for the simplest possible case, one--dimensional
coordinate space, i. e. two--dimensional phase space.  This is the
case where it is relatively easy to check the semiclassical
approximations.  We have done work in higher dimensions as well, but
we do not include it here, as it would only obscure the basic
relationships and further lengthen the paper.  The conclusions we have
reached are stated in section 7, and the reader who is already
familiar with the subject may jump to them now to get an overall view.
Because the pitfalls are numerous, however, we shall follow a slower
approach, a historical one in this introduction, and then a systematic
and detailed one in the body of the paper.

The study of semiclassical methods has two basic motivations. First,
it provides approximations to quantum mechanical quantities in terms
of classical ingredients. These approximations should be very good if
the typical classical actions are much larger than Planck's constant.
Interestingly, they are often fairly good even at very low quantum
numbers.  Second, semiclassical methods also help in understanding the
quantum mechanical processes themselves, providing a more intuitive
description. This description includes quantum mechanical
interference, since both amplitudes and phases can be calculated
semiclassically.

The semiclassical approximation for the evolution operator, or
propagator, in the coordinate representation has been known for more
than 70 years and was first written by Van Vleck \cite{Van28}. It is a
complex number with a modulus and a phase.  The main part of the phase
is the action of a classical trajectory joining a given initial
coordinate to a given final coordinate in a given time.  Finding such
trajectories is usually not a simple task.  It is known as ``the root
search problem'' and it gets more and more complicated as the number
of dimensions increases.  The modulus of the semiclassical propagator
is related to the second derivative of the action with respect to
these initial and final points.  It measures the dispersal of nearby
trajectories.  Gutzwiller, among others, revisited this problem around
1970 \cite{Gutz71,Gutz90}, focusing on non-integrable systems and
giving birth to the field of quantum chaos.  Much progress has been
made since then, particularly on the topological properties of Maslov
indices \cite{Robb91} and on the scars of periodic trajectories
\cite{Hell84,Bogo88}.

But there is another interesting representation for the evolution
operator, which seems at first sight to be more appropriate for
comparisons with classical mechanics.  This is the representation
using the coherent states of a harmonic oscillator. They are gaussian
states, localized in both coordinates and momenta, and therefore they
can be thought of as {\it quantum points} in phase space.  Although
the exact coordinate propagator and the exact coherent state
propagator are related by a simple change of representation, the
semiclassical approximations to them are quite different.  One of the
differences is that the classical hamiltonians with which the
trajectories are calculated are different.  Another is that the
classical trajectories for the coherent state propagator are usually
complex.  Both semiclassical propagators involve trajectories with
mixed initial and final conditions, hence both have the root search
problem.

This semiclassical coherent state propagator first appeared in the
work of Klauder \cite{Klau78,Klau79,Klau87a} without a detailed
derivation. Weissman \cite{Weis82b} extended the old semiclassical
correspondence relations to the case of coherent state variables and
presented a first derivation of the semiclassical propagator; his
derivation was based on the general semiclassical machinery rather
than on path integrals techniques.  The possibility of a rigorous
derivation using the latter is mentioned in several papers, but it
does not seem to have actually been published, to our knowledge.  The
properties of the propagator were studied, however, for a number of
fundamental quantum processes (see, e.g.,
\cite{Xavi96,Xavi96a,Xavi97,Gros98b}).  A recent application to the
semiclassical quantization of a system with classically mixed regular
and chaotic dynamics \cite{98ivr} demonstrated the power of this
approach.

There is another important difference between the coordinate and
coherent state representations.  Due to the overcompleteness of the
latter, the path integral for the coherent state propagator is not at
all unique, and this non-unicity reflects itself in a large
multiplicity of possible semiclassical propagators.  For the quantum
mechanical path integral, some good discussion of this variety was
given in a review by Klauder and Skagerstam \cite{Klau85}.  Yet more
ambiguousness arises because coordinates and momenta do not commute in
quantum mechanics, and therefore there is no unique general way of
associating a quantal hamiltonian with a classical one.  The net
result of all this is that many arbitrary decisions need to be made
whenever one contemplates doing a semiclassical approximation.

Without doubt, semiclassical approximations based on the propagator
have been highly successful in chemical, molecular, atomic and nuclear
physics.  In spite of this, however, problems are turning up more and
more often for which these methods are made very hard or inapplicable
by the root search difficulty.  These are usually problems in which
the underlying classical dynamics is chaotic, which means that the
number of contributing (real or complex) ``root trajectories'' can be
extremely large.  Consequently, people have attempted more and more to
avoid mixed initial and final conditions.  The ideal method is one in
which one is given a coordinate and a momentum {\em both at the
initial time}.  Then the classical trajectory is unique, and the wave
function evolves with time by following this unique thread.  Such a
method is called an ``initial value representation'' or IVR.  Much
work has been done on IVR's by many people, including Miller {\it et
al.}  \cite{Mil70,Sun97}, Levit and Smilansky \cite{Lev77}, Brumer
{\it et al.} \cite{Cam92,Pro95}.  These IVR's do not involve coherent
states.  Klauder \cite{Klau87b} also struggled with this problem with
coherent states.  Kay \cite{Kay94a,Kay94b,Kay97} compared several
IVR's numerically.  A very popular IVR, based on coherent states, is
the Herman--Kluk or HK propagator
\cite{Herm84,Herm86,Kluk86,Gros97,Guer98,Gros98a}.  For recent reviews see the
papers by Sep\'ulveda and Grossmann \cite{Sepu96,Gros99}.

Finally we should note that the coherent state representation is not
the only way to do quantum mechanics in phase space.  A fascinating
alternative is provided by the Wigner-Weyl representation
\cite{Wign32,Hill84,Berr89}. We shall not pursue this approach in this
paper and we refer to a recent review article by Ozorio de Almeida
\cite{Ozo98}.

{\bf Contents of this paper} ~:~ In section 2 we give a very complete
derivation of the semiclassical propagator in the coherent state
representation.  The result is \refg{glg119} or \refg{en4}.  We have
actually two different calculations of the path integral, a
step--by--step calculation in section 2 and a more general method in
the appendix.  The latter is used again (in section 3) to perform a
different integral.  Section 3 discusses the variety of possible path
integrals using coherent states and compares them.  Section 4 contains
the derivation of our IVR, which is \refg{3.24}.  Our original purpose
was to give a solid derivation of the HK formula but, when we were
finished, we found a result very different from theirs, and in much
better agreement with the expected behavior of such a formula.  Also
in section 4, we compare our IVR with Heller's IVR.  Both have equal
claims to being a correct semiclassical IVR, but they are different.
Section 5 returns to the HK propagator and points out the errors made
in two papers where it was derived.  It also explains why, in spite of
being an incorrect semiclassical formula, HK still works (sometimes
poorly) in some situations.  In section 6 we Fourier transform the
propagator from time to energy, which yields the Green's function in
the coherent state representation.  By looking for the poles of this
Green's function, we obtain the quantization rule for the energy
levels.  By looking at the residues of the poles, we obtain
approximate Husimi distributions for the stationary states. Finally
section 7 contains a summary of our results and our conclusions.


\section{The Semiclassical Coherent--State Propagator}
  \label{kap2}

In classical mechanics, it is convenient to describe the time evolution by
focusing on a trajectory in phase space.  One candidate for a similar quantity
in quantum mechanics is the operator which describes the time evolution in the
coherent state representation, the coherent state propagator.
In this section we shall construct the semiclassical limit of this
propagator. The result is \refg{glg119}.  For convenience, we shall confine
ourselves here to a single degree of freedom. The extension to higher
dimensional systems will be discussed in the future.  

\subsection{The Path Integral}
  \label{kap2_1}

The coherent states $|z\rangle$ of a harmonic oscillator of mass $m$ and
frequency $\omega$ are defined by
\begin{equation}
  \label{glg48}
  |z\rangle = \re^{-\frac{1}{2}|z|^2}\re^{z\hat{a}^\dagger}|0\rangle
\end{equation}
with $|0\rangle$ the harmonic oscillator ground state and
\begin{equation}
  \label{glg49}
  \hat{a}^\dagger = \frac{1}{\sqrt{2}}\left( \frac{\hat{q}}{b}-\ri 
              \,\frac{\hat{p}}{c} \right), \qquad
  z =  \frac{1}{\sqrt{2}}\left( \frac{q}{b}+\ri 
              \,\frac{p}{c} \right).
\end{equation}
In the above $\hat q$, $\hat p$, and $\hat{a}^\dagger$ are operators; $q$ and $p$
are real numbers; $z$ is complex.  The parameters
\begin{equation}
  \label{glg47}
  b = {(\hbar/ m \omega )}^{\frac{1}{2}}\qquad \mbox{and} \qquad
  c = {(\hbar m \omega )}^{\frac{1}{2}}
\end{equation}
define the length and momentum scales, respectively, and their product is
$\hbar$.
We shall need the wave function of a coherent state in the position
representation, which is
\begin{align}
  \label{glgmb1}
\langle x|z\rangle &= \pi^{-\frac{1}{4}} b^{-\frac{1}{2}}
                      \exp\left(-\frac{{(x-q)}^2}{2 b^2}\right)
                      \exp\left(\frac{i}{\hbar}p(x - q/2)\right) \nonumber\\
                   &= \pi^{-\frac{1}{4}} b^{-\frac{1}{2}}
 \exp\left[ -\frac{1}{2}(x/b-\sqrt{2}z)^2 + \frac{z}{2}(z-z^{*})\right] \; .
\end{align}

Now we consider a system with Hamiltonian operator $\hat{H}(t)$.  We
restrict ourselves to ``reasonable'' Hamiltonians, i.e. we assume that
$\hat{H}(t)$, written as a function of the creation and annihilator
operators $\hat{a}^\dagger$ and $\hat{a}$, can be expanded into a power
series of $\hat{a}^\dagger,\hat{a}$.  The matrix elements of the evolution
operator from time 0 to time $t$ in the basis of the coherent states
(\ref{glg48}) are
\begin{equation}
  \label{glg1}
  K(z'',t;z',0) = \langle z'' | \hat{T} \re^{-\frac{\ri }{\hbar}
                  \int_0^t \hat{H}(t')\rd t'} | z^\prime \rangle
\end{equation}
where $\hat{T}$ is the time--ordering operator. For time--independent
$\hat{H}$ this is simply
\begin{equation}
  \label{glg51}
  K(z'',t;z',0) = \langle z'' | \re^{-\frac{\ri }{\hbar}\hat{H}t} 
                  | z^\prime \rangle.
\end{equation}

In order to write the propagator as a path integral, we divide the time
interval $(0,t)$ into $N$ small intervals of length $\tau := t/N$, and we write
\begin{align}
  \label{glg39}
  K(z'',t;z',0) = \langle z_N | \prod_{j=0}^{N-1}
                  \hat{T} \exp\left\{ -\frac{\ri }{\hbar}
                  \int_{j\tau}^{(j+1)\tau} 
                  \hat{H}(t')\rd t'\right\} | z_0 \rangle
\end{align}
where, for simplicity of notation, we identify $|z_N\rangle \equiv |z''
\rangle$ and $|z_0\rangle \equiv |z' \rangle$.  If the time step $\tau$ is
small enough, the variable $\hat{H}(t')$ in the integral can be replaced by the
constant $\hat{H}(t_j)$, with some intermediate time $t_j \in [j\tau,
(j+1)\tau]$. Then the time--ordering operator becomes unnecessary and we get,
with large $N$,
\begin{align}
  \label{glg81}
  K(z'',t;z',0) \approx \langle z_N | \prod_{j=0}^{N-1}
                   \re^{-\frac{\ri }{\hbar}
                   \hat{H}(t_j)\tau} | z_0 \rangle.
\end{align}
Now we insert the unit operator, namely
\begin{equation}
  \label{glg21}
  \openone = \int|z\rangle\frac{\rm{d}^2z}{\pi}\langle z|
           \equiv \iint|z\rangle\frac{\rd x\rd y}{\pi}\langle z|
       \equiv \iint|z\rangle\frac{\rd z^\star\rd z}{2\pi\ri }\langle z|~,
\end{equation}
everywhere between adjacent propagation steps.  We denoted the real and
imaginary parts of $z$ by $x$ and $y$, respectively. The sign of the last
equation member on the right is actually undetermined, because the sign of the
Jacobian depends on the order in which the variables are taken.  To avoid any
possible confusion, we state here that, in all integrations, ${\rm
d}z^\star\rd z/2\pi\ri $ actually means $\rd x\rd y/\pi$.
After the insertions, the propagator becomes a $2(N-1)$--fold
integral over the whole phase space
\begin{align}
  \label{glg82}
  K(z'',t;z',0) &= \int \Bigl\{ \prod_{j=1}^{N-1}\frac{\rd^2z_j}{\pi}
                   \Bigr\} \prod_{j=0}^{N-1} \Bigl\{ \langle z_{j+1} | 
                   \re^{-\frac{\ri }{\hbar}
                   \hat{H}(t_j)\tau} | z_j \rangle 
                   \Bigr\} \nonumber\\
                &= \int \Bigl\{ \prod_{j=1}^{N-1}\frac{\rd z^\star _j\rd 
                   z_j}{2\pi\ri }\Bigr\} \re^{f(z^\star ,z)}
\end{align}
where $f$ is defined by
\begin{equation}
  f(z^\star ,z) := \ln \left[ \prod_{j=0}^{N-1} \Bigl\{ \langle z_{j+1} | 
                  \re^{-\frac{\ri }{\hbar}\hat{H}(t_j)
                  \tau} | z_j \rangle \Bigr\} \right]
\end{equation}
with $z := (z_0,z_1,\ldots,z_N)$ and its complex conjugate $z^\star :=
(z_0^\star ,z_1^\star ,\ldots,z_N^\star )$.
The reason for writing $z^\star$ separately from $z$ as arguments of $f$ is
that they must be considered independent variables, because we are integrating
over the two variables $x_j$ and $y_j$ for each $j$.  Eventually, when we carry
out the stationary exponent approximation, each of these two real variables
will be allowed to become complex, which will result in {\it four} real
variables for each $j$.
Following Klauder \cite{Klau78}, we transform the integrand
$\re^{f(z^\star,z)}$ as follows
\begin{align}
  \label{glg52}
  \re^{f(z^\star ,z)} 
               &\approx \prod_{j=0}^{N-1} \bigl \langle z_{j+1} 
                        \bigl| 1 - \frac{\ri \tau}{\hbar}
                        \hat{H}(t_j) \bigr| z_j 
                        \bigr \rangle \nonumber\\
                       &= \prod_{j=0}^{N-1} \langle z_{j+1} | z_j \rangle 
                        \left( 1 - \frac{\ri \tau}{\hbar}\frac{\langle 
                        z_{j+1}|\hat{H}(t_j)|z_j\rangle }
                        {\langle z_{j+1}|z_j\rangle } 
                        \right) \nonumber\\
                       &\approx \left[ \prod_{j=0}^{N-1} \langle z_{j+1}|z_j
                        \rangle \right] \cdot \exp \left\{\sum_{j=0}^{N-1}- 
                        \frac{\ri \tau }{\hbar}{\cal H}_{j+1,j}\right\}
\end{align}
with the abbreviation
\begin{equation}
  \label{glg40}
  {\cal H}_{j+1,j} \equiv {\cal H}(z_{j+1}^\star,z_j;t_j) 
            :=\frac{\langle z_{j+1}|\hat{H}(t_j)|z_j\rangle }
                   {\langle z_{j+1}|z_j\rangle } \; .
\end{equation}
Using the coherent state overlap formula
\begin{equation}
  \label{glg50}
 \langle z_{j+1} | z_j \rangle 
                        = \exp\left\{ -\frac{1}{2}|z_{j+1}|^2
                        + z_{j+1}^\star z_j 
                        - \frac{1}{2}|z_j|^2\right\}
\end{equation}
we write $\re^{f(z^\star ,z)}$ as
\begin{align}
  \label{glg2}
   \re^{f(z^\star ,z)}
                       &= \exp\sum_{j=0}^{N-1} \left\{ -\frac{1}{2}|z_{j+1}|^2
                        + z_{j+1}^\star z_j - \frac{1}{2}|z_j|^2-\frac{\ri 
                        \tau }{\hbar}{\cal H}_{j+1,j}\right\} \nonumber\\
                       &= \exp\sum_{j=0}^{N-1} \left\{ \frac{1}{2}(
                        z_{j+1}^\star - z_j^\star )z_j - \frac{1}{2}
                        z_{j+1}^\star (z_{j+1} - z_j)-\frac{\ri \tau }
                        {\hbar}{\cal H}_{j+1,j}\right\} .
\end{align}

Later the limit $N \rightarrow \infty$ (respectively $\tau \rightarrow 0$) will
be taken. Then the above summations will turn into integrals, and expressions
(\ref{glg81}) to (\ref{glg52}) would appear to be exact, were it not for the
well--known problems attached to the meaning of such functional integrals. We
see, however, that the treatment of time--dependent systems is (almost)
identical to that of time--independent ones.


\subsection{The Stationary Exponent Approximation}
  \label{kap2_2}

In the semiclassical limit of small $\hbar$ we can approximate the integral
(\ref{glg82}), with \refg{glg2}, by looking for the places where the exponent
$f$ is stationary and replacing it in their vicinity by a quadratic form of its
variables $(z^\star ,z)$.  We call this the stationary exponent approximation
or the gaussian approximation. 
In the literature it is often referred to as the
stationary phase approximation or the steepest descent approximation.  Strictly
speaking, neither of these two names is quite correct.  Our exponent $f$ is
complex, therefore it is not a phase.  And ``steepest descent'' refers to a
geometrical interpretation for the case of a single complex variable.

Our approximation method involves going into the complex plane for the
variables $x_j$ and $y_j$ which are intrinsically real.  Therefore it
is important to keep clearly in mind why we go through so many
developments where we treat them as complex.  The integrals we are
after are over the {\it real} variables $x_j,y_j$. Anything else we
may do to calculate them is just mathematical tricks.  The trick of
going into the complex plane works only with analytic functions.
Hence, when we have to do an integral over real $x_j$ and $y_j$, we
shall first make sure that the function is analytic, and then we shall
see what happens to this analytic function when we let the variables
be complex.  These words of caution may seem superfluous at this time,
but they will turn out to be crucial later, when the need for further
simplification arises.  As we shall see in Sec. 3, confusion about
this point seems to be what led Grossmann and Xavier \cite{Gros98a}
into error.  As a start, one should test this analyticity requirement
for the present integral (\ref{glg82}).  We already assumed in
subsec. 2.1 that the operator $\hat{H}$ could be approximated by a
polynomial in the operators $\hat{a}^\dagger$ and $\hat{a}$.  When we take
its matrix element between the bra $\langle z_{j+1}|$ and the ket
$|z_j\rangle$, we have to do an integral of the type
\begin{align}
  \label{glgmb2}
\langle z_{j+1}|\hat{H}(t_j)|z_j\rangle = \int dx \langle z_{j+1}|x\rangle
\left(\mbox{Polynomial in $x$ and $\frac{d}{dx}$}\right)\langle x|z_j\rangle
\end{align}
where the two wave functions are given by (\ref{glgmb1}) and its complex
conjugate.  The integral produces an analytic function of $z_{j+1}^\star$ and
$z_j$, with the additional factor
\begin{align}
  \label{glgmb3}
\exp\left[-\frac{1}{2}(z_{j+1}^\star z_{j+1} + z_j^\star z_j)\right]
\end{align}
which is the only place where the other two variables, $z_{j+1}$ and
$z_j^\star$, occur.  According to (\ref{glg40}), this must then be divided by
$\langle z_{j+1}|z_j\rangle$, which is given in (\ref{glg50}) and which
contains the same factor (\ref{glgmb3}), times another, never vanishing,
analytic function of $z_{j+1}^\star$ and $z_j$.  When the quotient is taken,
the factor (\ref{glgmb3}) cancels out.  Hence the ``effective Hamiltonian
function'' ${\cal H}_{j+1,j}$ or ${\cal H}(z_{j+1}^\star,z_j;t_j)$ of
\refg{glg40} is an analytic
function of the variables $z_{j+1}^\star$ and $z_j$ separately, and {\it it
does not contain the other two variables $z_{j+1}$ and $z_j^\star$ at all !}

Thus, the basic idea is to approximate the argument of the exponential in
\refg{glg2} by a second order Taylor expansion in the vicinity of the 
stationary
trajectory. The resulting quadratic form in the exponent leads to a gaussian
integral which can be done exactly.  There may be more than one classical
trajectory between the end points, each with its own contribution, which leads
to a sum.  For clarity during the derivation, this sum will not be written
explicitly.  We find the stationary points by requiring the vanishing of the
derivatives of $f$ with respect to $z$ and $z^\star$ separately, as mentioned
earlier
\begin{alignat}{3}
  \label{glg3}
    \frac{\partial f}{\partial z_j} 
       &=z_{j+1}^\star - z_j^\star - \frac{\ri \tau}{\hbar}\frac{\partial 
        {\cal H}_{j+1,j}}{\partial z_j}
       &=0 \; ; \qquad 
       &j = 1,\ldots,N-1 \nonumber\\
    \frac{\partial f}{\partial z_{j+1}^\star} 
       &= - z_{j+1} + z_j - \frac{\ri \tau}{\hbar}\frac{\partial 
        {\cal H}_{j+1,j}}{\partial z_{j+1}^\star}
       &=0 \; ;  \qquad 
       &j = 0,\ldots,N-2 \; .
\end{alignat}
We introduce new integration variables $\eta$ and $\eta^\star$ which describe
the deviations from the points of stationary exponent: $z \rightarrow z +
\eta\;,\ z^\star \rightarrow z^\star + \eta^\star$, with the boundary
conditions
\begin{equation}
  \label{glg41}
  \eta_0=\eta_0^\star=\eta_N=\eta_N^\star=0 \; .
\end{equation}

Now the exponent in \refg{glg2}
\begin{align}
  \label{glg104}
    f(z^\star + \eta^\star, z+\eta)
      &= \sum_{j=0}^{N-1}\Bigl\{\frac{1}{2}(z_{j+1}^\star +\eta_{j+1}^\star 
         - z_j^\star -\eta_j^\star )(z_j+\eta_j)\nonumber\\
      &\qquad - \frac{1}{2}(z_{j+1}^\star +\eta_{j+1}^\star)(z_{j+1}+
         \eta_{j+1} - z_j-\eta_j)\nonumber\\
      &\qquad -\frac{\ri \tau }{\hbar}{\cal H}_{j+1,j}(z_{j+1}^\star
         +\eta_{j+1}^\star ,z_j+\eta_j) \Bigr\}
\end{align}
will be expanded into a Taylor series in $(\eta^\star ,\eta)$ around the
stationary points $(z^\star,z)$ up to second order:
\begin{align}
  \label{glg4}
  \begin{split}
    f(z^\star + \eta^\star, z+\eta)
      &\approx \sum_{j=0}^{N-1} \Bigl\{ 
         \frac{1}{2}(z_{j+1}^\star - z_j^\star )z_j 
         -\frac{1}{2}z_{j+1}^\star (z_{j+1} - z_j)\\
      &\qquad -\frac{\ri \tau }{\hbar}{\cal H}_{j+1,j}(z_{j+1}^\star ,z_j) 
         +\frac{1}{2}(z_{j+1}^\star - z_j^\star )\eta_j\\
      &\qquad+\frac{1}{2}(\eta_{j+1}^\star-\eta_j^\star)z_j
         -\frac{1}{2}z_{j+1}^\star(\eta_{j+1}-\eta_j)\\
      &\qquad -\frac{1}{2}\eta_{j+1}^\star(z_{j+1}-z_j)
         -\frac{\ri \tau }{\hbar}\frac{\partial {\cal H}_{j+1,j}}{\partial 
         z_j}\,\eta_j -\frac{\ri \tau }{\hbar}\frac{\partial {\cal H}_{j+1,j}}
         {\partial z_{j+1}^\star}\,\eta_{j+1}^\star\\
      &\qquad +\frac{1}{2}(\eta_{j+1}^\star-\eta_j^\star)\eta_j - \frac{1}{2}
         \eta_{j+1}^\star(\eta_{j+1}-\eta_j)\\
      &\qquad - \frac{\ri \tau }{2\hbar}\left[ \frac{\partial^2
        {\cal H}_{j+1,j}}{\partial z_j^2}\,\eta_j^2
        +2\frac{\partial^2 {\cal H}_{j+1,j}}{\partial  
         z_{j+1}^\star \partial z_j}\,\eta_{j+1}^\star\,\eta_j 
         +\frac{\partial^2 {\cal H}_{j+1,j}}
         {\partial z_{j+1}^{\star 2}}\eta_{j+1}^{\star 2} \right] \Bigr\}
  \end{split}\\
  \begin{split}
  \label{glg4a}
        =f(z^\star,z) + \sum_{j=0}^{N-1} &\Bigl\{
      \quad  \left(z_{j+1}^\star-\frac{1}{2}z_j^\star-\frac{\ri \tau }
         {\hbar}\frac{\partial {\cal H}_{j+1,j}}{\partial z_j}\right)\eta_j - 
         \frac{1}{2}z_{j+1}^\star\eta_{j+1}\\
      &\quad +\left(z_j-\frac{1}{2}z_{j+1}-\frac{\ri \tau }{\hbar}
         \frac{\partial {\cal H}_{j+1,j}}{\partial z_{j+1}^\star}\right)
         \eta_{j+1}^\star - \frac{1}{2}z_j\eta_j^\star\\
      &\quad - \frac{1}{2}\eta_{j+1}^\star\eta_{j+1} + 
         \eta_{j+1}^\star\eta_j - \frac{1}{2}\eta_j^\star\eta_j\\
    &\quad -\frac{\ri \tau }{2\hbar}\Bigl[ \frac{\partial^2 {\cal H}_{j+1,j}}
         {\partial z_j^2}\eta_j^2 +2\frac{\partial^2 {\cal H}_{j+1,j}}
         {\partial z_{j+1}^\star \partial z_j}\,\eta_{j+1}^\star\,\eta_j 
         +\frac{\partial^2 {\cal H}_{j+1,j}}
         {\partial z_{j+1}^{\star 2}}\,\eta_{j+1}^{\star 2} \Bigr] \Bigr\} .
  \end{split}
\end{align}
The terms of first order (second and third line of \refg{glg4a}) vanish,
when the boundary conditions (\ref{glg41}) are taken into account,
because of the stationary exponent conditions (\ref{glg3}). Inserting
\refg{glg4a} into \refg{glg82} yields, in view of \refg{glg41},
\begin{multline}
  \label{glg6}
  K(z'',t;z',0) = \re^{f(z^\star,z)} 
         \mbox{\large $\displaystyle \int$} \left\{ \prod_{j=1}^{N-1} 
         \frac{\rd\eta_j^\star \rd\eta_j}{2\pi \ri }
     \right\}{\rm exp} \sum_{j=0}^{N-1}\Bigl\{ -\frac{\ri\tau}{2\hbar}\,
         \frac{\partial^2 {\cal H}_{j+1,j}}
         {\partial z_j^2}\,\eta_j^2-\frac{\ri\tau}{2\hbar}\,\frac{\partial^2 
         {\cal H}_{j+1,j}}{\partial z_{j+1}^{\star 2}}\,\eta_{j+1}^{\star 2}\\
     -\eta_j^\star \eta_j+ \left(1- \frac{\ri \tau }{\hbar}
         \frac{\partial^2 {\cal H}_{j+1,j}}{\partial z_{j+1}^\star 
         \partial z_j} \right) \eta_{j+1}^\star\eta_j \Bigr\} .
\end{multline}

\subsection{Performing the Integrals}
  \label{kap2.9}

At this point we start to carry out the integrations over the variables
$\eta_j,\eta_j^\star$. We have two ways of doing this.  In the body of the
paper, we perform successively $\iint\rd \eta_1^\star\rd \eta_1$, $\iint\rd
\eta_2^\star\rd \eta_2$,\ \ etc\dots, deriving eventually a recursion relation,
which becomes a nonlinear differential equation when we go to the limit of
continuous variables.  In the appendix , we do it by writing the multiple
integral in terms of the determinant of the quadratic form, we calculate this
determinant with a pair of recursion relations, which turn into linear
differential equations in the limit.  The two methods are quite different, but
they give the same result.  The reader who wants to save time does not have to
study both.  In the appendix, then, we go on to use the same method to
calculate a different path integral which arises in section 3.

Here, we calculate the integrals $\iint\rd\eta_j^\star\rd\eta_j$
by applying the general formula
\begin{align}
  \label{mb1}
  \int_{-\infty}^\infty \int_{-\infty}^\infty\frac{\rd x\rd y}{\pi}
         \re^{A_1x^2+A_2y^2+A_3xy+B_1x+B_2y} 
  = \frac{1}{\sqrt{A_1A_2-A_3^2/4}} \, 
         \re^{\frac{-B_1^2A_2-B_2^2A_1+B_1B_2A_3}{4A_1A_2 - A_3^2} }
\end{align}
which is correct if the integrations are done along the two real axes, as they
would be if $x$ and $y$ are the real and imaginary parts of $\eta$,
respectively.  Two comments need to be made here. First, the integral must
converge. If we call $-\mu_1$ and $-\mu_2$ the eigenvalues of the symmetric
matrix of the quadratic form in the exponent on the lefthand side, convergence
requires that $\mu_1$ and $\mu_2$ both have a nonnegative real part. Second,
the phase of the square root needs to be defined, since $A_1, A_2, A_3$ are
complex numbers. This square root is also equal to the product
$\sqrt{\mu_1}\sqrt{\mu_2}$, and the phase $\phi_i$ of each $\sqrt{\mu_i}$ must
be chosen to satisfy $-\pi/4\le\phi_i\le\pi/4$. This phase rule is extremely
important in determining the phase of the semiclassical propagator when one
works in configuration space. For the present case of the semiclassical
coherent--state propagator, it is less crucial because the phase can usually be
determined by appealing to continuity in time.

We must now rewrite this formula in terms of the variables $\eta$ and
$\eta^\star$ rather than $x$ and $y$. The transformation of variables is simple
enough, but the paths of integration are totally changed, and the associated
conditions need to be restated. To simplify notations a little, we call the
variables $u$ and $v$ instead of $\eta$ and $\eta^\star$, respectively. Then
the new formula is
\begin{align}
  \label{glg44}
  \iint\frac{\rd u\rd v}{2\pi\ri }
         \re^{a_1u^2+a_2v^2+a_3uv+b_1u+b_2v}
  = \frac{1}{\sqrt{a_3^2-4a_1a_2}} \, 
         \re^{\frac{+b_1^2a_2+b_2^2a_1-b_1b_2a_3}{a_3^2-4a_1a_2} } \; .
\end{align}
The integral is convergent if and only if the two numbers
$$
\mu_{1,2} = -a_3 \pm 2\sqrt{a_1a_2}~,
$$
which are the negatives of the eigenvalues of the symmetric matrix in the
exponent on the lefthand side of \refg{mb1}, {\em not} (\ref{glg44}), both have
a nonnegative real part.  The square root in \refg{glg44} is again the product
$\sqrt{\mu_1}\sqrt{\mu_2}$ and, once again, the phase $\phi_i$ of each
$\sqrt{\mu_i}$ must be chosen to satisfy $-\pi/4\le\phi_i\le\pi/4$.

For the $\eta_1^\star,\eta_1$ integration, the parameters are
\begin{equation*}
  a_1=-\frac{\ri\tau}{2\hbar}\frac{\partial^2 {\cal H}_{2,1}}
         {\partial z_1^2} \qquad 
  a_2=-\frac{\ri\tau}{2\hbar}\frac{\partial^2 {\cal H}_{1,0}}
         {\partial z_1^{\star 2}} 
     := X_1 \qquad 
  a_3=-1 \qquad 
\end{equation*}
\begin{equation}
  \label{mb2}
  b_1=\left(1- \frac{\ri \tau }{\hbar}\frac{\partial^2 {\cal H}_{2,1}}
{\partial z_2^\star\partial z_1} \right)\eta_2^\star \qquad 
  \mbox{and} \qquad b_2 = 0 \; .
\end{equation}
Convergence is assured since $-a_3$ is real, positive, and much larger than
$a_1$ and $a_2$ ($\tau$ is arbitrarily small). The correct branch of the
square root is the one whose phase is close to 0. The result is
\begin{align}
  \label{glg42}
  \begin{split}
  K(z_N,t;z_0,0) 
    &=\frac{\re^{f(z^\star,z)}}{\displaystyle 
         \sqrt{1+2\ri \frac{\tau}{\hbar}
         \frac{\partial ^2{\cal H}_{2,1}}{\partial z_1^2}\,X_1}}\cdot
         \mbox{\Large $\displaystyle \int$} 
         \left[ \prod_{j=2}^{N-1} \frac{d\eta_i^\star d\eta_i}
         {2\pi \ri }\right] \\ 
    &\qquad \exp \left\{ \frac{\displaystyle  \left( 
         1-\frac{\ri \tau}{\hbar} \frac{\partial^2{\cal H}_{2,1}}
         {\partial z_2^\star \partial z_1}\right)^2X_1\,\eta_2^{\star 2}}
     {\displaystyle 1 + 2\frac{\ri \tau}{\hbar}\frac{\partial^2{\cal H}_{2,1}}
         {\partial z_1^2}\,X_1} \right\} \\
    &\qquad \exp \Bigl\{ \sum_{j=2}^{N-1} -\frac{\ri \tau }{2\hbar}
         \frac{\partial^2 {\cal H}_{j+1,j}}{\partial z_j^2}\,\eta_j^2-
         \frac{\ri \tau }{2\hbar}\frac{\partial^2 {\cal H}_{j,j-1}}
         {\partial z_j^{\star 2}}\,\eta_j^{\star 2}-\eta_j^\star \eta_j\\
    &\qquad\qquad + \left(1- \frac{\ri \tau }{\hbar}
      \frac{\partial^2 {\cal H}_{j+1,j}}{\partial z_{j+1}^\star \partial z_j } 
         \right)\eta_{j+1}^\star\eta_j \Bigr\}.
  \end{split}
\end{align}
The second set of integrals, $\int\int\rd \eta_2^\star\rd \eta_2$, is
done again with \refg{glg44}, but now $a_2$ becomes more complicated and is
expressed as a function of $X_1$. This kind of behavior continues at each stage
of the integrations. If we call $X_j$ the value of $a_2$ when we do the
integrals $\int\int\rd \eta_j^\star\rd \eta_j$ according to
\refg{glg44}, then we find the following parameters
\begin{equation}
  \label{glg46}
  a_1=-\frac{\ri\tau}{2\hbar}\frac{\partial^2 {\cal H}_{j+1,j}}
         {\partial z_j^2} \qquad 
  a_3=-1 \qquad 
  b_1=\left(1- \frac{\ri \tau }{\hbar}\frac{\partial^2 {\cal H}_{j+1,j}}
{\partial z_{j+1}^\star\partial z_j} \right)\eta_{j+1}^\star \qquad 
  b_2 = 0
\end{equation}
while the formula for $a_2$ becomes a recursion relation for $X_j$
\begin{align}
  \label{glg7}
  X_0 &= 0 \qquad \mbox{(initial condition)} \nonumber\\
  X_j &= -\frac{\ri \tau }{2\hbar}\frac{\partial^2 {\cal H}_{j,j-1}}
         {\partial z_j^{\star 2}} + \frac{\displaystyle \left( 
         1-\frac{\ri \tau}{\hbar} \frac{\partial^2{\cal H}_{j,j-1}}
         {\partial z_j^\star \partial z_{j-1} }\right)^2}
         {\displaystyle 1 + 2\frac{\ri \tau}{\hbar}
         \frac{\partial^2{\cal H}_{j,j-1}}{\partial z_{j-1}^2}
         X_{j-1}}X_{j-1}~;\qquad j=1,\ldots,N-1~.
\end{align}
Once all integrations are done, the result is
\begin{equation}
  \label{glg43}
  K(z_N,t;z_0,0) = \re^{f(z^\star,z)}\prod_{j=1}^{N-1} \frac{1}
         {\sqrt{\displaystyle 1+2\ri \frac{\tau}{\hbar}
         \frac{\partial ^2{\cal H}_{j+1,j}}{\partial z_j^2}X_j}}
\end{equation}
with $X_j$ satisfying \refg{glg7}. Once again, the phase of each square root
should be chosen close to 0.

\subsection{Continuous Variables}
  \label{kap2.2}

The time has come to perform the limit $N \rightarrow \infty$ (respectively
$\tau \rightarrow 0$). This gets rid of the approximations associated with the
time discretization in \refg{glg2}, and the discrete recursion formula
(\ref{glg7}) becomes a solvable differential equation.  The stationary phase
conditions (\ref{glg3}) are in this limit identical to Hamilton´s equations
\begin{align}
  \label{glg8}
  \dot{z}^\star &= \frac{\ri }{\hbar}
                   \frac{\partial {\cal H}}{\partial z}\nonumber\\
  \dot{z}       &= -\frac{\ri }{\hbar}
                   \frac{\partial {\cal H}}{\partial z^\star}
\end{align}
where ${\cal H}(z^\star,z,t)$ is the limit of ${\cal H}_{j+1,j}$, \refg{glg40},
and is simply given by
\begin{equation}
\label{mb4}
{\cal H}(z^\star,z,t) = \langle z|\hat{H}(t)|z \rangle \; .
\end{equation}

The question of boundary conditions presents us with a grave problem at this
point. We have been approximating the propagator going in time $t$ from
coherent state $(q',p')$ to coherent state $(q'',p'')$, and our approximation
seems to involve the classical path between the two points in phase space. But
there is no such classical path, real or complex, between these two points,
generically speaking! The unique classical trajectory which goes through
$(q',p')$ at time 0 does not in general go through $(q'',p'')$ at time $t$. We
have too many boundary conditions!

One way out of this quandary was shown by \cite{Weis83}. Actually, $z_N$ and
$z_0^\star$ do not enter the equations of motion at all. Neither one occurs in
eqs. (\ref{glg2}) or (\ref{glg3}). Thus, the problem is really to find a
classical path going from $z_0$ to $z_N^\star$ in time $t$, and such a path
does exist, but it is usually complex, in spite of the fact that $(q_0,p_0)$
and $(q_N,p_N)$ are real. The path will be real only if $(q_0,p_0)$ and
$(q_N,p_N)$ happen to be on the same classical trajectory.
Given that the intermediate values of $q$ and $p$ will now be
both complex usually, it makes no sense to retain the notations $z$
and $z^\star$, which can only lead to confusion. We replace these by
the two complex variables $u$ and $v$, which are now manifestly independent
\begin{align}
  \label{glg9}
z &\rightarrow  u = \frac{1}{\sqrt{2}}\left(\frac{q}{b}
       +\ri \,\frac{p}{c} \right) \nonumber\\
z^\star &\rightarrow  v = \frac{1}{\sqrt{2}}\left(\frac{q}{b}
       -\ri \,\frac{p}{c} \right).
\end{align}
Note also the inverse formulae
\begin{align}
  \label{glg9a}
q &= \frac{b}{\sqrt{2}}(u+v)    \nonumber\\
p &= -\frac{\ri c}{\sqrt{2}}(u-v) \; .
\end{align}
The differential equations (\ref{glg8}) are now
\begin{align}
  \label{glg10}
  \ri \hbar\dot{u} &= + \frac{\partial {\cal H}}{\partial v}\nonumber\\
  \ri \hbar\dot{v} &= - \frac{\partial {\cal H}}{\partial u}
\end{align}
with the boundary conditions
\begin{align}
\label{mb12}
u(0) &= z'~, \qquad v(t) = {z''}^\star~, \nonumber  \\
v(0) &= \quad \mbox{ nothing special} \quad \neq {z'}^\star~,  \\
u(t) &= \quad \mbox{ nothing special} \quad \neq z''~. \nonumber
\end{align}
$v(0)$ and $u(t)$ come out of the calculation and do not have any simple
relation to $z'$ and $z''$, except in the special case when there exists a real
trajectory going from $z'$ to $z''$ in time $t$. Then and only then do we get
$v(0) = {z'}^\star$ and $u(t) = z''$. Otherwise, the end points of the
classical path are truly complex in phase space. To complete the definitions,
we rename these end points as follows
\begin{equation}
\label{mb7}
u(0)\equiv u',\qquad v(0)\equiv v',\qquad u(t)\equiv u'',\qquad v(t)\equiv v''.
\end{equation}
Given the change of variables, we note the following differentiation
rules, which follow from eqs. (\ref{glg9}) and (\ref{glg9a}) and will be
needed later
\begin{alignat}{3}
  \label{glg12}
  \frac{\partial}{\partial q} &= \frac{\partial v}{\partial q}\frac{\partial}
         {\partial v}+\frac{\partial u}{\partial q}\frac{\partial}{\partial u}
       &\, = \, &\frac{1}{\sqrt{2}\, b}\biggl(\frac{\partial}{\partial v} 
         +\frac{\partial}{\partial u}\biggr)\nonumber\\
  \frac{\partial}{\partial p} &= \frac{\partial v}{\partial p}\frac{\partial}
         {\partial v} +\frac{\partial u}{\partial p}\frac{\partial}{\partial u}
        &\, = \, &\frac{\ri }{\sqrt{2}\, c}\biggl(
         -\frac{\partial}{\partial v} +\frac{\partial}{\partial u}\biggr)
\end{alignat}
\begin{align}
  \label{glg13}
  \frac{\partial}{\partial v} &= \frac{1}{\sqrt{2}}\biggl(b\frac{\partial}
         {\partial q}+\ri c\frac{\partial}{\partial p} \biggr)\nonumber\\
  \frac{\partial}{\partial u} &= \frac{1}{\sqrt{2}}\biggl(b\frac{\partial}
         {\partial q}-\ri c\frac{\partial}{\partial p} \biggr) .
\end{align}

Let us rewrite $f(z^\star,z)$, \refg{glg2}, in the continuous limit and in
terms of the new variables. Since $f$ can also be written
\begin{equation*}
\tau \sum_{j=0}^{N-1}\left\{ \frac {z_{j+1}^\star - z_j^\star}{2\tau} z_j 
 -z_{j+1}^\star \frac{z_{j+1} - z_j}{2\tau}
-\frac{\ri }{\hbar}{\cal H}_{j+1,j}\right\}
\end{equation*}
one might think that, in the limit $\tau \rightarrow 0$, this would reduce to
\begin{equation}
\label{mb3}
\int_0^t \rd t' \left[ \frac{1}{2} (\dot{v} u - \dot{u} v)
-  \frac{\ri}{\hbar} {\cal H}(u,v,t') \right]
\end{equation}
but that would be a mistake. Recalling the earlier discussion, we realize that
\refg{mb3} assumes that $u(t) = u_N$, which is not equal to $z''$, and it also
assumes that $v(0) = v_0$, which is not equal to ${z'}^\star$. But in
expression (\ref{glg2}), $u_N$ or $z_N$ {\em was} $z''$, and $v_0$ or
$z_0^\star$ {\em was} ${z'}^\star$. To correct this mistake, we must take out
from the sum the two terms containing $u_N$ and $v_0$, namely $-\frac{1}{2} v_N
u_N$ and $-\frac{1}{2} v_0 u_0$, and replace them by their correct values,
namely $-\frac{1}{2}{|z''|}^2$ and $-\frac{1}{2}{|z'|}^2$ . Consequently, the
value of $f$ in the limit $\tau \rightarrow 0$ should be
\begin{equation}
\label{glg15}
f = \int_0^t \rd t' \left[ \frac{1}{2} (\dot{v} u - \dot{u} v)
            - \frac{\ri}{\hbar} {\cal H}(u,v,t') \right]  
+ \frac{1}{2} (v''u'' + v'u') - \frac{1}{2} ({|z''|}^2 + {|z'|}^2) \; .
\end{equation}

Next we rewrite the product in \refg{glg43}, performing the limit $N
\rightarrow \infty$ and using $\ln (1+x) = x + O(x^2)$~:
\begin{align}
  \label{glg14}
  \lim_{N\rightarrow\infty} \prod_{j=1}^{N-1} \left\{ 1+2\ri \,\frac
          {\tau}{\hbar}\frac{\partial ^2{\cal H}}{\partial u_j^2}\,X_j
          \right\}^{-\frac{1}{2}} 
      &= \lim_{N\rightarrow\infty} \exp\biggl\{ - \frac{1}{2}\sum_{j=1}^{N-1} 
          \ln \Bigl( 1+2\ri \frac{\tau}{\hbar}\frac{\partial ^2{\cal H}}
          {\partial u_j^2}\,X_j \Bigr)\biggr\}\nonumber\\
      &= \lim_{N\rightarrow\infty} \exp \biggl\{ - \ri \frac{\tau}
         {\hbar} \sum_{j=1}^{N-1}\frac{\partial ^2{\cal H}}{\partial u_j^2}X_j
          \biggr\}\nonumber\\
      &= \exp \biggl\{ - \frac{\ri }{\hbar} \int\limits_0^t \rd t'
          \, \frac{\partial^2 {\cal H}}{\partial u^2}(t') X(t') \biggr\} \; .
\end{align}
Altogether, the propagator is
\begin{align}
  \label{glg18}
  K(z'',t;z',0) 
    &= \exp\biggl\{-\frac{\ri }{\hbar} \int\limits_0^t \rd t'
          \frac{\partial^2 {\cal H}}{\partial u^2}(t') X(t') \biggr\}
          \exp \biggl\{ \int\limits_0^t \rd t'
          \left[ \frac{1}{2}\left(\dot{v}u - 
          \dot{u}v\right) -\frac{\ri }{\hbar}{\cal H} \right] \nonumber\\
    &\qquad\qquad + \frac{1}{2}\left(v'u' + v''u''\right) -
          \frac{1}{2}\left(|z'|^2 + |z''|^2\right) \biggr\} \; .
\end{align}

We still have to write the continuous form of the discrete recursion formula
(\ref{glg7}) for $X(t)$. With $\frac{(1-ax)^2}{1+bx}=1-(2a+b)x+O(x^2)$
\refg{glg7} gives
\begin{equation}
  \label{glg16}
  X_j \approx -\frac{\ri \tau}{2\hbar}\frac{\partial^2{\cal H}}{\partial 
          v_j^2} + X_{j-1} - \biggl( 2\frac{\ri \tau}{\hbar}
          \frac{\partial^2{\cal H}}{\partial u_{j-1}\partial v_j} +2
          \frac{\ri \tau}{\hbar} \frac{\partial^2{\cal H}}{\partial u_{j-1}^2} 
          X_{j-1}\biggr) X_{j-1} \; .
\end{equation}
In the limit $N \rightarrow \infty$ this leads to the nonlinear differential
equation
\begin{align}
  \label{glg17}
  \dot{X}(t) &= -\frac{\ri }{2\hbar}\frac{\partial^2{\cal H}}{\partial 
          v^2}-2\frac{\ri }{\hbar}\frac{\partial^2{\cal H}}{\partial 
          u\partial v}X(t) -2\frac{\ri }{\hbar} \frac{\partial^2{\cal H}}
          {\partial u^2} X^2(t)\\
\intertext{with the initial condition}
  \label{glg45}
  X(0) &= 0~.
\end{align}
To solve this, we consider small variations in the solutions of Hamilton's
equations \refg{glg10} around a given solution $v(t)$, $u(t)$:
\begin{align}
  \label{glg30}
  \delta \dot{u} &= -\frac\ri {\hbar}\frac{\partial^2{\cal H}}{\partial u
          \partial v} \,\delta u - \frac\ri {\hbar}\frac{\partial^2{\cal H}}
          {\partial v^2}\,\delta v\nonumber\\
  \delta \dot{v} &= +\frac\ri {\hbar}\frac{\partial^2{\cal H}}{\partial u^2} 
          \,\delta u + \frac\ri {\hbar}\frac{\partial^2{\cal H}}{\partial u
          \partial v} \,\delta v \; .
\end{align}
It can be seen that the solution of \refg{glg17} is $X = \frac{1}{2}
\frac{\delta u}{\delta v}$~. The time derivative is
\begin{align}
  \label{glg31}
  \dot{X}&=\frac{1}{2}\frac{\delta \dot{u}}{\delta v} - \frac{1}{2}
          \frac{\delta u}{\delta v^2} \,\delta \dot{v}\nonumber\\
         &=-\frac\ri {2\hbar}\frac{\partial^2{\cal H}}{\partial u\partial v}
          \frac{\delta u}{\delta v}-\frac\ri {2\hbar}\frac{\partial^2{\cal H}}
          {\partial v^2} - \frac{1}{2}\frac{\delta u}{\delta v}\left[
        \frac\ri {\hbar}\frac{\partial^2{\cal H}}{\partial u^2}\frac{\delta u}
          {\delta v}+\frac\ri {\hbar}\frac{\partial^2{\cal H}}{\partial u
          \partial v} \right]\nonumber\\
         &=-\frac\ri {2\hbar}\frac{\partial^2{\cal H}}{\partial v^2}-\frac\ri 
          {\hbar}\frac{\partial^2{\cal H}}{\partial u\partial v}\frac{\delta u}
          {\delta v}-\frac\ri {2\hbar}\frac{\partial^2{\cal H}}{\partial u^2}
          \left( \frac{\delta u}{\delta v} \right)^2\nonumber\\
         &=-\frac\ri {2\hbar}\frac{\partial^2{\cal H}}{\partial v^2}-2\frac\ri 
          {\hbar}\frac{\partial^2{\cal H}}{\partial u\partial v}X-2\frac\ri 
          {\hbar}\frac{\partial^2{\cal H}}{\partial u^2}X^2
\end{align}
which agrees with the differential equation (\ref{glg17}). The initial
condition $X(0) = 0$ can be satisfied by picking $\delta u(0)=\delta u'=0\,,\;
\delta v(0) = \delta v'$ arbitrary. The integrand in the first exponential of
\refg{glg18} can be transformed with the help of \refg{glg30}
\begin{align}
  \label{glg32}
  \frac\ri {\hbar}\frac{\partial^2{\cal H}}{\partial u^2}X 
         &= \frac\ri {2\hbar}\frac{\partial^2{\cal H}}{\partial u^2}
          \frac{\delta u}{\delta v}
          =\frac{1}{2}\frac{\delta \dot{v}}{\delta v}-\frac\ri {2\hbar}
          \frac{\partial^2{\cal H}}{\partial u\partial v}
          =\frac{1}{2}\frac{\rm{d}}{\rm{d}t}\ln \delta v - \frac\ri 
          {2\hbar}\frac{\partial^2{\cal H}}{\partial u\partial v}
\end{align}
so that the first exponent of \refg{glg18} is
\begin{align}
  \label{glg33}
  \exp \left\{ -\frac{\ri}{\hbar}\mbox{\large $\displaystyle 
          \int$}^t_{\!\!\!\!0} \rd t' \frac{\partial^2{\cal H}}{\partial u^2}
          (t' )X(t' ) \right\}
         &=\exp \left\{ -\frac{1}{2}\mbox{\large $\displaystyle 
          \int$}^t_{\!\!\!\!0} \rd t' \left[ \frac{\rd}{\rd t'}
          (\ln \delta v) - \frac{\ri}{\hbar}\frac{\partial^2{\cal H}}{\partial 
          u\partial v} \right]  \right\} \nonumber\\
         &=\sqrt{\frac{\delta v'}{\delta v''}}\exp\left\{ \frac{\ri}
          {2\hbar}\mbox{\large $\displaystyle \int$}^t_{\!\!\!\!0} \rd
          t' \frac{\partial^2{\cal H}}{\partial u\partial v} \right\} .
\end{align}
It is understood that $\delta v'/\delta v''$ is calculated with the initial
condition $\delta u' = 0$. The phase of the square root evolves continuously
with time, starting at 0 for $t=0$. Because all the numbers are generically
complex, $\delta v''(t)$ does not usually have zeroes; therefore the phase is
always well defined, barring an accident. In the end we obtain
\begin{align}
  \label{glg34}
  K(z'',t;z',0) 
         &= \sqrt{\frac{\delta v'}{\delta v''}}\exp\left\{ \frac{\ri}
          {2\hbar}\mbox{\large $\displaystyle \int$}^t_{\!\!\!\!0} \rd
        t' \frac{\partial^2{\cal H}}{\partial u\partial v} \right\}\nonumber\\
     \exp&\biggl\{\int\limits_0^t\rd t'\left[\frac{1}{2}\left(\dot{v}u - 
          \dot{u}v\right) -\frac{{\ri}}{\hbar}{\cal H} \right] 
         + \frac{1}{2}(v''u''+v'u') - \frac{1}{2}(|z'|^2 +|z''|^2)\biggr\} \; .
\end{align}

Just as a plausibility check, consider the limit $t \rightarrow 0$. In this
case, the complex path has $u''=u'=z'$, $v'=v''={z''}^\star$, and therefore
$K(z'',0;z',0)=\exp\{0\}\exp\{0+\frac{1}{2}v'u'+\frac{1}{2}v''u''
-\frac{1}{2}|z'|^2-\frac{1}{2}|z''|^2\}
=\exp\{-\frac{1}{2}|z'|^2+z'{z^\star}''-\frac{1}{2}|z''|^2\}
=\langle z''|z'\rangle$, which is the overlap of two coherent states.

\subsection{The Complex Action}
  \label{kap2.3}

In most discussions of hamiltonian mechanics, one gains much simplicity and
under\-standing by defining the `action', which is the quantity entering in
Hamilton's variational principle and in the Hamilton--Jacobi equation. In the
present problem, this is true also.  The action is complex in most cases, like
the trajectories themselves and like the energy. It is given by the formula
\begin{align}
  \label{glg83}
  S(v'',u',t) : 
         &=\int\limits_0^t \rd t' \left[\frac{\ri \hbar}{2}
          (\dot{u}v-\dot{v}u) - {\cal H}(u,v,t') \right]
          - \frac{\ri \hbar}{2} ( u''v'' + u'v' )\; .
\end{align}
The independent variables are $v''$ and $u'$, the two end variables which
define the classical trajectory, and the time $t$~; $u''$ and $v'$ must be
understood as functions of these three variables.  Note that $u$ and $v$ are
proportional to $1/\sqrt{\hbar }$, since $b$ and $c$ in \refg{glg9} are both
proportional to $\sqrt{\hbar}$. Therefore the only term in $S$ which depends on
$\hbar$ is $\int_0^t{\cal H}(u,v,t')\rd t'$. We shall show in the following
that this $S$ is indeed the correct action for the boundary conditions we have.
Given its definition, $S$ should be an analytic function of its variables most
of the time, since ${\cal H}(u,v)$ is analytic and the velocities $\dot{u}$ and
$\dot{v}$ can be written as derivatives of $\cal H$ using Hamilton's equations
(\ref{glg10}).  It would take an accident in the determination of the classical
trajectory from the boundary conditions to produce a singularity.  Similarly,
the functions $u''$ and $v'$ should be analytic in $v''$ and $u'$ most of the
time;  in fact, according to \refg{mb9}, they are essentially the partial
derivatives of $S$.

Suppose that we make small variations $\delta v''$, $\delta u'$, $\delta t$
in each of the independent variables. This induces variations $\delta u''$,
$\delta v'$ in $u''$ and $v'$. It also induces variations $\delta u(t')$,
$\delta v(t')$ in the trajectory itself. The consequent variation in $S$ is
\begin{align}
\label{mb5}
\delta S &= \int\limits_0^t \rd t'\left[\frac{\ri \hbar}{2}
(v\delta\dot{u}-u\delta\dot{v}-\dot{v}\delta u+\dot{u}\delta v)
-\frac{\partial {\cal H}}{\partial v}\delta v
-\frac{\partial {\cal H}}{\partial u}\delta u\right]  \nonumber   \\
&\qquad -\frac{\ri \hbar}{2}(v''\delta u''+u''\delta v''
 +v'\delta u'+u'\delta v') \\
&\qquad +\delta t\left[\frac{\ri \hbar}{2}(\dot{u}''v''-\dot{v}''u'')
-{\cal H}(u'',v'',t)\right] \; . \nonumber
\end{align}
We do two integrations by parts as follows
\begin{align}
  \label{mb6}
 \int_0^t \rd t' v\delta\dot{u} &= v''\delta u(t)-v'\delta u(0)
                                 - \int_0^t \rd t' \dot{v}\delta u \nonumber \\
-\int_0^t \rd t' u\delta\dot{v} &= -u''\delta v(t)+u'\delta v(0)
                                 + \int_0^t \rd t' \dot{u}\delta v~.
\end{align}
Since the lower limit 0 of the $t'$-integral is not changed, we have
\begin{equation*}
\delta u(0)=\delta u',\qquad \delta v(0)=\delta v' .
\end{equation*}
But the upper limit is changed from $t$ to $t+\delta t$, hence
\begin{equation*}
\delta u(t)=\delta u''-\dot{u}''\delta t,\qquad
\delta v(t)=\delta v''-\dot{v}''\delta t~.
\end{equation*}
Much simplification occurs when we carry this back into \refg{mb5}, and we
find for the variation in S
\begin{align}
\label{mb8}
\delta S &= \int\limits_0^t \rd t'\left[\left(\ri\hbar\dot{u}
-\frac{\partial {\cal H}}{\partial v}\right)\delta v+\left(-\ri\hbar\dot{v}
-\frac{\partial {\cal H}}{\partial u}\right)\delta u\right]  \nonumber   \\
 &\qquad -\ri\hbar(u''\delta v''+v'\delta u')-{\cal H}(u'',v'',t)\delta t \; .
\end{align}
The first line says that paths satisfying Hamilton's equations (\ref{glg10})
have a stationary action $S$ when the independent variables $v'',u',t$ are
held fixed. The second line says that, for such a classical trajectory, the
derivatives of $S$ with respect to these variables are
\begin{equation}
\label{mb9}
\frac{\partial S}{\partial v''}= -\ri\hbar u'',\qquad
\frac{\partial S}{\partial u'}= -\ri\hbar v',\qquad
\frac{\partial S}{\partial t}= -{\cal H}(u'',v'',t)~.
\end{equation}
In the special case of a time--independent hamiltonian, ${\cal H}(u'',v'')$
is the constant energy ${\cal E}$, and the last equality becomes
\begin{equation}
\label{mb10}
\frac{\partial S}{\partial t}= -{\cal E}~.
\end{equation}

We are now able to express most of the coherent--state propagator,
\refg{glg34}, in terms of the action function. By \refg{mb9} we have
\begin{equation}
\label{mb11}
  \frac{\delta v'}{\delta v''}
      =\frac{\ri }{\hbar}\frac{\partial^2 S}{\partial u'\partial v''} \; .
\end{equation}
Therefore the coherent--state propagator can be written
\begin{align}
  \label{glg119}
  K(z'',t;z',0) = \sum_\nu
         &\sqrt{\frac{\ri}{\hbar}\frac{\partial^2 S_\nu}{\partial 
          u' \partial v''}} \exp\left\{ \frac{\ri}
          {2\hbar}\mbox{\large $\displaystyle \int$}^t_{\!\!\!\!\!0} \rd t'
     \left( \frac{\partial^2{\cal H}}{\partial u\partial v}\right)_{\!\!\!\nu}
          \right\} \nonumber\\
         &\exp \left\{ \frac{\ri}{\hbar}S_\nu(v'',u',t) - \frac{1}{2} 
          \bigl( |z''|^2 + |z'|^2\bigr) \right\} .
\end{align}
The sum over $\nu$ represents the sum over all (complex) classical trajectories
satisfying the boundary conditions, since there may be more than one. We
already mentioned (see before \refg{glg3}) that we were going to suppress this
sum for purposes of clarity, and we shall do so in the future again.

There is one very important caveat.  The hamiltonian function ${\cal H}(u,v,t)$
entering eqs.{\hskip 2pt}(\ref{glg83}) and (\ref{glg119}) {\it is not the
original classical hamiltonian} of the problem, the quantal version of which we
used when we first wrote the path integral (\ref{glg1}).  As eq.{\hskip
2pt}(\ref{mb4}) shows, this ``script'' hamiltonian is a smoothed version of the
original one, obtained by folding it with a gaussian in phase space.  This is
one of the interesting features of formula (\ref{glg119}).  Another interesting
feature is the first exponential, which we shall write 
\begin{align}
\label{2.mb1} 
\re^{\frac{i}{\hbar}\cal I} 
\end{align} 
where the quantity $\cal I$, with dimensions of action, is defined by 
\begin{align} 
\label{2.mb2} 
{\cal I}(v'',u',t) := 
\frac{1}{2}\mbox{\large $\displaystyle\int$}^t_{\!\!\!\!\!0}
\rd t' \left( \frac{\partial^2{\cal H}} {\partial u\partial v}\right) \; .
\end{align} 
Though this $\cal I$ term was assuredly obtained in the past by various workers
in the field, most authors seem to have been in a big hurry to forget its
existence.  One simple and compelling reason for not doing so is the following:
when you calculate the harmonic oscillator (as we do in section 6), you get the
exact answer if you keep $\cal I$, but you don't if you drop it.

There will be more discussion of $\cal I$ in sections 3 and 4.  This, and only
this, is the result of making the standard semiclassical approximations on the
coherent state propagator (\ref{glg82}).  As far as we know, no formula
equivalent to (\ref{glg119}) has been in wide use before.  Both Herman and Kay
\cite{Herm86,Kay94a,Kay94b} claim, in words only, that a particular formula
follows from a semiclassical treatment of the coherent state propagator.  In
both cases the claim is not justified and the formula is incorrect.  In the
next section, we shall see how other formulae can be derived in which the two
special features above are either absent or reversed.  In sections 4 and 5 we
shall compare with two other formulae in the literature.

For an additional note, we use eqs.{\hskip 2pt}(\ref{glg9a}) to write the
integrand of $\cal I$ in terms of $q$ and $p$
\begin{align}
  \label{2.mb3}
\frac{\partial^2{\cal H}}{\partial u\partial v} =
\frac{b^2}{2} \frac{\partial^2\cal H}{\partial q^2} +
\frac{c^2}{2} \frac{\partial^2\cal H}{\partial p^2} \; .
\end{align}
Given formulae (\ref{glg47}), we see that $\cal I$ is of order $\hbar$.
Whenever $q$ and $p$ are real, $\cal H$ is real, therefore
$\partial^2{\cal H}/\partial u\partial v$ is real.  Hence, if the classical
trajectory happens to be real, $\cal I$ is real.

\subsection{The Tangent Matrix}
  \label{kap2.4}

For the applications of our semiclassical formula (\ref{glg119}) to be
developed in sections 4 and 5, we shall have to write the prefactor in
terms of the elements of the tangent matrix $M$, which connects small
displacements of the classical trajectory about the initial point at time zero
to the evolved displacements at time $t$.  Differentiating in \refg{mb9}, but
keeping the variable $t$ constant, we can obtain the connection between initial
and final displacements.  In matrix form it is
\begin{align}
  \label{en1}
           -\ri \hbar
           \begin{pmatrix}
             \delta u'' \\ \\ 
             \delta v'  
           \end{pmatrix}
   =       \begin{pmatrix}
             A_{uv} & A_{vv} \\ \\ 
             A_{uu} & A_{vu} 
           \end{pmatrix}
           \begin{pmatrix}
             \delta u' \\ \\ 
             \delta v'' 
           \end{pmatrix}
\end{align}
with the notation
\begin{align}
  \label{en1a}
A_{uv} = \frac{\partial^2 S}{\partial u'\partial v''} \qquad
A_{vv} = \frac{\partial^2 S}{\partial v''\partial v''} \qquad
A_{uu} = \frac{\partial^2 S}{\partial u'\partial u'} \qquad
A_{vu} = A_{uv} \; .
\end{align}
Solving for $\delta u''$ and $\delta v''$ in
terms of $\delta u'$ and $\delta v'$ yields
\begin{align}
  \label{en2}
           \begin{pmatrix}
             \delta u'' \\ \\ 
             \delta v''  
           \end{pmatrix}
  &=       \frac{1}{A_{uv}}
           \begin{pmatrix} 
             \frac{\ri}{\hbar} (A^2_{uv}-A_{uu} A_{vv}) & A_{vv} \\ \\
             -A_{uu} & -\ri\hbar
           \end{pmatrix} 
           \begin{pmatrix}
             \delta u' \\ \\ \delta v' 
           \end{pmatrix} \; .
\end{align}
If we call $M_{uu}, M_{uv}$, etc... the matrix elements of the tangent matrix
(Note: unlike $A_{uu}, A_{uv}$, etc..., these are {\it not} second derivatives)
this last equation should also be
\begin{align}
  \label{en2a}
           \begin{pmatrix}
             \delta u'' \\ \\ 
             \delta v''  
           \end{pmatrix}
  &=       \begin{pmatrix} 
             M_{uu} & M_{uv} \\ \\
             M_{vu} & M_{vv}
           \end{pmatrix} 
           \begin{pmatrix}
             \delta u' \\ \\ \delta v' 
           \end{pmatrix} \; .
\end{align}
Therefore we have
\begin{align}
  \label{en2b}
\frac{\ri}{\hbar}\frac{\partial^2 S}{\partial u'\partial v''}=\frac{1}{M_{vv}}
\end{align}
and we rewrite the propagator (\ref{glg119}) as
\begin{align}
  \label{en4}
           K(z'',t;z',0) 
  =        \sum_\nu \frac{1}{\sqrt{(M_\nu)_{vv}}} \re^{\frac{i}{\hbar}\cal I}
           \exp \left\{ \frac{\ri}{\hbar}S_\nu(v'',u',t) - \frac{1}{2} 
           \bigl( |z''|^2 + |z'|^2\bigr) \right\} \; .
\end{align}
The square root in the prefactor has an undetermined sign.  To know which sign
is correct, one must start from $t=0$ when the square root is simply unity, and
proceed by continuity along the trajectory, which can be done since $M_{vv}$
never vanishes.  We shall meet an important example of this in subsection 6.1.

Another useful representation of the tangent matrix is in terms
of the scaled variables $q/b$ and $p/c$~:
\begin{align}
  \label{glg4.1}
\left( \begin{array}{l}
  \delta q''/b \\
  \delta p''/c 
  \end{array} \right) = 
\left( \begin{array}{ll}
  m_{qq} & m_{qp} \\
  m_{pq} & m_{pp} 
  \end{array} \right)
\left( \begin{array}{l}
  \delta q'/b \\
  \delta p'/c 
  \end{array} \right)
  \; .
\end{align}
The relation between the matrix elements above and those in eq.{\hskip
2pt}(\ref{en2a}) is:
\begin{align}
  \label{enm}
 \begin{split}
M_{uu} &= \frac{1}{2}(m_{qq}+m_{pp}+\ri m_{pq}-\ri m_{qp}) \\
M_{uv} &= \frac{1}{2}(m_{qq}-m_{pp}+\ri m_{pq}+\ri m_{qp}) \\
M_{vu} &= \frac{1}{2}(m_{qq}-m_{pp}-\ri m_{pq}-\ri m_{qp}) \\
M_{vv} &= \frac{1}{2}(m_{qq}+m_{pp}-\ri m_{pq}+\ri m_{qp}) \; .
 \end{split}
\end{align}
 

%
\section{Ambiguities in the Choice of Path Integral}
  \label{kap3}

There is more than one way of representing the propagator by a path integral in
phase space in terms of coherent states.  The way we chose in section 2 is just
the one adopted by most workers in the field.  Although each one of these
different path integrals would give the same answer in an exact quantum
mechanical calculation, they may differ when semiclassical approximations are
made.  In this section we shall discuss a number of such alternatives, and the
implications of these ambiguities for the validity of the approximations.

We begin with a qualitative remark.  We introduced the path integral by
splitting the propagator (\ref{glg1}) into many segments, each with
infinitesimal time--interval $\tau$, and then introducing the unit operator
(\ref{glg21}) between each pair of adjacent segments.  But the basis
$|z\rangle$ in terms of which the unit operator is written is vastly
overcomplete (see \cite{Vou97} for a recent discussion with references), and
therefore there is an infinite number of ways of writing the unit operator in
terms of the $|z\rangle$'s.  The way chosen for \refg{glg21} is only one of
them.  We shall not pursue this approach to the ambiguous choice, but it does
demonstrate the existence of an enormous arbitrariness.  Instead, we shall
discuss two other aspects of the problem.  The first of these, through a
different derivation of the path integral, leads to a different ``effective
hamiltonian''.  The second deals with the arbitrariness in operator ordering
when one goes from a classical hamiltonian to a quantum mechanical one.  It
will turn out that both of them will lead us to reconsider the significance of
the first exponential (\ref{2.mb1}) in the semiclassical propagator
(\ref{glg119}).

\subsection{Alternative Forms of the Path Integral}
  \label{kap3.1}

In their introduction to their overview of coherent states \cite{Klau85},
Klauder and Skagerstam (KS) discuss two ways of arriving at a path integral,
which they call ``Path Integral~--~First Form'' (p.60) and
``Path Integral~--~Second Form'' (p.69).  The first form is the one we gave in
subsection 2.1.  The second form starts from the ``diagonal representation'' of
the hamiltonian operator, namely
\begin{align}
  \label{5mb1}
\hat{H} = \int|z\rangle h(z)\frac{\rd^2z}{\pi}\langle z| \; .
\end{align}
Given that we assumed (see second paragraph of subsec.{\hskip 5pt}2.1) that
$\hat{H}$ was either a polynomial in $p$ and $q$ or a converging sequence of
such polynomials, this diagonal representation always exists.  The notation
$h(z)$ for the function in eq. (\ref{5mb1}) is that of KS; we shall change it
to $H_2(z^\star,z)$.  This will be contrasted with the first--form hamiltonian
function which we called $\cal H$ in eq. (\ref{mb4}), and which in this
section we are going to call $H_1(z^\star,z)$,
assuming no explicit time--dependence. Equation (\ref{5mb1}) is now rewritten
\begin{align}
  \label{5mb2}
\hat{H} = \int|z\rangle H_2(z^\star,z)\frac{\rd^2z}{\pi}\langle z| \; .
\end{align}
According to KS, the connection between $H_1$ and $H_2$ is
\begin{gather}
  \label{5mb3}
H_1(z^\star,z) = \int\frac{\rd^2z'}{\pi}\left|\langle z|z'\rangle\right|^2
 H_2({z'}^\star,z') = \int\frac{\rd^2z'}{\pi}\re^{-|z-z'|^2}H_2({z'}^\star,z')
   \\
  \label{5mb4}
H_2(z^\star,z) = \left(\exp\;-\frac{\partial^2}{\partial z^\star
   \partial z}\right)H_1(z^\star,z) \; .
\end{gather}
In addition to these two effective hamiltonian functions, there is also the
original classical hamiltonian $H_C(z^\star,z)$.  All three are different.

A very common type of classical hamiltonian is $H_C(q,p)=p^2/2m+V(q)$ where,
once again, we assume that the function $V(q)$ can be approximated by a
polynomial.  For such a hamiltonian, the choice of quantum mechanical operator
$\hat{H}$ is straightforward.  Then we can work out the connection between
$H_C$, $H_1$, and $H_2$ for simple monomials.  If we call $x$ and $y$ the real
and imaginary parts of $z$, respectively, this connection is given by the table
\begin{tabbing}
\ \kern 27mm\=\kern 40mm\=\kern 40mm\=\kern 41mm\=\kern 9mm\+\kill
$H_2$   \>$H_C$   \>$H_1$   \\
\   \\
1     \>1     \>1     \\
$x$    \>$x$   \>$x$     \\
$x^2-\frac{1}{4}$\>$x^2$  \>$x^2+\frac{1}{4}$\>(3.T)\\
$x^3-\frac{3}{4}x$\>$x^3$ \>$x^3+\frac{3}{4}x$ \\
$x^4-\frac{3}{2}x^2+\frac{3}{16}$\>$x^4$  \>$x^4+\frac{3}{2}x^2
       +\frac{3}{16}$ \\
etc\dots\>etc\dots\>etc\dots
\end{tabbing}
with an identical table for monomials of $y$.  We see that, up to cubic terms,
we have $H_C=\frac{1}{2}(H_1+H_2)$, but this does not remain true for higher
powers.  In a qualitative way, one can think of $H_1$ as a smoothing of $H_C$,
and of $H_2$ as an unsmoothing.  This idea becomes precise for monomials
containing purely $x$ or purely $y$.  In the first case, we have
\begin{align}
  \label{5mb6}
\begin{split}
H_C(x)&=\sqrt{\frac{2}{\pi}}\int\rd x'\re^{-2(x-x')^2}H_2(x')  \\
H_1(x)&=\sqrt{\frac{2}{\pi}}\int\rd x'\re^{-2(x-x')^2}H_C(x')
\end{split}
\end{align}
showing that $H_C$ is a smoothing of $H_2$, and $H_1$ a smoothing of $H_C$.
The relations for pure functions of $y$ are identical.  If the monomial in
$H_C$ contains both $x$ and $y$, then the problem of operator ordering arises; 
we shall discuss it in the next subsection.

We shall now sketch the derivation of the semiclassical propagator for the
second form of path integral.  For simplicity, we exclude an explicit
time--dependence in $\hat{H}$.  With fewer details, we repeat the steps of
subsecs.{\hskip 2pt}2.1, 2.2, and 2.3.  We use the notation $K_1$ for the
propagator of the first form and $K_2$ for that of the second form.  The
starting point is \refg{glg81}.  Following KS (their eq. (6.11)), we write each
of the $N$ exponentials as
\begin{align}
  \label{5mb7}
\re^{-\frac{\ri}{\hbar}\hat{H}\tau}\approx\int|z_j\rangle
   \re^{-\frac{i\tau}{\hbar}H_2(z_j^\star,z_j)}
   \frac{\rd^2z_j}{\pi}\langle z_j| \; .
\end{align}
To facilitate the comparison between $K_2$ and $K_1$, it is convenient to write
$\re^{-\frac{\ri}{\hbar}\hat{H}t}$ as the product of $N-1$ factors like
(\ref{5mb7}) rather than $N$.  In other words, we take $(N-1)\tau=t$, which
makes no difference in the limit of large $N$ and infinitesimal $\tau$.  Then
$j$ in eq. (\ref{5mb7}) goes from $1$ to $N-1$.  The complete propagator is
\begin{align}
  \label{5mb8}
K_2(z_N,t;z_0,0)&=\langle z_N|\text{ $(N-1)$ factors similar to (\ref{5mb7}) }
                |z_0\rangle  \nonumber \\
&=\int\prod_{j=1}^{N-1}\frac{\rd^2z_j}{\pi}\,\re^{-\frac{i\tau}{\hbar}
  H_2(z_j^\star,z_j)}\prod_{j=0}^{N-1}\langle z_{j+1}|z_j\rangle \; .
\end{align}
Contrast this with the first--form propagator, obtained from eqs. (\ref{glg82})
and (\ref{glg52})
\begin{align}
  \label{5mb9}
K_1(z_N,t;z_0,0)=
  \int\prod_{j=1}^{N-1}\frac{\rd^2z_j}{\pi}\prod_{j=0}^{N-1}\re^{-\frac
  {i\tau}{\hbar}H_1(z_{j+1}^\star,z_j)}\langle z_{j+1}|z_j\rangle \; .
\end{align}
Whereas in $K_1$ the two arguments of $H_1$ belong to two adjacent times in the
mesh, the two arguments of $H_2$ in $K_2$ belong to the same time.  This
difference is important and results in different semiclassical propagators
after one does the stationary exponent approximation.  The stationarity
conditions are found to be
\begin{align}
  \label{5mb10}
\begin{split}
z_{j+1}^\star - z_j^\star - \frac{\ri\tau}{\hbar}\frac{\partial}{\partial z_j}
   H_2(z_j^\star,z_j)&=0\; ;\qquad  j = 1,2,\ldots,N-1  \\
-z_j + z_{j-1} - \frac{\ri\tau}{\hbar}\frac{\partial}{\partial z_j^\star}
   H_2(z_j^\star,z_j)&=0\; ;\qquad  j = 1,2,\ldots,N-1 \; .
\end{split}
\end{align}
Contrast this with the stationarity conditions for $K_1$, which are
\begin{align}
  \label{5mb11}
\begin{split}
z_{j+1}^\star - z_j^\star - \frac{\ri\tau}{\hbar}\frac{\partial}{\partial z_j}
   H_1(z_{j+1}^\star,z_j)&=0\; ;\qquad  j = 1,2,\ldots,N-1  \\
-z_j + z_{j-1} - \frac{\ri\tau}{\hbar}\frac{\partial}{\partial z_j^\star}
   H_1(z_j^\star,z_{j-1})&=0\; ;\qquad  j = 1,2,\ldots,N-1 \; .
\end{split}
\end{align}
The differences are subtle, but real.  In the continuum limit, both sets of
equations become the classical equations of motion, but for hamiltonian $H_1$
in the case of $K_1$, and for hamiltonian $H_2$ in the case of $K_2$.  In both
cases, neither $z_0^\star$ nor $z_N$ appear in the equations, hence the
trajectory is determined purely by $z_0$ and $z_N^\star$ and is complex.  The
two actions $S_1$ and $S_2$, coming from different hamiltonians, are different.

Now we must calculate the prefactor.  As in subsec.{\hskip 2pt}2.2, the
exponent in (\ref{5mb8}) is taken at the points $z_j^\star+\eta_j^\star,\
z_j+\eta_j$ and expanded to second order in the vicinity of the stationary
points $z_j^\star, z_j$.  The zeroth order gives an expression calculated on
the classical trajectory, the first order vanishes by the stationarity
conditions, and the second order is a quadratic form leading to a gaussian
integration.  The quadratic form for $K_2$ is
\begin{align}
  \label{5mb12}
- \frac{\ri\tau}{\hbar}\sum_{j=1}^{N-1}\left[ \frac{1}{2}\eta_j^2\frac
{\partial^2}{\partial z_j^2} + \eta_j^\star\eta_j\frac{\partial^2}
{\partial z_j^\star\partial z_j} + \frac{1}{2}\eta_j^{\star2}\frac{\partial^2}
{\partial z_j^{\star2}} \right]H_2(z_j^\star,z_j)
- \sum_{j=1}^{N-1}\eta_j^\star\eta_j + \sum_{j=1}^{N-2}\eta_{j+1}^\star\eta_j
\; .
\end{align}
According to eq.(\ref{glg6}) the quadratic form for $K_1$ is
\begin{align}
  \label{5mb13}
\begin{split}
&- \frac{\ri\tau}{\hbar}\sum_{j=1}^{N-1}\frac{1}{2}\left[ \eta_j^2\frac
{\partial^2}{\partial z_j^2}H_1(z_{j+1}^\star,z_j) + \eta_j^{\star2}\frac
{\partial^2}{\partial z_j^{\star2}}H_1(z_j^\star,z_{j-1})\right] \\
&- \frac{\ri\tau}{\hbar}\sum_{j=1}^{N-2}\eta_{j+1}^\star\eta_j\frac{\partial^2}
{\partial z_{j+1}^\star\partial z_j}H_1(z_{j+1}^\star,z_j)
- \sum_{j=1}^{N-1}\eta_j^\star\eta_j + \sum_{j=1}^{N-2}\eta_{j+1}^\star\eta_j
\; .
\end{split}
\end{align}
The comparison of $K_2$ with $K_1$ is easy if this gaussian integration is
performed by the determinant method presented in the appendix of this paper.
The details are given there and the result is the following.  The semiclassical
$K_2$ is given by eq. (\ref{glg119}) again, with two changes: (1) Of course,
$H_1$ is replaced by $H_2$, and $S_1$ by $S_2$; (2) The first exponent is
$-\frac{i}{\hbar}\cal I$ instead of $+\frac{i}{\hbar}\cal I$~.

In those cases where the semiclassical approximation is expected to be good,
one can then hope that the two changes cancel each other approximately.  Both
path integrals are exact originally and therefore they should give the same
propagator.  But one semiclassical propagator contains $H_1$ and the other
contains $H_2$, which necessarily leads to different results unless the
hamiltonians are purely quadratic.  If both approximations are good, it must
mean that the difference is cancelled by another one: the change in sign of the
first exponent.

Out of these considerations comes the justification for a procedure which has
been used in some of the past literature: leave out the first exponential in
(\ref{glg119}), or replace it by unity, and do the classical calculations using
the original classical hamiltonian $H_C$ instead of $H_1$ (or $\cal H$).  This
follows from the fact that, as eqs.{\hskip 4pt}(\ref{5mb6}) demonstrate, $H_C$
can be thought of as more or less half--way between $H_1$ and $H_2$.  It should
then be associated with a first exponent which is also half--way between
first--form and second--form, and that means zero.  Another justification was
given in \cite{Kurc89}, where it is shown that this is the way of getting the
exact first--correction term in an expansion in powers of $\hbar$.  Hence the
procedure is correct in the extreme semiclassical limit, the limit of high
quantum numbers.  It may not be as good for low energies, comparable with the
energies of low excited states.  We shall return to this question in a future
publication.  In our opinion, the best procedure for low energies is to take at
face value the result (\ref{glg119}) of the first--form path integral, using
hamiltonian $H_1$ and including the first exponential.  This is because $H_1$
is the smoothest of the three hamiltonians, and therefore the stationary
exponent approximation has the best chance of being good in this case.  There
is one final question one might ask: besides the first form and the second form
of path integrals, can one find a continuum of integrals which interpolate
smoothly between these two?  The answer is yes.  Instead of the diagonal
form(\ref{5mb7}) for the infinitesimal propagator, one can write it in
non--diagonal form thus
\begin{align}
  \label{5mb14}
\re^{-\frac{\ri}{\hbar}\hat{H}\tau} = \int\!\!\!\int|z''\rangle
\frac{\rd^2z''}{\pi} \langle z''|\re^{-\frac{\ri}{\hbar}
\hat{H}\tau}|z'\rangle \frac{\rd^2z'}{\pi} \langle z'| \; .
\end{align}
We can approximate the matrix element as in subsec. 2.1, which gives
\begin{align}
  \label{5mb15}
\re^{-\frac{\ri}{\hbar}\hat{H}\tau} \approx \int\!\!\!\int|z''\rangle
\frac{\rd^2z''}{\pi} \langle z''|z'\rangle
\re^{-\frac{i\tau}{\hbar}H_1({z''}^\star,z')}
\frac{\rd^2z'}{\pi} \langle z'| \; .
\end{align}
After multiplying together many such expressions, one has to perform many
integrals, twice as many as before.  But half of the variables do not occur in
the $H_1$ functions, and the integrations over them are straightforward, the
stationary exponent method being exact.  This leads back to the first form of
path integral.  However, nothing prevents us from mixing the two recipes
(\ref{5mb7}) and (\ref{5mb15}) in any proportions whatever, generating all
possible interpolations between the first and the second form.  We shall not
pursue this approach further.

\subsection{Ambiguities in Operator Ordering}
  \label{kap3.2}

In the above discussion, many of the statements were precise only if each
monomial in the hamiltonian contained only $q$ or only $p$.  We shall now
generalize to arbitrary monomials, which brings up the question of operator
ordering.  Given a classical $H_C(q,p)$, there are many corresponding
$\hat{H}$'s, because $q$ and $p$ do not commute and their order matters.  Hence
there are many quantum mechanical problems and therefore many different
results, the differences becoming smaller as $\hbar$ becomes smaller.  The
number of possible orderings, or combinations of orderings, is infinite, but
three of them stand out.  {\it The normal ordering} consists in rewriting $H_C$
in terms of $z$ and $z^\star$, replacing in each monomial $z$ by $a$ and
$z^\star$ by $a^\dagger$ (see eq.{\hskip 2pt}(\ref{glg49})), and writing all
creation operators $a^\dagger$ to the left of all annihilation operators $a$,
making each monomial look like $c_{mn}a^{\dagger m}a^n$.  {\it The antinormal
ordering} does the same thing, but it writes the creation operators to the
right of the annihilation operators, thus $c_{mn}a^na^{\dagger m}$.  The third
kind, called {\it Weyl ordering or symmetric ordering}, has several equivalent
definitions.  One says to write all possible orderings of the monomial, and
then take the average of them all.  Another definition of the Weyl operator
$\hat{A}_W$ corresponding to a classical function $A(q,p)$ is as follows.
First, write $A(q,p)$ as a double Fourier transform
\begin{align}
  \label{5mb16}
A(q,p) = \int\!\!\!\int\rd\alpha\rd\beta\,B(\alpha,\beta)
\re^{\ri(\alpha q + \beta p)} \; .
\end{align}
Then the Weyl operator is obtained by changing $q$ and $p$ into operators in
this formula
\begin{align}
  \label{5mb17}
\hat{A}_W = \int\!\!\!\int\rd\alpha\rd\beta\,B(\alpha,\beta)
\re^{\ri(\alpha\hat{q} + \beta\hat{p})} \; .
\end{align}
Some easily--read references are \cite{Hill84}, \cite{Kurc89}, and
\cite{Voro89}.

Thus we have three different ways of associating an operator with a function in
classical phase space.  We can call them $\hat{A}_N$, $\hat{A}_A$, and
$\hat{A}_W$.  Of the three, $\hat{A}_W$ is the most ``reasonable'' one in the
classical limit.  Conversely, there are three ways of associating a classical
function with an arbitrary operator $\hat{A}$, which are the inverses of the
three transformations above.  The most interesting one (see \cite{Hill84})
turns out to be the inverse of the Weyl transformation.  It is
\begin{align}
  \label{5mb18}
A_C(q,p) = \int\rd s\,\re^{\frac{\ri}{\hbar}ps}\left\langle q-\frac{s}{2}\left|
   \hat{A}\right|q+\frac{s}{2}\right\rangle \; .
\end{align}
This is called the Wigner transformation.  $A_C(q,p)$ is called the Weyl symbol
of the operator $\hat{A}$ and is often denoted by $A_W(q,p)$.  The other
two inverse transformations also have names.  For the transformation which
associates a normal--ordered operator to a classical function, the inverse
transformation is called the $Q$--transformation, and the classical function is
called the $Q$--symbol of the operator
\begin{gather}
  \label{5mb19}
\begin{split}
A(q,p)\quad\Longrightarrow\quad\text{normal ordering}\quad\Longrightarrow
\quad\hat{A}_N \\
A_Q(q,p)\quad\Longleftarrow\quad\text{$Q$--transformation}\quad\Longleftarrow
\quad \hat{A} \; .
\end{split}
\end{gather}
Similarly, the inverse of antinormal ordering is called the
$P$--transformation, and the classical function is the $P$--symbol of the
operator
\begin{gather}
  \label{5mb20}
\begin{split}
A(q,p)\quad\Longrightarrow\quad\text{antinormal ordering}\quad\Longrightarrow
\quad \hat{A}_A \\
A_P(q,p)\quad\Longleftarrow\quad\text{$P$--transformation}\quad\Longleftarrow
\quad \hat{A} \; .
\end{split}
\end{gather}
There is a very simple explicit expression for the $Q$--symbol
\begin{align}
  \label{5mb21}
A_Q(q,p) = \langle z|\hat{A}|z\rangle \; .
\end{align}
For the $P$--symbol there is nothing as easy, but it is given implicitly by the
requirement
\begin{align}
  \label{5mb22}
\hat{A} = \int|z\rangle A_P(q,p)\frac{\rd^2z}{\pi}\langle z| \; .
\end{align}
Of the three symbols, the $Q$--symbol, which is analytic in both $z$ and
$z^\star$, is the smoothest.  The $P$--symbol is the most likely to be
singular.  The Weyl symbol is in between.

Coming back to our subject, we see now that $H_1$ is the $Q$--symbol associated
with the operator $\hat{H}$, $H_2$ is the $P$--symbol, and the function which
we have called $H_C(q,p)$ in the past should be the Weyl symbol $H_W$.  As we had
defined it for the pure monomials of $q$ or $p$, it is indeed the Weyl symbol.
And everything we have said about $H_C$ so far will remain true in general,
provided we define $H_C$ as the Weyl symbol $H_W$ of $\hat{H}$.

Let us summarize the results.  We have three effective classical hamiltonians
associated with the quantum mechanical $\hat{H}$.  Their smoothness decreases
in the order $H_1$, $H_W$, $H_2$.  This is evident from the relations
\begin{align}
  \label{5mb23}
\begin{split}
H_1(q,p)&=\sqrt{\frac{2}{\pi}}\int\rd^2z'\re^{-2|z-z'|^2}H_W(q',p')  \\
H_W(q,p)&=\sqrt{\frac{2}{\pi}}\int\rd^2z'\re^{-2|z-z'|^2}H_2(q',p') \; .
\end{split}
\end{align}
Going in the opposite direction, the relations are
\begin{align}
  \label{5mb24}
\begin{split}
H_2(q,p)&= \left(\exp\;-\frac{1}{2}\frac{\partial^2}{\partial z^\star
   \partial z}\right)H_W(q,p)  \\
H_W(q,p)&= \left(\exp\;-\frac{1}{2}\frac{\partial^2}{\partial z^\star
   \partial z}\right)H_1(q,p) \; .
\end{split}
\end{align}
In the extreme semiclassical limit, i.e. for large quantum numbers,
the best formula for the propagator is (\ref{glg119}) calculated with
$H_W$ throughout and omitting the first exponential (\ref{2.mb1}).
For low energy, the best formula is (\ref{glg119}) as it stands,
i.e. calculated with $H_1$ and with the first exponential.  The third
formula, (\ref{glg119}) calculated with $H_2$ and with $\cal I$
replaced by $-\cal I$ (see after \refg{5mb13} and also the last
sentence of the appendix), is valid too, though we do not know in what
circumstances it might be expected to be better.  All three
propagators have claims to being called ``the'' semiclassical
propagator in phase space, although there is no simple derivation of
the propagator with the Weyl Hamiltonian from path integrals. They are
all different since the classical hamiltonians are different.  All
three are exact for the harmonic oscillator and the free particle.
But it would be wrong to calculate with $H_1$ without including the
$\cal I$ term, or with $H_2$ without including $-\cal I$, or with
$H_W$ including $\cal I$.

\subsection{Powers of $\hbar$}

In the following sections we shall use our semiclassical formula for the
coherent state propagator in a number of different situations.  In particular,
we shall derive an Initial Value Representation for the propagator in section 4
and the Green's function in section 6.  In order to perform these calculations
consistently, we must state precisely what is the `philosophy' of our
approximation, i.e., what terms must be kept and what terms can be discarded in
calculations involving this semiclassical propagator.  In order to do so, we
have to understand the difference between calculations with the Weyl
Hamiltonian, which does not involve $\hbar$ explicitly, and those with
$H_1\equiv{\cal H}$ or $H_2$.  The discussion boils down to understanding the
semiclassical formulae in terms of powers of $\hbar$.

We start by recalling the nature of the stationary phase approximation
(SPA) for a simple one-dimension integral. Let 
\begin{align}
\label{sec27a}
A = \int_{-\infty}^{+\infty} g(x) \; \re^{\frac{i}{\hbar}f(x)} \; \rd x \; .
\end{align}
Assume that $f$ has a single stationary point at $x=x_0$ and that $g(x)$ is a
slowly varying function of $x$. The SPA can be applied if $\hbar$ is small. It
amounts to expanding $f(x)$ to second order about $x_0$ while keeping $g(x)$
constant in the vicinity of $x_0$. The integration is then reduced to that of a
Gaussian, which is straightforward. The result is
\begin{align}
\label{sec27b}
A \approx A_0 :=\sqrt{\frac{2\pi \hbar}{|f^{(2)}|}} \; g(x_0)  \; 
\displaystyle{\re^{\frac{i\pi s}{4} + \frac{i}{\hbar}f(x_0)}}
\end{align}
where $f^{(2)}$ is the second derivative of $f$ and $s$ is its sign.  This
approximation neglects third and higher derivatives of $f$, as well as all
derivatives of $g$.

By keeping more derivatives, we can calculate the next non-zero contribution to
the integral.  This is done in appendix B.  The result is that $A$ can be
written as
\begin{align}
\label{sec27c}
A &= A_0 \left[1+i \hbar R(x_0) + {\cal O}(\hbar^2)\right]
\end{align}
where $R(x_0)$, given explicitly in eq.{\hskip 2pt}(\ref{sec27k}), involves the
first and second derivatives of $g$ and the third and fourth derivatives of
$f$, all computed at $x=x_0$.  If instead we were to write these corrections in
the exponential, the result would look like this
\begin{align}
\label{sec27ca}
 A = \sqrt{\frac{2\pi \hbar}{|f^{(2)}|}} \; g(x_0)  \; 
\displaystyle{\re^{\frac{i\pi s}{4}} \; 
\re^{\frac{i}{\hbar}\left[f(x_0)+\hbar^2 R(x_0) + {\cal O}(\hbar^3)\right]}}
\; .
\end{align}
Therefore, since what we do all along is a quadratic exponent approximation, we
do not have any hope of being able to calculate completely and accurately
either the terms of order $\hbar$ in the integrals themselves, as
(\ref{sec27c}) shows, or the terms of order $\hbar^2$ in quantities occurring
in an exponent with overall coefficient $i/\hbar$, as shown in (\ref{sec27ca}).
However, since we actually do have terms of this sort in our results, more
discussion needs to take place.

The path integral calculation of the semiclassical propagator involves many
integrals similar to (\ref{sec27a}), which are evaluated by an appropriate
quadratic exponent approximation. The role of $f(x)$ is played by the action
and that of $x_0$ by the stationary trajectory. In the coordinate or momentum
representations, neither the action nor the stationary trajectory depend on
$\hbar$. In these cases, as in the one-dimensional integral (\ref{sec27a}),
corrections beyond the stationary phase approximation are of order $\hbar$ with
respect to the semiclassical formula, as in eq.{\hskip 2pt}(\ref{sec27c}).  In
the coherent state representation, however, both the action {\it and} the
stationary trajectory depend on Planck's constant via ${\cal H}$. Therefore,
the quadratic approximation already involves `nonclassical' $\hbar$-dependent
terms, and an expansion of the propagator in powers of $\hbar$ similar to
(\ref{sec27c}) becomes somewhat confusing. Interestingly, it is shown in
appendix C that the non-classical terms in the action $S$ cancel those in
${\cal I}$ up to first order in $\hbar$. Therefore, the non-classical terms in
the `effective phase' $S+{\cal I}$ are of order $\hbar^2$, which is beyond the
precision of the approximation. It is not clear at this point whether these
extra terms improve or not the semiclassical formula with respect to a pure
`Weyl' calculation. We shall come back to this point in section 6 when we
discuss semiclassical quantization rules.

In view of these results, this is what we should do in future calculations to
be consistent with our semiclassical procedure.  We must distinguish two parts
to the integrand.  One is the exponent that we make stationary.  This is
always multiplied by a factor $i/\hbar$, and it usually consists of some kind
of action.  The other part, which is everything else, we call the prefactor.
Then:

\begin{enumerate}

\item In the stationary exponent, we can drop terms containing third or higher
order derivatives of ${\cal H}$ or $S$ with respect to $u$ and $v$, since they
have not been taken into account in the quadratic approximation.  We should
keep all terms of order $\hbar$.  They are the ones that are expected to cancel
out.  All terms of order $\hbar^2$ can be discarded.

\item In the prefactor, we do not need to expand at all.  Improving the result
by expanding the prefactor is an illusion.  In addition, the prefactor may have
a phase, which of course we should keep.

\end{enumerate}
 
%
\section{An Initial Value Representation}
  \label{kap4}

The biggest difficulty in numerical applications using the coherent
state propagator is the root--searching problem.  This explains the
recent popularity of the initial--value--representations (IVR), which
avoid this problem.  In this section we shall transform the propagator
(\ref{glg119}) into a semiclassical IVR propagator with the same
initial and final states as the Herman-Kluk (HK) propagator
\cite{Herm84}.  We had expected that the result would be the HK
propagator itself, but it did not turn out that way: our IVR
propagator differs radically from HK's.  Along the way, we shall point
out the mistake made by Grossmann and Xavier \cite{Gros98a} in their
attempt at a similar derivation. We shall return to the HK propagator
in section 5, where we compare it to our IVR and point out that its
original derivation in \cite{Herm84} also contains a
mistake very similar to that made by Grossmann and Xavier.

The basic purpose of an IVR formula is the following.  We are given at time 0 a
wave function $\langle x|\psi(0)\rangle$ in the usual configuration--space
representation.  We want to calculate its evolution, i.e. we want to know
\begin{align}
  \label{ba1}
      \langle x|\psi(t)\rangle 
   =  \langle x|K(t)|\psi(0)\rangle
\end{align}
again in configuration--space representation, $K(t)$ being the usual Feynman
propagator.  We want to do this semiclassically, in terms of integrals over
classical trajectories, {\it and we want these trajectories to be specified
purely in terms of the initial values of their coordinates.}  The alternative,
mixed values of the coordinates, some being initial and some final, leads to
unacceptable ``root--search'' difficulties.  Among the many IVR formulae that
have been proposed, it has been reported \cite{Kay94b} that the Herman--Kluk
expression \cite{Kluk86} is particularly easy to use and gives particularly
good results.  To use this formula, one must first transform the initial wave
function to the coherent state representation (a.k.a. the Bargmann
representation) by doing the integral
\begin{align}
  \label{ba2}
      \langle z'|\psi(0)\rangle 
   =  \int\langle z' | x \rangle \rd x\langle x|\psi(0)\rangle \; .
\end{align}
Then the propagation is carried out with a mixed propagator, which has coherent
state coordinates initially and configuration coordinates finally,
\begin{align}
  \label{ba3}
      \langle x|\psi(t)\rangle 
   =  \int\langle x|K(t)|z'\rangle\frac{\rd^2z'}{\pi}
      \langle z'|\psi(0)\rangle \; .
\end{align}

The mixed propagator $\langle x|K(t)|z'\rangle$ is the quantity calculated by
HK.  Our purpose is also to calculate it, but we obtain a different result
(subsections 4.1 and 4.2).  Our result is the one that follows naturally when
one makes the type of semiclassical approximations that were made in
\refkap{kap2}.  Moreover, it has the essential property of conserving the
normalization of the initial wave function, which the HK propagator does not
have, as we shall see.  Since the HK propagator has been so popular in the
past, it is obviously desirable to make some numerical comparisons between the
two formulae.  In section 5 we shall discuss the HK propagator in some detail
and show some simple one--dimensional comparisons, for which the new formula is
far superior.  More tests are needed, including some for two--dimensional
problems.

It should be clear that the mixed propagator $\langle x|K(t)|z'\rangle$ is
simply a description of the time--evolution of a gaussian wave--packet.  But a
semiclassical approximation for this was proposed long ago by Heller
\cite{Hell75}.  Obviously we need to compare it with our result.  We do this in
subsection 4.3.  Once again the two formulae are different, but this time there
is a strong resemblance, which we discuss in the light of the ambiguities
encountered in section 3.  Neither result is ``better'' than the other: they
have different regions of validity.

Finally, let us recall that the research on IVR formulae was originally
motivated by the fact that the semiclassical propagator in coordinate space,
long known as the Van Vleck formula, led to unpleasant root--search problems.
In subsection 4.4 we examine the relationship between our and Heller's IVR with
the Van Vleck propagator.

\subsection{A mixed representation}
  \label{kap4.1}

Our starting point is the coherent state propagator of \refg{glg119}.
Our expression for the mixed propagator is then
\begin{align}
  \label{ba4}
      \langle x|K(t)|z'\rangle
  &=  \int\langle x|z''\rangle
      \frac{\rd ^2z''}{\pi}\langle z''|K(t)|z'\rangle \nonumber \\
  &=  \int\frac{\rd p''\rd q''}{2\pi\hbar}
      \pi^{-\frac{1}{4}}b^{-\frac{1}{2}}
      \exp\biggl\{-\frac{(x-q'')^2}{2b^2}
      + \frac{\ri}{\hbar}p''
      \left(x-\frac{q''}{2}\right)\biggr\}      \\
  &\re^{\frac{i}{\hbar}{\cal I}(v'',z',t)}
      \sqrt{\left.\frac{\ri}{\hbar}\:\frac{\partial^2 S}
      {\partial u' \partial v''}\right|_{u'=z',v''={z''}^\star}}
      \exp \biggl\{ \frac{\ri}{\hbar}S(v'',z',t)
      - \frac{1}{2} \bigl( |z''|^2 + |z'|^2\bigr)\biggr\} \nonumber \; .
\end{align}
Before one does the integration over $\rd p''\rd q''$, it is good to
remind oneself of the philosophy spelled out in the second paragraph of
subsection \ref{kap2_2}.  The integrals to be done are actually two real
integrals over
$\rd p''$ and $\rd q''$, and if the functions are analytic in $p''$ and
$q''$, then it is all right to deform the contour into the 4--dimensional space
of complex  $p''$ and complex $q''$.  But there are still two integrals and,
therefore, in this  4--dimensional space the integration runs over a 
2--dimensional surface.  Applying these thoughts to the last member of eq.
(\ref{ba4}), we see that the argument of the exponential in the first
line is obviously analytic, and we can rewrite it in terms of $u''$ and $v''$
without problem.  In the second line we have
\begin{align}
  \label{ba5}
      |z''|^2 
   =  \frac{1}{2}\left(\frac{{q''}^2}{b^2}+\frac{{p''}^2}{c^2}\right)
\end{align}
which is analytic and can also be written $u''v''$.  Also in the second line,
we have the function $S(v'',z',t)$ and its second derivative.  For them, the
argument of analyticity and change of variables was made at an earlier stage
already, when the coherent state propagator was derived in \refkap{kap2}.
Recall that $S(v'',u',t)$ is a complex function of the real variables
$p'',q'',p',q'$ which depends on these variables only through the combinations
$v''$ and $u'$.  The fact that it is an action calculated along a complex
trajectory is really not relevant to doing the integrals.  The thing which {\it
is} relevant is that it is analytic in $v''$ and $u'$, and therefore, if we
have to do integrals over $p''$, $q''$, $p'$, or $q'$, we may continue them
into the complex planes of these variables.  Everything that was just said
about $S(v'',u',t)$ can be repeated word for word about ${\cal I}(v'',u',t)$,
which also occurs in the last line of (\ref{ba4}).  Finally, another very
relevant thing is that we may wish to talk about the derivatives of $S$, and
therefore we give them names as follows:
\begin{align}
  \label{ba6}
  \begin{split}
      \frac{\ri}{\hbar}\frac{\partial S}{\partial v''}
  &=  U''(v'',u',t) \\
      \frac{\ri}{\hbar}\frac{\partial S}{\partial u'}
  &=  V'(v'',u',t) \; .
  \end{split}
\end{align}
Then, in terms of $u''$ and $v''$, which are more convenient variables than
$p''$ and $q''$, here are the integrals we have to do
\begin{align}
  \label{ba7}
  \begin{split}
  &   \langle x|K(t)|z'\rangle
   =  \int\frac{\rd u''\rd v''}{2\pi\ri}
      \pi^{-\frac{1}{4}}b^{-\frac{1}{2}}\:
     {\rm e}^{\frac{i}{\hbar}{\cal I}(v'',z',t)}      
      \sqrt{\left.\frac{\ri}{\hbar}\:\frac{\partial^2 S}
      {\partial u' \partial v''}\right|_{u'=z',v''={z''}^\star}} \\
  &   \qquad\quad\exp\biggl\{ -\frac{x^2}{2b^2} +\frac{\sqrt{2}}{b}xu''
      -\frac{1}{2}{u''}^2 -u''v'' +\frac{\ri}{\hbar}S(v'',z',t)
      - \frac{1}{2}|z'|^2 \biggr\} \; .
  \end{split}
\end{align}

We are going to do these integrals by the stationary exponent approximation.
First, we must look for the stationary point of the exponent.  We do not
include the $\cal I$ term in the exponent, because that would involve 
calculating third order derivatives of ${\cal H}$ (see section 3.3).  To
find a point in 4 dimensions, one needs 4 real equations, or 2 complex ones.
If we call $\Gamma$ the exponent in the second line of \refg{ba7}, these two
complex equations are
\begin{align}
  \label{ba8}
  \begin{split}
      \frac{\partial\Gamma}{\partial v''}
  &   \equiv -u''+ U''(v'',u',t)
   =  0 \\
      \frac{\partial\Gamma}{\partial u''}
  &   \equiv \frac{\sqrt{2}}{b}x-u''-v''
   =  0 \; .
  \end{split}
\end{align}
There are two crucial comments to be made here.  One is that $S$ does not
depend on $u''$ at all, and therefore the second equation does not contain any
derivative of $S$.  In fact the second equation, according to \refg{glg9}, says
very simply
\begin{align}
  \label{ba9}
      q'' = x \; .
\end{align}
The second crucial comment is that $u''$ and $U''$ are not the same.  One is
the independent variable $u''$~.  The other, $U''$, is a function of the other
independent variable $v''$~.  The fact that $U''$ is obtained by calculating a
certain complex trajectory is interesting, but irrelevant to the integration
problem: it is some function of $v''\:$.  On the other hand, $u''$ is an
independent variable; it can be anything, irrespective of the values of $v''$
and $u'$~.  Therefore the first equation, taken by itself, defines a
2--dimensional surface in 4--dimensional complex space.  One of the points on
this surface, and only one, is the end point of the real trajectory which
begins at $p',q'$~.  To find the stationary point, one must combine this
equation with the second one.  These two important facts are the ones that were
missed by Grossmann and Xavier \cite{Gros98a}.  They totally ignored the
existence of the second equation.  And they claimed that the first equation
said that the end point of the real trajectory was the stationary point.  This
is not so, unless one happens to pick $x$ equal to the value of $q$ for this
end point.  For all other choices of $x$, the stationary point is the end point
of a {\it complex} trajectory.  Its $q''$ is real and equal to $x$~.  But its
$p''$ is complex.  We give its value in \refg{3.23a}.

We could take this result and use the stationary exponent approximation in the
vicinity of the complex stationary trajectory.  There are two defects to this
approach.  One is that finding this complex trajectory is once again a
root--search problem, since it is specified by one initial coordinate $z'$ and
one final coordinate $x$~.  The other defect is the inconvenience of having to
do classical mechanics in the complex domain.  It is obviously preferable, if
one can, to set up the practical applications in a way that involves the actual
calculation only of real trajectories.  This can be done, and at the same time
the wished--for IVR character is recovered.  The idea is that the complex
trajectories which are in the vicinity of a real trajectory are a little bit
like the gaussian which is in the vicinity of a stationary point when you do
stationary phase or steepest descent integration.  In the latter case, most of
the contribution to the integral comes from the vicinity of the stationary
point, so that you can approximate the exponent in this vicinity by a
quadratic, even though it is not really a quadratic.  You do a Taylor expansion
of the exponent near the stationary point and you keep only the first two terms
(actually, the first term vanishes).  Similarly, when summing over the whole
bunch of classical trajectories, we expect most of the contribution to the
integral to come from the vicinity of the real trajectory, so that we can
approximate the complex trajectories nearby by doing some sort of Taylor
expansion to second order.  The assumption is that the contribution of a
complex trajectory falls off gaussian--like as it gets farther away from the
real trajectory.  This assumption is correct at least for simple systems like
the free particle and the harmonic oscillator and it was shown numerically to
hold for the quartic oscillator as well \cite{Xavi96}.

Returning to \refg{ba7}, we see that three terms in it prevent us from doing
the integral exactly: the two terms containing the function $S$, one in the
exponential and one in the prefactor, and the term containing $\cal I$.  In
accordance with the ideas expressed above, the $S$ in the exponential, which is
the value of the action for the complex classical trajectory, will be expanded
to second order in powers of $v''-v_r$, where $v_r$ (or ${z_r}^\star$) refers
to the final point of the {\it real} classical trajectory issued from $z'$,
thus
\begin{align}
  \label{3.4}
      \frac{\ri}{\hbar}S(v'',z',t) 
  \approx   \frac{\ri}{\hbar}S(v_r,z',t) + u_r(v''-v_r) 
      + \frac{1}{2}\gamma(v''-v_r)^2
\end{align}
with
\begin{align}
  \label{3.5a}
      u_r
   =  \left.\frac{\ri}{\hbar}\frac{\partial S}{\partial v''}
      \right|_{u'=z',v''=v_r}={v_r}^\star
\end{align}
and
\begin{align}
  \label{3.5}
      \gamma 
   =  \left.\frac{\ri}{\hbar}\:\frac{\partial^2 S}
      {\partial {v''}^2}\right|_{u'=z',v''=v_r}
   =  \frac{M_{uv}}{M_{vv}} 
   =  \frac{m_{qq}+\ri m_{qp}+\ri m_{pq}-m_{pp}}
      {m_{qq}+\ri m_{qp}-\ri m_{pq}+m_{pp}} \; .
\end{align}
Here we have used eqs.{\hskip 2pt}(\ref{en2}) and (\ref{enm}) relating second
derivatives of the action and elements of the tangent matrix.  A similar
expansion to first order in $v''-v_r$ can be done in the first equation
(\ref{ba8}), which allows us to calculate $p''$, the momentum at the end point
of the complex trajectory
\begin{align}
  \label{3.23a}
p'' \approx p_r + \ri\:\frac{1-\gamma}{1+\gamma}\:\frac{c}{b}\left(x-q_r\right) \; .
\end{align}

As for the second derivative of $S$ under the square root in the prefactor, it
presents us with a bit of a problem.  We know it (with its factor ${\rm
i}/\hbar$) to be equal to the inverse of $M_{vv}$, taken at $u' (=z')$ and
$v''$.  If we were to expand this in the vicinity of the real trajectory, we
would be taking derivatives of the tangent matrix, which is itself made up of
second derivatives of $S$, and going to higher order than anyone ever goes in
this kind of semiclassical argument.  Common practice would say: just replace
it by its value at the stationary point.  Unfortunately, as we already know,
the stationary point is {\it not} the real trajectory.  It is a complex
trajectory, hopefully rather close to the real one but not simple.  Moreover,
if we tried to do that, we would be back into the pitfall of mixed initial and
final conditions.  Hence the only reasonable thing to do for this presumably
weakly--varying term is to use its value for the real trajectory, which is the
place that will turn out to give the maximum contribution anyway.  (Of course
there are phase-space points $q''$ and $p''$ for which the difference $v''-v_r$
is not small. But at these points the propagator itself should be negligible).
The very same argument holds for the $\cal I$ term, since it already 
involves second derivatives of $\cal H$.
Hence we shall replace it by the value it takes for the real classical
trajectory issued from $(q',p')$, and we shall call it ${\cal I}_r$.

Here is the complete formula now, in terms of the variables $u''$ and
$w'':= v''-v_r$
\begin{align}
  \label{3.6}
  \begin{split}
      \langle x|K(t)|z'\rangle 
   =  \pi^{-\frac{1}{4}}b^{-\frac{1}{2}}\re^{\frac{i}{\hbar}{\cal I}_r}
      {M_{vv}}^{-\frac{1}{2}}\exp\left[ -\frac{x^2}{2b^2}+\frac{\ri}
      {\hbar}S(z_r^\star,z',t)-\frac{1}{2}|z'|^2 \right]  \\
      \int\frac{\rd u''\rd w''}{2\pi\ri}\exp\left[ -\frac{1}{2}{u''}^2
      +\frac{1}{2}\gamma{w''}^2 -u''w'' 
      +\left(\frac{\sqrt{2}}{b}x-v_r\right)u''+u_rw'' \right]
  \end{split}
\end{align}
where it is understood that $M_{vv}$ is taken for the real trajectory.  The
integral is now a pure gaussian and we can do it using formula (\ref{glg44}),
with
\begin{align}
  \label{3.7}
  \begin{split}
      a_1
   =  -\frac{1}{2}\;,\qquad 
      a_2
   =  \frac{1}{2}\gamma\;,\quad&\quad 
      a_3
   =  -1\;,\qquad
      b_1
   =  \frac{\sqrt{2}}{b}x-v_r\;,\qquad 
      b_2
   =  u_r\;,  \\
  &   a_3^2 - 4a_1a_2 
   =  1 + \gamma \; .
  \end{split}
\end{align}
The two quantities
\begin{align}
  \label{3.8}
      \mu_{1,2}
   =  -a_3\pm2\sqrt{a_1a_2}
   =  1\pm\ri\sqrt{\gamma}
\end{align}
must have a non--negative real part, which requires
\begin{align}
  \label{3.9}
      \left|{\rm Im}\sqrt{\gamma}\right|
   \le 
      1 \; .
\end{align}
We have shown, with a fair amount of algebra, that this condition is indeed
satisfied.  According to \refg{glg44}, the value of the gaussian integral,
i.e. the second line of \refg{3.6}, is
\begin{align}
  \label{3.10}
      \frac{1}{\sqrt{1+\gamma}}\exp\frac{1}{1+\gamma}\left[
      -\frac{{u_r}^2}{2} +u_r\left(\frac{\sqrt{2}}{b}x
      -v_r\right)+\frac{\gamma}{2}\left(\frac{\sqrt{2}}{b}x
      -v_r\right)^2 \right] \; .
\end{align}
All we need to do now is to simplify the result.

The new prefactor $(1+\gamma)^{-1/2}$, multiplied together with the other
prefactor $(M_{vv})^{-1/2}$, yields $(M_{vv}+M_{uv})^{-\frac{1}{2}}$, 
which is the same as $(m_{qq}+\ri m_{qp})^{-\frac{1}{2}}$.
Now let us rewrite $S(z_r^\star,z',t)$ in terms of the usual action of
Hamilton. Equation (\ref{glg83}) says
\begin{align}
  \label{3.12}
      \frac{\ri}{\hbar}S(z_r^\star,z',t)-\frac{1}{2}|z'|^2 
   =  \int_0^t\rd t'\left[ \frac{1}{2}(\dot{v}u-\dot{u}v)
      -\frac{\ri}{\hbar}H(u,v,t')\right] +\frac{1}{2}|z_r|^2 \; .
\end{align}
We rewrite the first term in the bracket in terms of $p$ and $q$:
\begin{align}
  \label{3.13}
      (\dot{v}u-\dot{u}v) 
  &=  \frac{1}{2}\left(\frac{\dot{q}}{b}-\ri\frac{\dot{p}}{c}\right)
      \left(\frac{q}{b}+\ri\frac{p}{c}\right)
      -\mbox{complex conjugate}  \nonumber \\
  &=  \ri\left(\frac{\dot{q}}{b}\frac{p}{c}-\frac{q}{b}\frac{\dot{p}}{c}
      \right) 
   =  \frac{\ri}{\hbar}(\dot{q}p-\dot{p}q)
\end{align}
since $bc=\hbar$~. With an integration by parts we have
\begin{align}
  \label{3.14}
      \int_0^t\!\rd t'(\dot{q}p-\dot{p}q) 
  &=  2\int_0^t\!\rd t'\,\dot{q}p\:- pq\Bigr|_0^t   \nonumber  \\
  &=  2\int_0^t \!p\rd q\;-p_rq_r +p'q'
\end{align}
where $q_r$ and $p_r$ are the final points of the real trajectory starting at
$(q',p')$.  As for the last term $\frac{1}{2}|z_r|^2$ of \refg{3.12}, it
will be convenient to write it $\frac{1}{2}u_rv_r$~.  Expression (\ref{3.12})
is therefore
\begin{align}
  \label{3.15}
      \frac{\ri }{\hbar}S(z_r^\star,z',t)-\frac{1}{2}|z'|^2 
   =  \frac{\ri }{\hbar}S_H -\frac{\ri }{2\hbar}p_rq_r
      +\frac{\ri }{2\hbar}p'q' +\frac{1}{2}u_rv_r
\end{align}
where $S_H$ is Hamilton's action
\begin{align}
  \label{3.16}
      S_H 
   =  \int_i^f(p\rd q-{\cal H}\rd t')
\end{align}
for the real trajectory.
Carrying this into \refg{3.6}, whose second line is \refg{3.10}, we find for
the complete exponent in $\langle x|K(t)|z'\rangle$, in addition to the
${\cal I}_r$ term:
\begin{align}
  \label{3.17}
\begin{split}
&-\frac{x^2}{2b^2} +\frac{\ri }{\hbar}S_H
-\frac{\ri }{2\hbar}p_rq_r+\frac{\ri }{2\hbar}p'q'+\frac{1}{2}u_rv_r \\
&\qquad+\frac{1}{1+\gamma}\left[ -\frac{u_r^2}{2} -u_rv_r+\frac{\gamma}{2}v_r^2
+\frac{\sqrt{2}}{b}x(u_r-\gamma v_r) +\gamma\frac{x^2}{b^2} \right] \; .
\end{split}
\end{align}
This is where the serious work of simplification begins.

We start by gathering all the terms containing $x$ and we complete the square
in $x$, which gives
\begin{align}
  \label{3.19}
  \begin{split}
      -\frac{{x}^2}{2b^2}\frac{1-\gamma}{1+\gamma}
 &  + \frac{x}{b\sqrt{2}}\frac{2(u_r-\gamma v_r)}{1+\gamma} \\
 &  = -\frac{1-\gamma}{1+\gamma}\left( \frac{x}{b\sqrt{2}} -
      \frac{u_r-\gamma v_r}{1-\gamma} \right)^2 +
      \frac{(u_r-\gamma v_r)^2}{1-\gamma^2} \; .
  \end{split}
\end{align}
The first term on the right hand side of \refg{3.19}, written in terms of $p_r$
and $q_r$, becomes
\begin{align}
  \label{3.21}
      -\frac{1-\gamma}{1+\gamma}
 &    \left( \frac{x}{b\sqrt{2}} 
      -\frac{q_r}{b\sqrt{2}} -\ri \,\frac{1+\gamma}{1-\gamma}\:
      \frac{p_r}{c\sqrt{2}} \right)^2  \nonumber\\
 &  = -\frac{1}{2}\:\frac{1-\gamma}{1+\gamma}\left(\frac{x-q_r}{b}
      \right)^2 +\frac{\ri }{\hbar}p_r(x-q_r)
      +\frac{1}{2}\:\frac{1+\gamma}{1-\gamma}\:\frac{p_r^2}{c^2} \; .
\end{align}
Return now to expression (\ref{3.17}) and gather all terms quadratic in
$(u_r,v_r)$, or in $(q_r,p_r)$, including the last term on the right hand side
of (\ref{3.19}):
\begin{align}
  \label{3.22}
      -\frac{\ri }{2\hbar}q_rp_r +\frac{1}{1+\gamma}\left[ 
      -\frac{u_r^2}{2} -u_rv_r +\frac{1+\gamma}{2}u_rv_r
      +\frac{\gamma}{2}v_r^2
      +\frac{(u_r-\gamma v_r)^2}{1-\gamma} \right] \; .
\end{align}
When this is written solely in terms of $q_r$ and $p_r$, much simplification
occurs and one is left with the single term
\begin{align}
  \label{3.23}
      -\frac{1}{2}\:\frac{1+\gamma}{1-\gamma}\:\frac{p_r^2}{c^2}
\end{align}
which cancels the last term of (\ref{3.21}). 

This is the end of the simplifications.  The final formula is
\begin{align}
  \label{3.24}
  \begin{split}
         \langle x|K(t)|z'\rangle 
  &=     \frac{\pi^{-\frac{1}{4}}b^{-\frac{1}{2}}}
         {\sqrt{m_{qq}+\ri m_{qp}}}\;\re^{\frac{i}{\hbar}{\cal I}_r}\;\exp
         \Biggl[-\frac{1}{2}\,\frac{1-\gamma}{1+\gamma}
         {\left(\frac{x-q_r}{b}\right)}^2 \\
  &      \qquad\qquad\qquad\qquad +\frac{\ri}{\hbar}
         \left\{p_r\left(x-q_r\right)
         +\frac{1}{2}p'q'+S_H\right\}\Biggr] \; .
  \end{split}
\end{align}
\textit{This} formula, and \textit{not} the HK formula, is the logical
consequence of transforming the semiclassical coherent state propagator to a
mixed representation by applying the standard semiclassical approximations.

\subsection{Some properties of the mixed propagator}
  \label{kap4.5}

According to eq.{\hskip 2pt}(\ref{3.24}) the mean position of the packet
$\langle\hat{q}\rangle$ and its mean momentum $\langle\hat{p}\rangle$ are given
by $q_r$ and $p_r$, the real classical trajectory originating at $q',p'$.  Its
squared width $\langle(\hat{q}-q_r)^2\rangle$ comes out of the real part of the
coefficient of the gaussian, which should be and is negative.  Since we have
\begin{align}
  \label{3.26}
      \frac{1-\gamma}{1+\gamma} 
  &=   \frac{m_{pp}-\ri m_{pq}}{m_{qq}+\ri m_{qp}} 
   =  \frac{1-\ri (m_{pp}m_{qp}+m_{qq}m_{pq})}{m_{qq}^2+m_{qp}^2}
\end{align}
we find
\begin{align}
  \label{3.26a}
\langle(\hat{q}-q_r)^2\rangle = \frac{b^2}{2}\left( m_{qq}^2+m_{qp}^2 \right)
   =  \Delta q_{class}^2
\end{align}
where $\Delta q_{class}^2$ is the classical spreading of a gaussian initial
ensemble corresponding to the initial phase space distribution
$|\langle p,q|p',q'\rangle|^2$.  In similar fashion we have
\begin{align}
  \label{3.26b}
\langle(\hat{p}-p_r)^2\rangle = \frac{b^2}{2}\left( m_{pq}^2+m_{pp}^2 \right)
   =  \Delta p_{class}^2 \; .
\end{align}
We see incidentally that the denominator of the prefactor in (\ref{3.24}) can
never vanish, and therefore the mixed propagator can never be singular, because
the determinant of the tangent matrix must equal unity.  This follows from
symplecticity, which in one dimension is the same as Liouville's theorem.

Another straightforward calculation yields the normalization of the packet
\begin{align}
  \label{3.33}
      N(t)
  &=  \int \rd x|\langle x|K(t)|z'\rangle |^2\nonumber\\
  &=  \frac{1}{\sqrt{\pi}b}\sqrt{m_{qq}^2+m_{qp}^2}\int \rd x\exp\left\{ 
      -\frac{1}{2}\left[ \frac{1-\gamma}{1+\gamma}
      +\frac{1-\gamma^\star}{1+\gamma^\star} \right]\frac{(x-q_r)^2}{b^2}
      \right\}
   =  1 \; .
\end{align}
Normalization is conserved and equal to unity at all times, which is as it
should be.

It may not be superfluous to mention once again that the sign of the square
root in \refg{3.24} is to be determined by continuous displacement along the
trajectory, given that this square root is unity at $t=0.$


\subsection{Comparison with Heller's IVR}
  \label{kap4.6}

Now we proceed to the comparison with Heller's approximation mentioned earlier.
This approximation is also known as ``the thawed gaussian approximation'' or
TGA.  A different derivation of the TGA was given later by Kay \cite{Kay94a}
and used by him in numerical comparisons with other approximations
\cite{Kay94b}.  Although Heller does not give a final formula, while Kay does,
we shall stick to Heller's presentation, as we do not find Kay's arguments
convincing: they seem to be based on convenience and (very limited) numerical
agreement, rather than solid basic principles.  Heller's paper assumes a
hamiltonian of the form $p^2/2m+V(q)$.  Here we shall consider a general
hamiltonian $H(p,q)$ and we shall carry the derivation all the way to an
explicit formula, which Heller does not do.  The result will be that Heller's
approximation leads to a formula identical to our \refg{3.24} except for two
differences: (1) the classical hamiltonian that must be used to compute the
trajectories, instead of being the smoothed $\cal H$ or the $Q$--symbol defined
in section 3, is the Weyl symbol $H_W$ of the quantum mechanical operator; (2)
the term $\re^{\frac{i}{\hbar}{\cal I}_r}$ is absent.  Hence the discussion of
section 3 returns to the fore: there exist indeed different approximations,
all legitimately derived, and the question becomes one of deciding under what
circumstances one or the other can be expected to be better.  We already saw in
section 3 that, in the extreme semiclassical limit, when $\hbar$ becomes very
small, Heller's approximation is expected to be best \cite{Kurc89}.  At low
energy, on the other hand, we expect (\ref{3.24}) to be best.  We gave a
qualitative argument for this in section 3, but we have also performed several
numerical comparisons, which we reserve for a future publication, as this paper
is probably too long already.

Heller's idea is to assume that the original wave--packet $\langle x|z'\rangle$
remains gaussian as it propagates, but that the parameters of the gaussian
change with time thus
\begin{align}
  \label{3.37}
\langle x|K(t)|z'\rangle = \exp\frac{i}{\hbar}\left[
\alpha(t)\left(x-q(t)\right)^2 + p(t)\left(x-q(t)\right) + \beta(t)\right]\;.
\end{align}
He further assumes that $q(t)$ and $p(t)$, which are the expectation values of
the operators $\hat{q}$ and $\hat{p}$, follow a real classical trajectory for
some classical hamiltonian $H(q,p)$.  On the other hand he takes $\alpha(t)$
and $\beta(t)$ to be complex, so that there are 6 undetermined real functions
in the formula.  The problem is to determine them.  The exact quantum
mechanical packet obeys the Schr\"odinger equation with a certain quantum
hamiltonian operator $\hat{H}(\hat{q},\hat{p})$.  What should be the connection
between $H$ and $\hat{H}$?  The most straightforward assumption, and the one
that works for the simple problems of elementary quantum mechanics, is to say
that they are Weyl--related, i.e. $\hat{H}$ is the Weyl operator associated
with $H$, and $H$ is the Weyl symbol associated with $\hat{H}$.  But we do not
want to make this assumption explicitly:  it should come out of the
calculation, and it will.

Heller argues that, since the wave--packet is small, the only part of the
quantum hamiltonian that matters is the part which, in phase space, refers to
the vicinity of the classical trajectory, i.e. the region where the
wave--packet is appreciably different from 0.  In that small region it is
permissible to expand the quantal hamiltonian up to second order in the
variables $\hat{q}-q$ and $\hat{p}-p$.  This ``local quadratic expansion'' is
the embodiment in this case of the semiclassical approximation, quite similar
to the stationary exponent approximation in section 2 and to the second order
expansion in \refg{3.4}.  Hence we expand the hamiltonian operator as follows
\begin{align}
  \label{3.38}
\hat{H}(\hat{q},\hat{p}) &= H(q,p)  \nonumber \\
&+ H_q(\hat{q}-q) + H_p(\hat{p}-p)  \\
&+ \frac{1}{2}\left\{ H_{qq}(\hat{q}-q)^2 + H_{pp}(\hat{p}-p)^2
+H_{qp}\left[ (\hat{q}-q)(\hat{p}-p) + (\hat{p}-p)(\hat{q}-q)
\right]\right\}  \nonumber
\end{align}
where $\hat{q}$ is the operator `multiplication by $x$', $\hat{p}$ is
$-i\hbar\partial/\partial x$, and $H_q,H_p,H_{qq},H_{pp},H_{qp}$ are the first
and second derivatives of $H(q,p)$, which are functions of $q$ and $p$ and do
not contain any operators.  The crucial point to note here is that, if
formula (\ref{3.38}) is true, the Weyl symbol of the operator
$\hat{H}(\hat{q},\hat{p})$ is simply $H(q,p)$, without any additional terms.
This comes out from repeated applications of the Wigner transformation
(\ref{5mb18}), where one does the integrals by using the standard properties of
the Dirac delta--function and its derivatives.  Denoting the Weyl symbol with
the subscript $W$, one finds
\begin{alignat}{4}
  \label{3.65}
      (\hat{q})_W
  &=  q
  &   (\hat{p})_W
  &=  p  \nonumber\\
      (\hat{q}^2)_W
  &=  q^2
  &   (\hat{p}^2)_W
  &=  p^2  \\
      (\hat{q}\hat{p})_W
  &=  qp + \frac{1}{2}i\hbar \qquad
  &   \qquad (\hat{p}\hat{q})_W
  &=  pq - \frac{1}{2}i\hbar  \nonumber \\
  &\qquad\qquad\qquad\qquad\text{etc}\ldots \; . \nonumber
\end{alignat}
Note that the $Q$--symbol of the right hand side of (\ref{3.38}) would be very
different.  For instance, while the Weyl symbol of $(\hat{q}-q)^2$ vanishes,
its $Q$--symbol is the squared width of the packet.  Thus we have chosen
$H(q,p)$ to be the Weyl symbol of $\hat{H}$, and this is not a trivial choice.
It is however a purely arbitrary choice for now.  The crucial moment will come
when we prove that $q(t)$ and $p(t)$ obey Hamilton's equations for this
particular classical $H(q,p)$.  This moment is close at hand.  Meanwhile, we
note that the last bracket $[\cdots]$ of (\ref{3.38}) can be written
\begin{align}
2(\hat{q}-q)(\hat{p}-p)-i\hbar  \; . \nonumber
\end{align}

Now we try to satisfy the Schr\"odinger equation $i\hbar\partial\psi/\partial
t = \hat{H}\psi$.  For $\psi(t)$ we substitute the wave--packet (\ref{3.37})
and for $\hat{H}$ we substitute (\ref{3.38}).  We find
\begin{align}
  \label{3.39}
\frac{1}{\psi}i\hbar\frac{\partial\psi}{\partial t} &\equiv -\dot{\alpha}
(x-q)^2 + 2\alpha\dot{q}(x-q) - \dot{p}(x-q) + p\dot{q} - \dot{\beta}
   \nonumber  \\
\frac{1}{\psi}\hat{H}\psi &\equiv H(q,p) + H_q(x-q) + \frac{1}{2}H_{qq}(x-q)^2
+ 2H_p\alpha(x-q)  \\
&\qquad + H_{pp}\left[2\alpha^2(x-q)^2-i\hbar\alpha\right]
+ H_{qp}\left[2\alpha(x-q)^2-\frac{1}{2}i\hbar\right]  \nonumber \;.
\end{align}
Matching powers of $(x-q)$, we get the following 3 complex equations to
determine the 6 unknown real functions
\begin{align}
  \label{3.41}
\dot{\alpha} &= -2\alpha^2H_{pp} -2\alpha H_{qp} -\frac{1}{2}H_{qq}  \\
  \label{3.42}
2\alpha\dot{q}-\dot{p} &= 2\alpha H_p+H_q  \\
  \label{3.43}
\dot{\beta} &= p\dot{q}-H(q,p)+i\hbar\alpha H_{pp}+\frac{1}{2}i\hbar H_{qp}\;.
\end{align}
We start with the imaginary part of (\ref{3.42}), which gives
\begin{align}
  \label{3.44}
\dot{q} = H_p \; .
\end{align}
Then the real part of the same equation says
\begin{align}
  \label{3.45}
\dot{p} = -H_q \; .
\end{align}
Hence $q(t)$ and $p(t)$ follow a classical trajectory of $H$, provided
that the latter is the Weyl symbol of $\hat{H}$.  Two equations
remain, (\ref{3.41}) and (\ref{3.43}).

To solve (\ref{3.41}) we notice that this equation becomes identical to that
for $X(t)$, \refg{glg17}, if we identify $v$ with $q$, $u$ with $p$ and
$\frac{i}{\hbar} {\cal H}$ with $H$. Following the calculation that leads to
\refg{glg31}, we find immediately
\begin{align}
\alpha = \frac{1}{2} \frac{\delta p}{\delta q}
\end{align}
where $\delta q$ and $\delta p$ are deviations from the classical trajectory,
which satisfy 
\begin{align}
\delta \dot{q} &= H_{qp} \delta q + H_{pp} \delta p \\
\delta \dot{p} &= -H_{qq} \delta q - H_{pq} \delta p \; .
\end{align}
Using the tangent matrix \refg{glg4.1} we get
\begin{align}
\alpha = \frac{1}{2} \; \frac{\frac{c}{b} m_{pq} \delta q' + m_{pp} \delta p'}
  {m_{qq} \delta q' + \frac{b}{c} m_{qp} \delta p'} \; = \; 
  \frac{1}{2} \; \frac{\frac{c}{b} m_{pq}  + 2 m_{pp} \alpha'}
  {m_{qq} + 2 \frac{b}{c} m_{qp} \alpha'}
\end{align}
where
\begin{align}
\label{alphaini}
\alpha' = \frac{1}{2} \frac{\delta p'}{\delta q'} = i\hbar/2b^2 = ic/2b \; .
\end{align}
Therefore 
\begin{align}
\label{alphasol}
\alpha &=   \frac{c}{2b} \; \frac{ m_{pq}  + i m_{pp}}{m_{qq} + i m_{qp}} 
        =  \frac{ic}{2b} \; \frac{ m_{pp}  - i m_{pq}}{m_{qq} + i m_{qp}} 
            \nonumber \\
       &=  \frac{ic}{2b} \; \frac{ 1-\gamma}{1+\gamma} \; .
\end{align}
Equation (\ref{3.43}) can now be integrated.  It can be written
\begin{align}
\dot{\beta} &= p \dot{q} - H(q,p) + 
  \frac{i \hbar}{2} \frac{\delta p}{\delta q} H_{pp} +  \frac{i \hbar}{2} H_{qp} 
	\nonumber \\
  &= L + \frac{i \hbar}{2\delta q} \left(\delta p H_{pp}+\delta q H_{qp} \right) 
    = L + \frac{i \hbar}{2\delta q} \delta \dot{q} \nonumber \\
  &= L + \frac{i \hbar}{2} \frac{\rd}{\rd t} (\log{\delta q})
\end{align}
where $L$ is the Lagrangian. Integrating both sides gives
\begin{align}
\label{beta1}
\beta &= S_H + \frac{i \hbar}{2} \log{\delta q} 
\end{align}
up to an additive constant, $S_H$ being Hamilton's action (for the Weyl
hamiltonian), and with
\begin{align}
\label{dqalp}
\delta q = m_{qq} \delta q' + \frac{b}{c} m_{qp} \delta p' =
\delta q' ( m_{qq}  + \frac{2b}{c} m_{qp} \alpha') \; .
\end{align}
The initial value $\delta q'$ is determined by 
$\beta(0)=q'p'/2 = i\hbar/2 \log{\delta q'}$, which gives
\begin{align}
\label{dqini}
\delta q' = \re^{-i q'p'/\hbar} \; .
\end{align}
Substituting \refgs{dqalp}, (\ref{dqini}), and (\ref{alphaini}) into 
(\ref{beta1}) gives
\begin{align}
\label{beta}
\beta &= S_H + \frac{i \hbar}{2} 
  \log{\left[\delta q'(m_{qq}+ \frac{2b}{c} \alpha' m_{qp})\right]}\nonumber\\
 &= S_H + \frac{i \hbar}{2} \log{\left[ \re^{-i q'p'/\hbar}(m_{qq}+i m_{qp})
   \right]}   \nonumber \\
 &= S_H + \frac{p'q'}{2} + \frac{i \hbar}{2} \log{(m_{qq}+i m_{qp})}
\end{align}
or
\begin{align}
\label{ebeta}
\re^{\frac{i \beta}{\hbar}} = \frac{1}{\sqrt{m_{qq}+i m_{qp}}}
  \re^{\frac{i}{\hbar}(S_H + p'q'/2)}
\end{align}
up to a multiplicative constant.

With these expressions for $\alpha$ and $\beta$, one sees immediately
that the Heller wave--packet (\ref{3.37}) is identical to ours
(\ref{3.24}), except for the different hamiltonian (Weyl's) and the
absence of the ${\cal I}_r$ exponential. Equation (\ref{3.33}) shows
that the normalization of the Heller packet is conserved at all times.

\subsection{Recovering Van Vleck's Formula from the IVR}

The semiclassical limit of the evolution operator in the coordinate
representation is given by the well known Van Vleck formula \cite{Van28}. It is
desirable, therefore, that other semiclassical representations of this operator
reduce to Van Vleck's when transformed back to coordinates. In this subsection
we shall calculate
\begin{align}
 \label{s6.1}
K(x'',t;x',0) := \langle x''|K(t)|x' \rangle =
  \int\langle x''|K(t)|z'\rangle\frac{\rd^2z'}{\pi}
      \langle z'|x'\rangle
\end{align}
for both mixed propagators, Heller's and ours.  The integration over $q'$ and
$p'$ will be performed by the stationary exponent approximation. We shall see
that Heller's approximation recovers Van Vleck's formula exactly. Our
semiclassical approximation recovers it only in the limit of small $\hbar$ (see
discussion in section 3), but this is also the only limit in which the Van
Vleck approximation can be justified.

We start by inserting the Heller propagator in eq.{\hskip 2pt}(\ref{s6.1}).
Using expression (\ref{glgmb1}) for $\langle x|z \rangle$ we get
\begin{align}
  \label{s6.2h}
K_{\text{Heller}}(x'',t;x',0) &= \int \frac{\rd q'\rd p'}{2\pi\hbar} 
\frac{\pi^{-\frac{1}{2}} b^{-1}}
{\sqrt{\left(m_{qq}+\ri m_{qp}\right)}} \nonumber\\
 &\times \mbox{exp}
\left[-\frac{1}{2}\,\left(\frac{1-\gamma}{1+\gamma}\right)\,
{\left(\frac{x''-q_r}{b}\right)}^2
+\frac{\ri}{\hbar}\left\{p_r\left(x''-q_r\right)
+\frac{1}{2}p'q'+S_H\right\}\right] \nonumber\\
 &\times   \exp\left[-\frac{{(x'-q')}^2}{2 b^2} \;
        - \frac{i}{\hbar}p'(x' - q'/2)\right] \; .
\end{align}
We call $\xi=\xi(q',p')$ the exponent in the second and third lines of
(\ref{s6.2h}). The stationary conditions are $\partial \xi/\partial q' =0$
and $\partial \xi/\partial p' =0$. We write $\xi$ explicitly first:
\begin{align}
\label{s6.3h}
\xi = -\frac{1}{2b^2}\left[\left(\frac{1-\gamma}{1+\gamma}\right)(x''-q_r)^2+
(x'-q')^2\right]
+ \frac{i}{\hbar}\left[p_r(x''-q_r)+p'(q'-x')+S_H \right] \; .
\end{align}
The derivatives are: 
\begin{align} 
\label{s6.4h} 
\frac{\partial \xi}{\partial q'} &= \frac{1}{b^2}\left[\left(\frac{1-\gamma}
{1+\gamma}\right)(x''-q_r)\frac{\partial q_r}{\partial q'}+(x'-q')\right] + 
\frac{i}{\hbar}\left[\frac{\partial p_r}{\partial q'}(x''-q_r)
-p_r\frac{\partial q_r}{\partial q'}  \right. \nonumber \\ 
 & \qquad\qquad  \left. + p' + \frac{\partial S_H}{\partial q'} +
\frac{\partial S_H}{\partial q_r}\frac{\partial q_r}{\partial q'}\right] 
\nonumber\\
 &= \frac{1}{b^2}\left[(x''-q_r)\left(\left(\frac{1-\gamma}{1+\gamma}\right)
    m_{qq}+ i m_{pq}\right)+(x'-q')\right] \nonumber \\
 &= \frac{1}{b^2}\left[\left(\frac{1}{m_{qq}+im_{qp}}\right)(x''-q_r)+(x'-q')
 \right]
\end{align} 
where we have used $\frac{\partial S_H}{\partial q'}=-p'$, 
$\frac{\partial S_H}{\partial q_r}=p_r$, and eqs.(\ref{glg4.1}). Similarly 
we find
\begin{align} 
\label{s6.5h} 
\frac{\partial \xi}{\partial p'} &= \frac{i}{\hbar}
\left[\left(\frac{1}{m_{qq}+im_{qp}}\right)(x''-q_r)-(x'-q')\right] \; .
\end{align} 
The stationary conditions are satisfied if $q'=x'$ and
$q_r(q',p',t)=x''$.  This last equation defines $p'$ implicitly, so
that the contributing trajectory is the one that leaves $x'$ at time
zero and reaches $x''$ at time $t$.  The value of $\xi$ computed at
the stationary trajectory is simply $i S_H /\hbar$. In order to
perform the integrals we need the second order derivatives of
$\xi$. The algebra is straightforward and the results
are:
\begin{align} 
\label{s6.15} 
\xi_{qq} = -\frac{1}{b^2} \; \frac{2 m_{qq}+i m_{qp}}{m_{qq}+i m_{qp}}
\end{align}
\begin{align} 
\label{s6.16} 
\xi_{qp}=\xi_{pq}= -\frac{1}{\hbar} \; \frac{m_{qp}}{m_{qq}+i m_{qp}}
\end{align}
\begin{align} 
\label{s6.17} 
\xi_{pp} = -\frac{i}{c^2} \; \frac{m_{qp}}{m_{qq}+i m_{qp}} \quad .
\end{align}
Notice that all derivatives of $\xi$ were calculated keeping the elements
of the monodromy matrix fixed. Their variations would involve
the computation of third or higher order derivatives of $S$.

Inserting all these expressions into eq.{\hskip 2pt}(\ref{s6.2h}), we find that
the coordinate propagator becomes
\begin{align}
  \label{s6.18}
K_{\text{Heller}}(x'',t;x',0) &= 
\pi^{-\frac{1}{2}} b^{-1} \frac{1}{\sqrt{m_{qq}+\ri m_{qp}}} 
\; \re^{i S_H/\hbar} \nonumber\\
&\times \int \frac{\rd Q\rd P}{2\pi\hbar} \mbox{exp}
\left[\frac{1}{2} (\xi_{qq} Q^2 + 2\xi_{qp} Q P + \xi_{pp} P^2) \right]
  \nonumber\\
&= \pi^{-\frac{1}{2}} b^{-1} \frac{1}{\sqrt{m_{qq}+\ri m_{qp}}} \; \times 
\frac{\re^{i S_H/\hbar}}{\hbar \sqrt{\xi_{qq}\xi_{pp}-\xi_{qp}^2}}
\end{align}
where $Q$ and $P$ are the variations of $q'$ and $p'$ from the stationary
point.  The quantity under the second square root in the last line is the
determinant of the quadratic form in the second line.  By eqs.{\hskip
2pt}(\ref{s6.15}) to (\ref{s6.17}) it is equal to
\begin{align}
  \label{s6.19}
\xi_{qq}\xi_{pp}-\xi_{qp}^2 = \frac{2 i m_{qp}}{\hbar^2 (m_{qq} + \ri m_{qp})}
\end{align}
and the final result is
\begin{align}
  \label{s6.20}
K_{\text{Heller}}(x'',t;x',0) &= \frac{1}{b\sqrt{2\pi i m_{qp}}} \;
    \re^{i S_H /\hbar} 
\end{align}
which is Van Vleck's famous expression.

Given the close similarity between Heller's wave--packet and ours, an identical
calculation with ours will obviously give the result
\begin{align}
  \label{s6.21}
   K(x'',t;x',0) &= \frac{1}{b\sqrt{2\pi i m_{qp}}} \; 
  \re^{i (S_H+{\cal I})/\hbar} 
\end{align}
all classical quantities being calculated with the smoothed hamiltonian
${\cal H}$ instead of the Weyl $H$.  This reduces to the Van Vleck formula for
small $\hbar$, and it might well be better for larger $\hbar$, but we have no
evidence for this at the moment. 

\section{Comparison with the Herman-Kluk Propagator}
 \label{newchap5}

In addition to Heller's thawed gaussian approximation, there is in the
literature a different initial value representation formula for the mixed
propagator $\langle x|K(t)|z\rangle$.  This is the Herman-Kluk formula
\cite{Herm84}, derived by convoluting the Van Vleck propagator with coherent
states and performing the resulting integrals by the method of stationary
phase.  This formula has been used many times in the last few years.

We already pointed-out in section 4.1 the mistake of Grossmann and Xavier
\cite{Gros98b} in their tentative derivation of the HK formula from the
semiclassical coherent-state propagator (\ref{glg119}). In fact, in their
original paper \cite{Herm84}, Herman and Kluk make a similar mistake in their
evaluation of the stationary phase integrals. When performing these integrals
they find (as we found in section 4.1) that the stationary trajectory is
complex and given by mixed boundary conditions (equations (16) and (17) of
\cite{Herm84}). However, instead of expanding the exponent of the integrand for
this complex trajectory about a nearby real trajectory, they make a change of
variables to initial position and momentum and assume these new variables to be
real. Therefore, from the point of view of semiclassical analysis, the HK
formula is incorrect. In subsection 5.2 we shall discuss why, in spite of that,
it may still work.

Here is the Herman-Kluk formula, for the mixed representation introduced in
(\ref{ba3}), taken from \cite{Kluk86} with some adjustment of notations.
\begin{align}
  \label{3.25}
\begin{split}
\langle x|K(t)|z'\rangle_{HK} &= \pi^{-\frac{1}{4}}b^{-\frac{1}{2}}
\sqrt{\frac{1}{2}\left(m_{pp}+m_{qq}-\ri m_{qp}+\ri m_{pq}\right)} \\
&\qquad\quad\mbox{exp}
\left[-\frac{1}{2}\,{\left(\frac{x-q_r}{b}\right)}^2
+\frac{\ri}{\hbar}\left\{p_r\left(x-q_r\right)
+\frac{1}{2}p'q'+S_H\right\}\right] \; .
\end{split}
\end{align}
Once again, the ${\cal I}_r$ term is absent and the classical hamiltonian used
by HK is the Weyl hamiltonian.  Besides these, there are two other differences
between \refg{3.24} and \refg{3.25}, both of them quite important at first
sight.  One is in the coefficient of the gaussian exponent, the other is in the
prefactor.  The coefficient $(1-\gamma)/(1+\gamma)$ of the gaussian in
(\ref{3.24}) was essential for obtaining the right semiclassical widths
(\ref{3.26a}) and (\ref{3.26b}), but it is absent in HK!  The difference in the
prefactors is also astonishing.  One formula has the square root occurring in
the denominator and the other in the numerator.  But actually the two
prefactors are related by
\begin{align}
  \label{3.30}
 \sqrt{\frac{1}{2}\left(m_{pp}+m_{qq}-\ri m_{qp}+\ri m_{pq}\right)}
 = \frac{1}{\sqrt{m_{qq}+\ri m_{qp}}} \sqrt{\frac{1+\gamma}{1-|\gamma|^2}}
\end{align}
which shows that both differences between the two formulae imply that HK set
$\gamma$ equal to 0.  This is quite generally incorrect, except for the plain
harmonic oscillator, for which both formulae give identical results.

There is another major difference between formulae (\ref{3.24}) and
(\ref{3.25}), the normalization, which is for (\ref{3.25})
\begin{align}
  \label{3.34}
      N_{HK}(t)
  &=  \frac{1}{2\sqrt{\pi}b}\sqrt{(m_{qq}+m_{pp})^2+(m_{pq}-m_{qp})^2}
      \int \rd x\exp\left\{ -\frac{(x-q_r)^2}{b^2}
      \right\}\nonumber\\
  &=  \frac{1}{\sqrt{1-|\gamma(t)|^2}} \; .
\end{align}
This normalization is not conserved since $\gamma$ is usually time-dependent.
This is a grave flaw which shows up in almost every example.  For the free
particle, for instance, (\ref{3.24}) is exact, while (\ref{3.25}) fails
completely in describing the spreading of the wave--packet.

It may be argued with some validity that, since the $|z'\rangle$'s form an
overcomplete set, the fact that the propagation of each $|z'\rangle$ according
to HK is wrong does not necessarily mean that a wrong result will always occur
when a $|z'\rangle$--integral is performed, as in \refg{ba3}. This is in
fact a crucial point concerning the applicability of the HK formula, and
we shall discuss it in detail in subsection 5.2.

\subsection{A numerical example}

We show here a detailed numerical illustration of the differences
between HK and our semiclassical IVR formula.  We consider the
scattering of a particle with initial wave function $\langle
x|z'\rangle$ by a potential barrier. We choose the following
Hamiltonian to test our results:
\begin{align}
 \begin{split}
  \label{n1}
   H &=  \frac{p^2}{2}  + V_0 \left[ \re^{\alpha(x-A)} + \re^{-\alpha(x+A)} 
    \right] \\ 
  &=  \frac{p^2}{2} + 2\; V_0 \re^{-\alpha A} \cosh{\alpha x}     
 \end{split}
\end{align}
where $V_0$, $\alpha$ and $A$ are parameters. The first term in the 
potential function represents an exponential wall located
at $x=+A$. The second exponential {\em closes} 
the system at the left end with a second wall at $x=-A$.

The wave-packet will be launched from $x = 0$ with positive
momentum. If $A$ is large compared to the coherent state width $b$, the 
packet will not see this left wall in its way towards the first collision.
However, the fact that the motion is bound simplifies the quantum mechanical
treatment, avoiding the complications of the continuous spectrum. We
have set $A=5$ and $\alpha = V_0 = 1$ for the potential and $q'=0$, $p'=1$
and $b=0.3$ for the initial coherent state $|z' \rangle$. Planck's constant
was set to $\hbar=0.05$, which gives $c=\hbar/b \approx 0.167$.

For short times the particle experiences almost no force, being well 
described by the free particle propagation. In this approximation the
classical trajectory is just
\begin{align}
 \begin{split}
  \label{n2}
   q &=  q' + p' t \\ 
   p &=  p'
 \end{split}
\end{align}
All the quantities entering the semiclassical formulae can be computed
immediately and the result is
\begin{align}
 \begin{array}{lll}
  \label{n3}
   m_{qq} = m_{pp} = 1 & m_{pq}=0 & \quad m_{qp}=ct/b \\
   S_H = {p'}^2 \; t/2 - c^2t/4 & & \quad {\cal I}_r = c^2t/4 \\
   \gamma = itc/(itc+2b) & & 
 \end{array}
\end{align}
Substituting these expressions in \ref{3.24} gives
\begin{align}
  \label{n4}
  \begin{split}
         \langle x|K(t)|z'\rangle 
  &=     \frac{\pi^{-\frac{1}{4}}b^{-\frac{1}{2}}}
         {\sqrt{1+itc/b}}\;\exp
         \Biggl[-\frac{1}{2}\,\frac{1}{1+itc/b}
         {\left(\frac{x-q'-p't}{b}\right)}^2 \\
  &      \qquad\qquad\qquad +\frac{\ri}{\hbar}
         \left\{p'\left(x-q'-p't\right)
         +\frac{1}{2}p'q'+{p'}^2t/2\right\}\Biggr] 
  \end{split}
\end{align}
which coincides with the exact quantum mechanical result. The HK
formula, on the other hand gives
\begin{align}
  \label{n5}
  \begin{split}
         \langle x|K(t)|z'\rangle_{HK}
  &=     \pi^{-\frac{1}{4}}b^{-\frac{1}{2}}
         \sqrt{1-itc/2b}\;\exp
         \Biggl[-\frac{1}{2}\,
         {\left(\frac{x-q'-p't}{b}\right)}^2 \\
  &      \qquad\qquad\qquad +\frac{\ri}{\hbar}
         \left\{p'\left(x-q'-p't\right)
         +\frac{1}{2}p'q'+{p'}^2t/2\right\}\Biggr] \; .
  \end{split}
\end{align}
This shows that HK not only describes poorly the width of the evolved packet 
but it also produces a non-physical increase in the total probability.

For times of the order of $A/p'$ the free particle approximation is
no longer valid and we have to solve the problem numerically. The exact
quantum mechanical solution of 
$H |\Psi_n \rangle = E_n |\Psi_n \rangle $ for the Hamiltonian (\ref{n1})
was performed by diagonalizing $H$ using as basis states 
$\langle x|\phi_n \rangle = \frac{1}{\sqrt{L}} \sin{(n\pi x/2L + n\pi/2)}$ 
where the width $L$ is chosen such that $V(L)=E_{max}$.
This is in fact the most accurate method (when you can use it).
$E_{max}$ is an 
upper limit to the eigen-energies to be calculated and it is related to the 
number $N$ of basis states used in the diagonalization by 
$E_{max}=N^2 \pi^2 \hbar^2/(8L^2)$. This guarantees that the basis
states span the energy interval from $0$ to $E_{max}$ and that the
part of the potential with $-L \le x \le L$ has $V(x) < E_{max}$.
We have used $N=400$ in our computations, which gives $E_{max} \approx
9.4$ and $L \approx 7.2$. With this choice the first 260 energy levels
converge with at least 5 digits (as compared to a larger diagonalization),
spanning an energy interval from $0$ to $5.5$. The initial wave packet 
$\langle x|z' \rangle$ can be easily expressed in terms of the 
$|\phi_n \rangle$ basis and, therefore, in the basis
$|\Psi_n \rangle$ of eigenstates of $H$.

The semiclassical calculation of both HK and our semiclassical formula
needs only one single trajectory starting from $(q',p')$ and its tangent
matrix to compute the whole function  $\langle x|K(t)|z'\rangle$. In our
formula this trajectory is a solution of Hamilton's equations 
for the smoothed Hamiltonian  $\langle v|\hat{H}|u \rangle$, which is given by
\begin{align}
  \label{n6}
   {\cal H}  = \frac{p^2}{2} + 2\; V_1 \re^{-\alpha A} \cosh{\alpha x} + 
\hbar^2 b^2 /4
\end{align}
where $V_1 = V_0 \exp{(b^2 \alpha^2/4)}$.
The results are shown in Figures 1a to 1d. Figure 1a shows the
square modulus of the wave function for the times
$t=0$ and $t=4$. The gaussian at t=0 represents the initial
coherent state. The dashed line shows the (properly scaled) potential 
function and the dashed vertical bar shows the location of the classical 
turning point. The lines at $t=4$ show the time evolved wave-function 
according to the exact calculation (solid), our formula (dashed) and the HK 
approximation (dotted). Notice the increase in the height of the HK gaussian,
in opposition to the exact evolution, which spreads and decreases. Our
semiclassical formula is in very good agreement with the exact calculation.
For $t=6$, Figure 1b, one has the impression that HK is not 
so bad, since the wave packet height increases again. That is, however, only
a fortuitous occurrence, as demonstrated in Figure 1c, for $t=8$ and Figure
1d for $t=10$. For longer times the peak of the HK gaussian increases more 
and more, whereas the exact wave packet height decreases to compensate for 
the spreading. At times of the order of $25$ (not shown) the exact 
propagator (and our semiclassical formula) has a peak of height around 
$0.2$ while HK's peak is around $8$. At this time the width in HK is also
completely wrong.

\begin{figure}[ht]
  \begin{center}
  \includegraphics[height=100mm,clip]{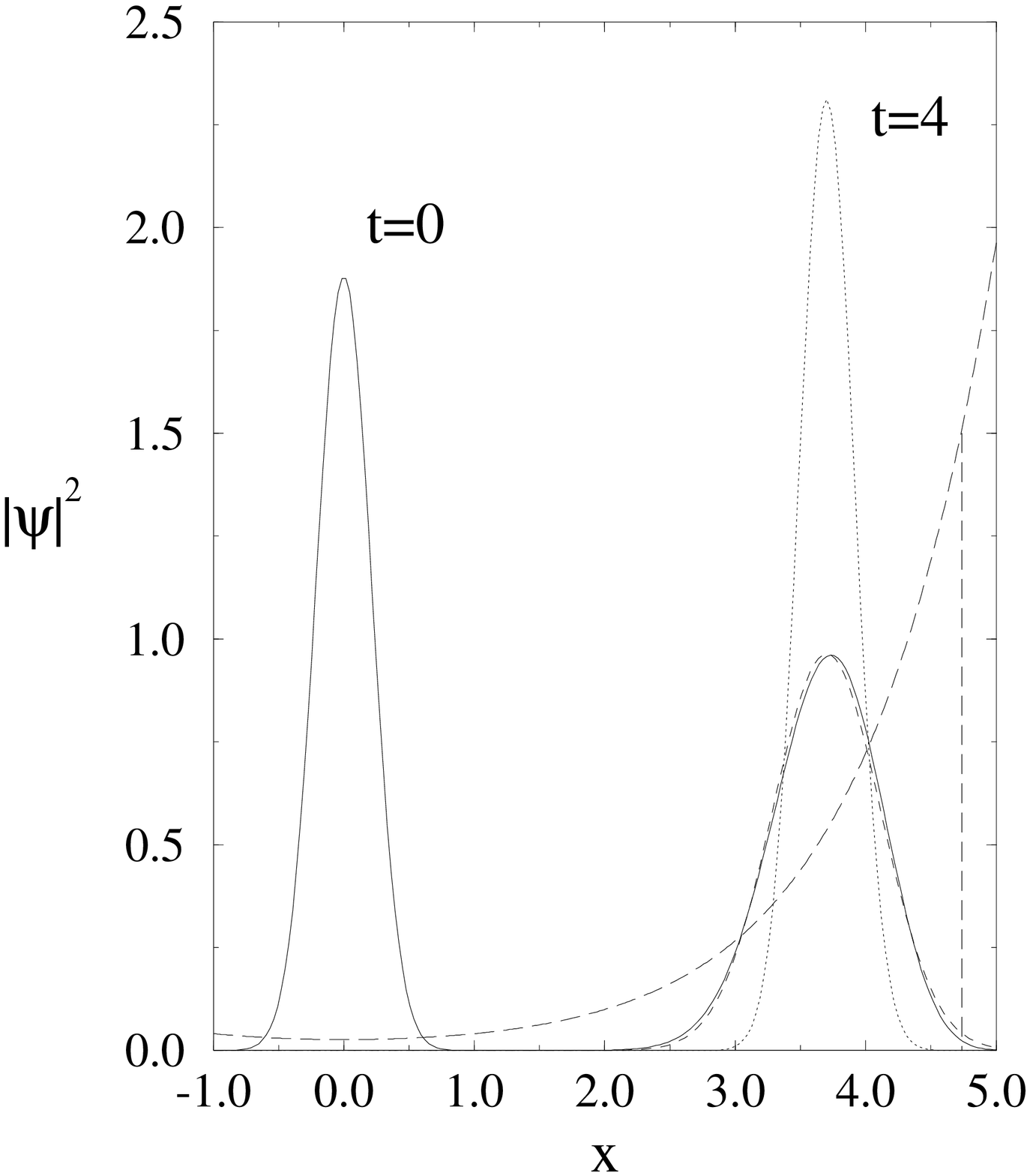}
  \includegraphics[height=100mm,clip]{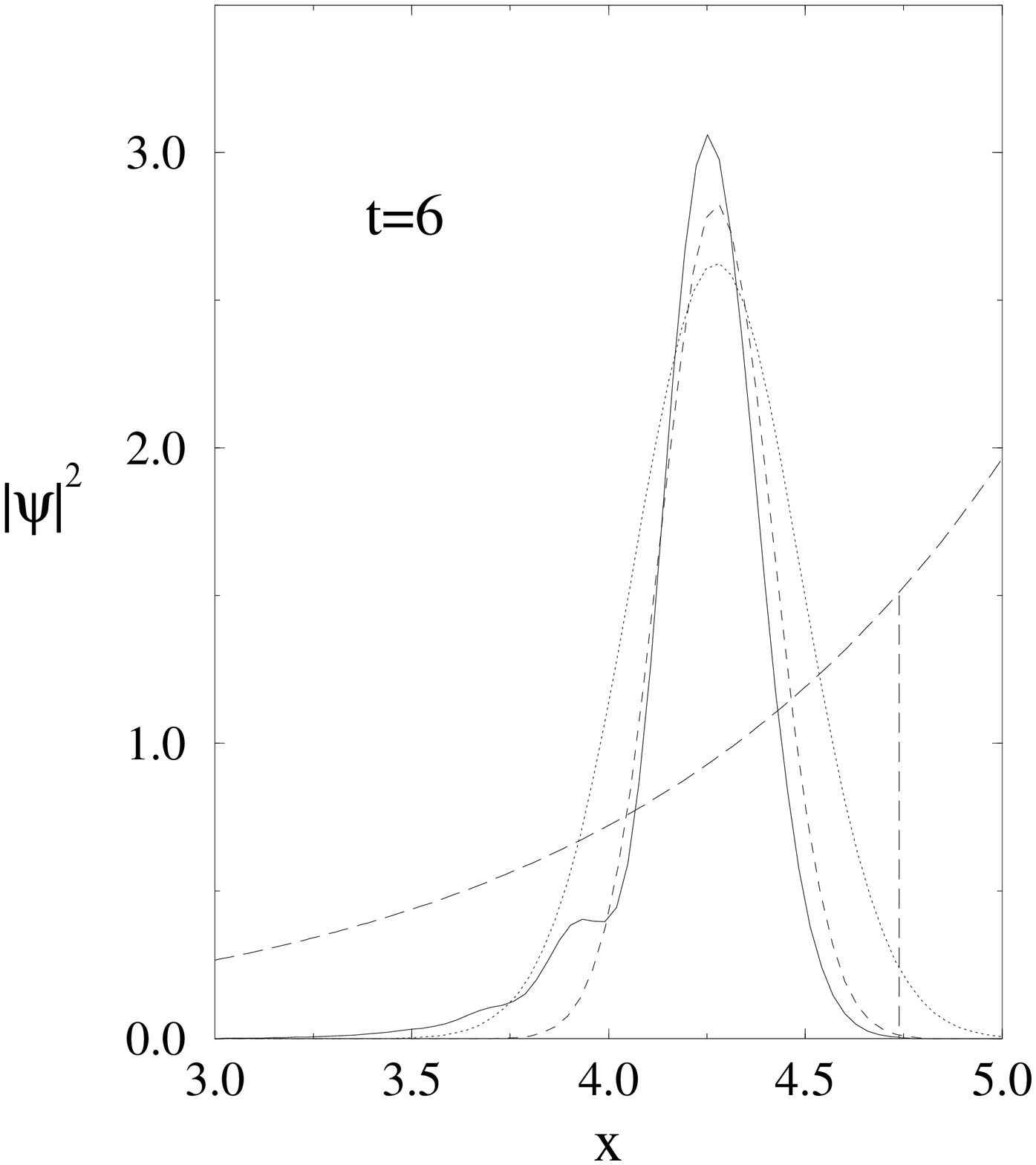}
  \includegraphics[height=100mm,clip]{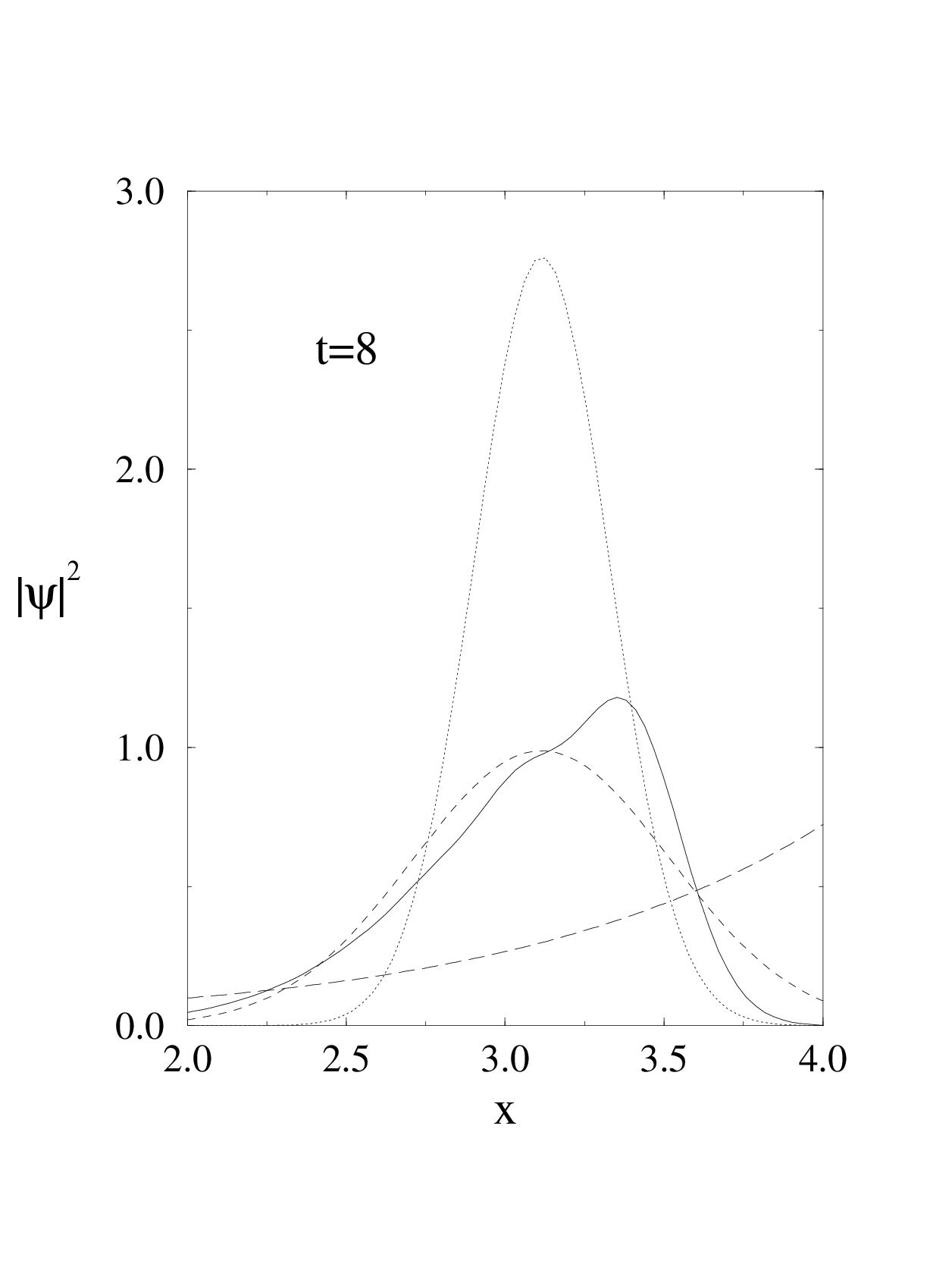}
  \includegraphics[height=100mm,clip]{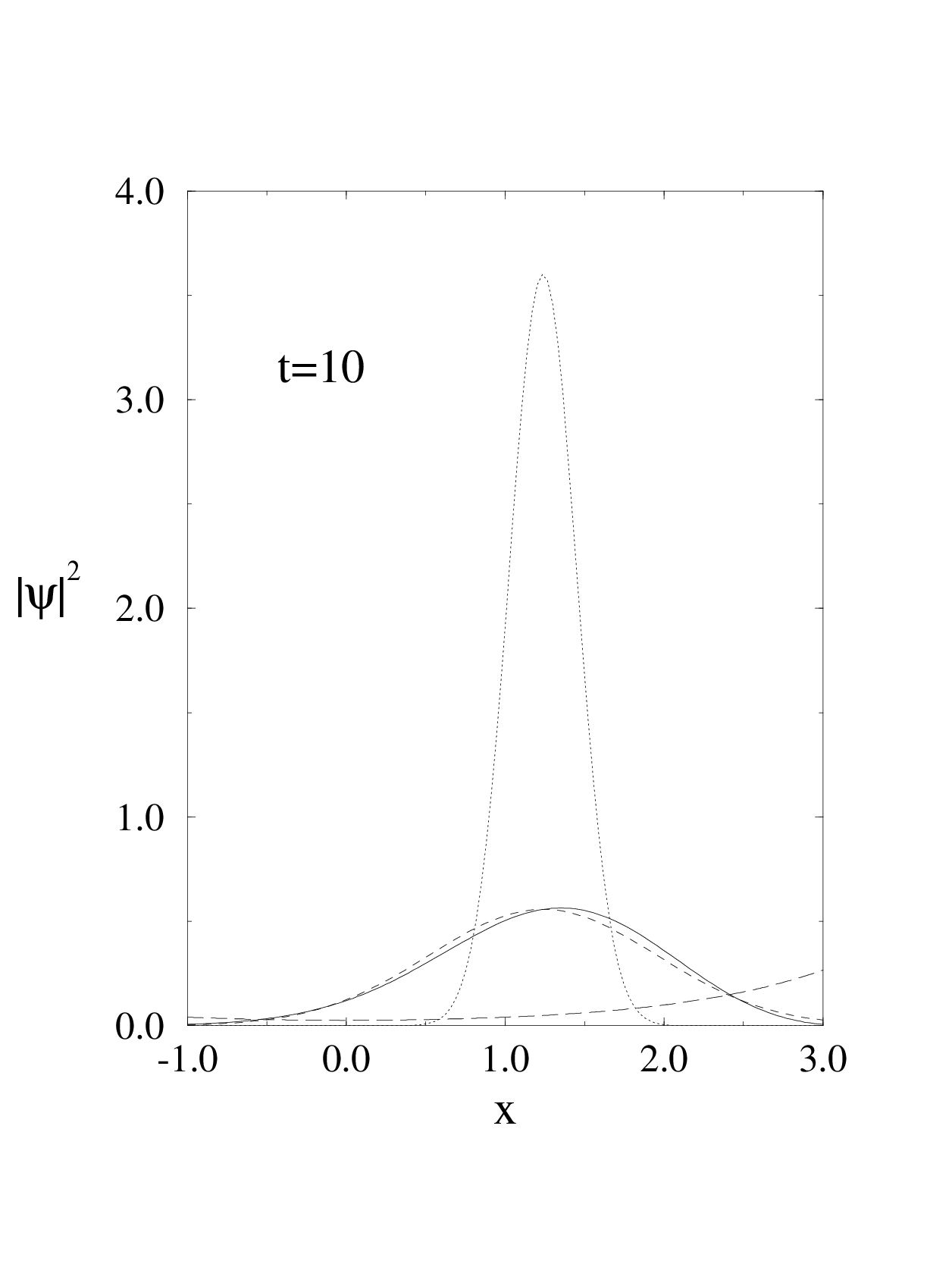}
  \end{center}
  \caption[]{\it \label{fig1}Square modulus of the time evolved coherent 
  state launched from q'=0 and p'=1. The dashed line shows the potential 
  function and the dashed bar shows the location of the classical turning 
  point. The solid line shows the exact propagation, the dashed line shows
  our semiclassical formula and the dotted line shows the HK approximation.
  (a) initial wave packet at $t=0$ and exact and semiclassical calculations 
  at $t=4$. 
  (b) exact and semiclassical calculations at $t=6$;
  (c) exact and semiclassical calculations at $t=8$;
  (d) exact and semiclassical calculations at $t=10$.}
\end{figure}

\subsection{Van Vleck's formula and the HK propagator}

We saw in the previous subsection that the propagation of gaussian
states via Herman-Kluk is very imperfect. It works only for harmonic
potentials or for very short propagation times. In spite of this, the
HK formula became rather popular in the last five years, and was
applied to several problems in Chemical Physics. Among those we cite
the photodissociation of $CO_2$ \cite{Wal95}, the colinear scattering
of $H_2$ by $H$ \cite{Gar96}, and the non-adiabatic dynamics in
pyrazine molecules \cite{Thos00}.  Several theoretical articles were
also published, testing the HK formula in model systems
\cite{Kay94b,Mai00} or proposing modifications in the formula for increasing
its accuracy \cite{Herm97,Thos00}. All these applications deal, not with the
plain mixed propagator $\langle x|K(t)|z'\rangle$, but with integrals over the
phase space variables $q'$ and $p'$, such as in eq.(\ref{ba3}).

The question is then how a poor time evolution of gaussian
wave-packets, as in HK, may produce acceptable results after
integration over $q'$ and $p'$.  In this subsection we shall clarify
this point and, at the same time, show why HK is so expensive
computationally \cite{Wal95} and sometimes just does not work
{\cite{Mcc00}. First of all we shall calculate the coordinate
propagator (\ref{s6.1}) using the semiclassical formula of
Herman-Kluk for $\langle x''|K(t)|z'\rangle$. Once again the
integration over $q'$ and $p'$ will be performed by the
stationary exponent approximation, just as we did in subsection
4.4. We shall see that HK does recover Van Vleck's formula
\cite{Van28} exactly!  And that is the reason why sometimes it works.  At the
end of this subsection we shall comment on this result and discuss why
HK does not lead to long--time convergence when the integrals over $q'$ and
$p'$ are performed numerically.  We shall also point out why the HK prefactor
diverges for chaotic trajectories when the correct prefactor should,
instead, go to zero.

We use the same notation as in subsection (4.4) and follow the same steps. 
The coordinate propagator for HK can be written
\begin{align}
  \label{s6.2b}
K_{HK}(x'',t;x',0) &= \int \frac{\rd q'\rd p'}{2\pi\hbar} 
\pi^{-\frac{1}{2}} b^{-1} 
\sqrt{\frac{1}{2}\left(m_{pp}+m_{qq}-\ri m_{qp}+\ri m_{pq}\right)} \,
\re^{\xi(q',p')}
\end{align}
where the exponent $\xi(q',p')$ and its derivatives are:
\begin{gather}
\label{s6.3}
\xi = -\frac{1}{2b^2}[(x''-q_r)^2+(x'-q')^2] + 
\frac{i}{\hbar}[p_r(x''-q_r)+p'(q'-x')+S_H] \\
\label{s6.4} 
\frac{\partial \xi}{\partial q'} 
 = \frac{1}{b^2}[(x''-q_r)(m_{qq}+i m_{pq})+(x'-q')] \\
\label{s6.5} 
\frac{\partial \xi}{\partial p'} 
 = \frac{i}{\hbar}[(x''-q_r)(m_{pp}-i m_{qp})-(x'-q')] \; .
\end{gather}
The stationary conditions are satisfied if $q'=x'$ and $q_r(q',p',t)=x''$ and
the value of $\xi$ computed at this trajectory is $i S_H/\hbar$. The
second order derivatives are
\begin{gather} 
\label{s6.6} 
\xi_{qq} = -\frac{1}{b^2} [1+m_{qq}(m_{qq}+i m_{pq})] \\
\label{s6.7} 
\xi_{qp} = \xi_{pq} = -\frac{i}{\hbar}[1- m_{qq} (m_{pp}-i m_{qp})] = 
-\frac{1}{\hbar} m_{qp} (m_{qq}+i m_{pq}) \\
\label{s6.8} 
\xi_{pp} = -\frac{ib}{c\hbar} m_{qp} (m_{pp}-i m_{qp}) \; .
\end{gather}

Inserting all these expressions into \refg{s6.2b}, we find that the
coordinate propagator becomes
\begin{align}
  \label{s6.9}
K_{HK}(x'',t;x',0) &= \pi^{-\frac{1}{2}} b^{-1} 
\sqrt{\frac{1}{2}\left(m_{pp}+m_{qq}-\ri m_{qp}+\ri m_{pq}\right)} \;
\re^{i S_H/\hbar} \nonumber\\
&\times \int \frac{\rd Q\rd P}{2\pi\hbar} 
 \mbox{exp}\left[\frac{1}{2} (\xi_{qq} Q^2 + 2\xi_{qp} Q P +
 \xi_{pp} P^2) \right] \nonumber\\
&= \pi^{-\frac{1}{2}} b^{-1} 
\sqrt{\frac{1}{2}\left(m_{pp}+m_{qq}-\ri m_{qp}+\ri m_{pq}\right)}\;\times
\frac{\re^{i S_H/\hbar}}{\hbar \sqrt{\xi_{qq}\xi_{pp}-\xi_{qp}^2}}
\end{align}
Once again the quantity under the square root in the denominator is
the determinant of the quadratic form in the second
line, which is
\begin{align}
  \label{s6.10}
\xi_{qq}\xi_{pp}-\xi_{qp}^2
= \frac{i}{\hbar^2} m_{qp}(m_{pp}+m_{qq}-\ri m_{qp}+\ri m_{pq}) \;.
\end{align}
The final result is
\begin{align}
  \label{s6.11}
K_{HK}(x'',t;x',0) = \frac{1}{b\sqrt{2\pi i m_{qp}}} \; \re^{i S_H/\hbar} 
\end{align}
and it coincides with the Van Vleck formula (\ref{s6.20}).

We can now understand why HK may sometimes work, even if it is not a correct
semiclassical formula, and we can also understand the origin of its main
drawbacks. The latter include the very slow convergence of the integrals over
$q'$ and $p'$ when done numerically \cite{Wan00,Mcc00}, the lack of
normalization \cite{Herm84}, and the blow--up of the prefactor for chaotic
trajectories \cite{Mai00}. In fact, numerical convergence of the integrals over
$q'$ and $p'$ is often achieved only after resorting to smoothings
\cite{Herm97,Mcc00,Thos00,Sun00}, and the results have not always been
satisfactory.

The reason why HK does sometimes give good results is precisely that it is able
to recover Van Vleck.  In fact, Kay \cite{Kay94a} derived the HK formula by
making an ansatz for the propagator and demanding that this ansatz satisfy the
basic condition of agreeing with Van Vleck.  In words, if one insists in
keeping the width of the propagated wave packet constant, then one {\it must}
arrive at the HK pre-factor if one wants to get Van Vleck when integrating over
$q'$ and $p'$. The price paid for doing so is non--conservation of the norm, a
high price to say the least. Therefore, if HK is to be used at all, it has to
be under an integral. However, even when integrated, HK leads to very slow
convergence and oscillatory behavior, especially at long times. To see why this
happens, we note that the square root of the determinant of the quadratic form
in the stationary integral (\ref{s6.9}) is a measure of the phase-space area
$\Delta q' \Delta p'$ that matters in the integration. We have
\begin{align}
\label{s6.22}
(\Delta q' \Delta p')_{HK} \approx
\frac{1}{\sqrt{\left|\xi_{qq}\xi_{pp} - \xi_{qp}^2\right|}} =
 \frac{\hbar}{\sqrt{|m_{qp}|}} \; \frac{1}
{\sqrt{|m_{pp}+m_{qq}-\ri m_{qp}+\ri m_{pq}|}}
\end{align}
As we saw in subsec. 5.1, $\sqrt{|m_{pp}+m_{qq}-\ri m_{qp}+\ri m_{pq}|}$
increases with time, for our numerical example as well as for the free
particle.  This means that the relevant phase-space zone of initial
trajectories that one needs to sample, gets smaller with time. The physical
interpretation of this is very simple: any initial swarm of trajectories
spreads as time passes. However, only those trajectories in the neighborhood of
the one connecting $x'$ to $x''$ contribute significantly.  The size of this
neighborhood is determined by the size of the propagated wavepacket. In the
case of HK the propagated packet keeps its width fixed at $\Delta q \Delta p =
b c = \hbar$. For long times the initial spread of trajectories that {\em end}
inside this small region shrinks very fast. This is why the numerical
integration of eq.(\ref{s6.2b}) by sampling initial trajectories is bound not
to converge for long times, since very few trajectories are going to be picked
up in the relevant region.

Compare this with the relevant phase-space spread for the integration with
our formula
\begin{align}
\label{s6.23}
(\Delta q' \Delta p') \approx
 \frac{1}{\sqrt{\left|\xi_{qq}\xi_{pp} - \xi_{qp}^2\right|}} =
 \frac{\hbar}{\sqrt{2m_{qp}}} \; \sqrt{|m_{qq}+\ri m_{qp}|} \;\; .
\end{align}
We know that $\sqrt{|m_{qq}+\ri m_{qp}|}$ increases with time, since the
wave packet spreads, and so does the phase space region of contributing
trajectories, ensuring an integration which must be more efficient. 

We also mention that the application of semiclassical formulae to
chaotic systems is known to be difficult due to the exponential
proliferation of contributing trajectories from $x'$ to $x''$ for
long times. This is compensated in part by the exponentially small
contribution of each individual orbit to the propagator. The
Herman-Kluk prefactor, however, assigns an already divergent
contribution to each of these trajectories, leaving no hope for
accurate results \cite{Mai00}.

As a final comment we note that the smoothing technique introduced by
Herman \cite{Herm97} has the role of cutting off the contributions
from trajectories whose action have a large first order variation. In
our semiclassical formula this cutoff is performed automatically by
the prefactor.  This is in fact exactly what one expects from a {\em
stationary phase} integral.
 

\section{The energy representation}
  \label{kap7}

In the special case of time--independent systems, the Fourier transform in time
of the coherent state propagator, which we shall refer to as the (coherent
state) Green's function, has poles at the quantized energy levels. For its
diagonal elements ($z'=z''$) the residues are the so--called Husimi
distributions, i.e. the absolute squares of the eigenfunctions in the coherent
state representation, which are also called the Bargmann wave functions. In
this section we shall derive the semiclassical expressions for the poles and
residues of the Green's function.  We shall obtain from them the semiclassical
quantization rule for the energy levels and the semiclassical Husimi
distributions.


\subsection{The monodromy matrix in one dimension}

When the trajectory is a periodic orbit, with the initial point
returning to itself after the {\it period} $T$, the tangent matrix $M$
of section 2.6 is called the monodromy matrix.  In the next subsections
it will be necessary to know the element $M_{vv}$ for a real periodic
orbit traversed $n$ times, i.e. for $t=nT$ .  We shall calculate it
here.  First we shall review a general formula for arbitrary canonical
variables, then we shall switch back to the $(u,v)$ variables.

Let $Q$ and $P$ be canonical variables and $H(Q,P)$ the
Hamiltonian. We want to find the monodromy matrix $M$ for periodic
orbits in the $(Q,P)$ representation. $M$ has 4 elements, which we
shall determine from 4 linear equations.  For the first 2 equations we
apply $M$ to $(\dot{Q},\dot{P})$, the velocity vector in phase space.
This should yield this vector again, because a small displacement
along the trajectory maps onto itself.  Thus
\begin{align}
  \label{glo149}
      M    \begin{pmatrix}
             \dot{Q} \\
             \dot{P}
           \end{pmatrix}
      =    \begin{pmatrix}
             \dot{Q} \\
             \dot{P}
           \end{pmatrix} \; .
\end{align}
For the other 2 equations we consider a small displacement
$(\delta Q,\delta P)$ non--colinear with $(\dot{Q},\dot{P})$, and we assume for
now that the Hamiltonian is not harmonic, so that the period $T$ of a periodic
orbit depends on its energy.  Then $(\delta Q,\delta P)$ points to another
periodic orbit, with a different period $T + \delta T$ and a different energy
$E + \delta E$.  When we apply the $M$ matrix to $(\delta Q,\delta P)$, since
the propagation time is only $T$, the result falls short of the original
displacement by $\delta T\times$(velocity), hence
\begin{align}
  \label{glo150}
      M    \begin{pmatrix}
             \delta Q \\
             \delta P
           \end{pmatrix}
      =    \begin{pmatrix}
             \delta Q \\
             \delta P
           \end{pmatrix}
- \delta T \begin{pmatrix}
             \dot{Q} \\
             \dot{P}
           \end{pmatrix} \; .
\end{align}
Obviously, given the 4 equations (\ref{glo149}) and (\ref{glo150}), $M$ must
have the form
\begin{align}
  \label{glo151}
     M = \openone + N
\end{align}
where $N$ must obey the 4 equations
\begin{align}
  \label{glo152}
      N    \begin{pmatrix}
             \dot{Q} \\
             \dot{P}
           \end{pmatrix}
    &=  0   \\
  \label{glo153}
      N    \begin{pmatrix}
             \delta Q \\
             \delta P
           \end{pmatrix}
    &= - \delta T \begin{pmatrix}
                     \dot{Q} \\
                     \dot{P}
                  \end{pmatrix} \; .
\end{align}
From \refgs{glo152} one sees that $N$ must have the form
\begin{align}
  \label{glo154}
      N =  \begin{pmatrix}
             \dot{P}a & -\dot{Q}a \\
             \dot{P}b & -\dot{Q}b         
           \end{pmatrix} 
\end{align}
with $a$ and $b$ still unknown.  From \refgs{glo153} one calculates $a$
and $b$ as
\begin{align}
  \label{glo155}
a = \frac{\delta T}{\delta E} \dot{Q} \qquad
b = \frac{\delta T}{\delta E} \dot{P}
\end{align}
where $\delta E$, the energy difference, actually appears as 
$\dot{Q}\delta P - \dot{P}\delta Q$, which is $\delta E$ by Hamilton's
equations.  The final form of the monodromy matrix in the $(Q,P)$
representation is therefore
\begin{align}
  \label{glo156}
  M  =   \openone + \frac{\rd T}{\rd E}
           \begin{pmatrix}
             \dot{Q} \dot{P} & -\dot{Q}^2 \\ \\
             \dot{P}^2 & -\dot{Q} \dot{P} 
           \end{pmatrix}
\end{align}
where $\rd T/\rd E$ is the derivative of $T(E)$, the period of the orbit as a
function of its energy.  Although we restricted ourselves to the
anharmonic case, it is obvious that this formula holds for the harmonic case
also, for which $\rd T/\rd E$ vanishes.

We shall now apply this to the variables $u$ and $v$.  But the latter are not
quite true canonical variables because of the factor $\ri \hbar$ in the
equations of motion (\ref{glg10}). To avoid this source of confusion we define
temporary canonical variables $U=\sqrt{\ri \hbar} u$ and $V=\sqrt{\ri \hbar} 
v$.
Then the general result (\ref{glo156}) certainly applies to $U$ and $V$.  When
we reintroduce $u$ and $v$, we find for the monodromy matrix in the $(u,v)$
representation
\begin{align}
  \label{glg162}
           M(u,v,T)
  =        \openone + \ri \hbar \frac{\rd T}{\rd E} \Gamma 
\end{align}
where $\Gamma$ is given by
\begin{align}
  \label{glg163}
           \Gamma 
  =        \begin{pmatrix}
             \dot{u} \dot{v} & -\dot{u}^2 \\
             \dot{v}^2 &  -\dot{u} \dot{v}
           \end{pmatrix} \; .
\end{align}
It is easy to check that, because $\Gamma^2 =0$, the monodromy matrix
for $n$ traversals of the orbit is just
\begin{align}
           M(u,v,nT)
  =        M(u,v,T)^n 
  =        \openone + n \ri \hbar \frac{\rd T}{\rd E} \Gamma .
\end{align}
In the next subsections we need only one matrix element, which we denote by
\begin{align}
\label{glg148}
           (M^n)_{vv} 
  =        1 - n \ri \hbar \frac{\rd T}{\rd E} 
           \dot{u} \dot{v} .
\end{align}

\subsection{Prologue: The Green's Function for the Harmonic Oscillator}

The Feynman-type propagator $K_F(z'',t;z',0)$ in the coherent state
representation, like other such propagators in quantum mechanics,
is discontinuous at $t=0$.  One distinguishes between the ``forward
propagator'', which is defined by \refg{glg1} for $t>0$ but vanishes for $t<0$,
and the ``backward propagator'' which vanishes for $t>0$.  The Green's function
is the Fourier transform of the forward propagator, hence it is obtained by
integrating the time only in the interval $t\in [0,\infty]$. We are interested
in its diagonal elements, which we denote by
\begin{align}
  \label{glg141}
           G(z,E+\ri\gamma):=\frac{1}{\ri\hbar}\int_0^{\infty}{\rm d}t
           K(z,t;z,0) \re^{\ri (E+\ri\gamma)t/\hbar} \; .
\end{align}
The small positive quantity $\gamma$ is introduced for convergence; at the end
it will be set equal to 0.  When we insert the semiclassical propagator
(\ref{en4}), $G$ becomes
\begin{align}
  \label{glg141a}
           G(z,E+\ri\gamma)=\frac{1}{\ri\hbar}\sum_\nu\int_0^{\infty}{\rm d}t
\frac{1}{\sqrt{(M_\nu)_{vv}}} \exp\left[\phi_\nu(t)\right]
\end{align}
with
\begin{align}
  \label{glg142}
           \phi_\nu (t) 
  =        \frac{\ri}{\hbar} \left[S_\nu(v'',u',t) + {\cal I}_\nu(v'',u',t) 
         +\frac{\ri\hbar }{2} 
        \left( |z''|^2 + |z'|^2\right)  + (E+\ri\gamma)t\right] \; .
\end{align}

Instead of the variables $v''$ and $u'$, we could also have written
\refg{glg142} in terms of the single variable $z$, since $u'=z$ and
$v''=z^\star$. But this does not mean that the trajectory going from
$z$ to $z$ in some arbitrary time $t$ is periodic.  As we have already
discussed (see before \refg{mb7}), the trajectory is usually complex
and neither $u$ nor $v$ match at the end points. However, for the case
of interest here, namely one-dimensional bound systems, there exists
for each phase-space point $z$ a (minimal) time $T=T(z)$ for which the
orbit through $z$ is periodic and real. Then the end points do match
and $T$ is the period.  Obviously this particular time $t=T$ will play
a special role in the integration of \refg{glg141}, and so will its
multiples $t=nT$, which correspond to repeated traversals of the
periodic orbit. Therefore, after evaluating the integral
(\ref{glg141}) by the stationary exponent method, we shall expand the
stationary time $t_0=t_0(z,E)$ about the classical period $nT(z)$.  
The resulting Green's function will be a good approximation to the 
actual Green's function $G(z,E)$ only for arguments $z$ and $E$ satisfying 
$t_0(z,E)\approx nT(z)$.

Before we carry out the calculation of $G(z,E)$ for a general Hamiltonian, it
is instructive to find out the result of these approximations, stationary
exponent plus expansion about $t_0=nT$, for the harmonic oscillator.  This will
help us understand the nature of the approximations.

We consider an oscillator of unit mass and frequency $\omega$, and we take the
parameter $b$ of \refg{glg47} to be $\sqrt{\hbar/\omega}$.  Then
the semiclassical propagator is easily shown to be
\begin{align}
\label{ho1}
K(z,t)= \re^{-i\omega t/2} \; \exp{\left[|z|^2 (\re^{-i\omega t}-1)\right]},
\end{align}
which is also the exact result. The first exponential is the prefactor
$M_{vv}^{-1/2}$.  This $K$ is now carried into \refg{glg141}, which becomes
\begin{align}
\label{ho2}
 G(z,E+\ri\gamma)=\frac{1}{\ri\hbar} \int_0^{\infty} \rd t \re^{-i\omega t/2}
      \re^{\ri (E+\ri\gamma)t/\hbar + |z|^2 (\re^{-i\omega t}-1)} \; .
\end{align}
The whole of the second exponent is proportional to $1/\hbar$, when $|z|^2$ is
written in terms of $q$ and $p$.  But there is no $\hbar$ in the prefactor.
Hence we are exactly in the situation mentioned at the end of section 3.3.  We
look for the time $t_0$ at which the second exponent is stationary.
We find that $t_0$ is given by the condition
\begin{align}
\label{ho3}
E+\ri\gamma = \hbar \omega \re^{-i\omega t_0} |z|^2 \; .
\end{align}
The solutions of this equation can be written $t_0=T_0+nT$, where $n$ is an
integer, $T=2\pi/\omega$, and $T_0=T_0(z,E)$ is pure imaginary (in the limit
$\gamma\to0$).  Thus there is an infinite number of times at which the second
exponent is stationary.  For each $t_0$ we expand this exponent to second order
in the vicinity and we perform the gaussian integral.  Then we add all these
results together to obtain the following $G$
\begin{align}
\label{ho4}
 G(z,E+\ri\gamma) & =\frac{1}{\ri\hbar} \sum_{n=1}^{\infty}   
  \; \re^{-\frac{i\omega}{2}(nT+T_0)}
   \re^{\frac{\ri}{\hbar}(E+\ri\gamma)(nT+T_0) + |z|^2 (\re^{-i\omega T_0}-1)}
\int_0^{\infty} \rd t \; \re^{
    -\frac{\omega^2 |z|^2}{2}(t-t_0)^2 \re^{-i\omega T_0} } \nonumber \\
 &=\frac{1}{\ri\hbar |z|} \sqrt{\frac{2\pi}{\omega^2 \re^{-i\omega T_0}}}
    \sum_{n=1}^{\infty} 
   \re^{\frac{2\pi\ri n}{\hbar\omega}(E+\ri\gamma-\hbar\omega/2)} \; 
   \re^{|z|^2 (\re^{-i\omega T_0}-1)+\ri(E+\ri\gamma-\hbar \omega/2)T_0/\hbar}
   \nonumber \\
 &= \frac{1}{\ri\hbar |z|} \sqrt{\frac{2\pi}{\omega^2 \re^{-i\omega T_0}}} \;
 \frac{\re^{\frac{2\pi\ri}{\hbar\omega}(E+\ri\gamma-\hbar\omega/2)}}
 {1-\re^{\frac{2\pi\ri}{\hbar\omega}(E+\ri\gamma-\hbar\omega/2)}} \; 
 \re^{|z|^2 (\re^{-i\omega T_0}-1)+\ri(E+\ri\gamma-\hbar \omega/2)T_0/\hbar} 
  \; .
\end{align}
In the above we started the sum over $n$ at $n=1$.  Why not $n=0$?
The answer is that it does not make any difference, as long as we use
this theory only to calculate the energies of the stationary states
and their Husimi distributions.  This is discussed in the paragraph
following \refg{gree} in the next subsection.

The poles of (\ref{ho4}) can now be found in the limit $\gamma\to 0$. They
are given by the condition $E=:E_m=\hbar \omega(m+1/2)$ with integer $m$.  The
residue for $E=E_m$, which is the Husimi distribution, is
\begin{align}
\label{ho5}
\rho_{\rm Husimi}(z)=  \frac{1}{ \sqrt{2\pi |z|^2 \; \re^{-i\omega T_0}}} \; 
\re^{|z|^2 (\re^{-i\omega T_0}-1)+\ri m \omega T_0} \; .
\end{align}
Using (\ref{ho3}) for $T_0$ with $E=E_m$ and $\gamma=0$, we obtain 
$\re^{-i\omega T_0} |z|^2=m+1/2$ and (\ref{ho5}) becomes 
\begin{align}
\label{ho6}
\rho_{\rm Husimi}(z) =  \frac{1}{\sqrt{2\pi (m+1/2)}} \; 
\re^{m+1/2-|z|^2}\; \left(\frac{|z|^2}{m+1/2}\right)^m \;.
\end{align}
This result should be compared with the exact Husimi distribution
\begin{align}
\label{hoex}
\rho_{\rm Husimi}^{exact}(z)  = \frac{1}{m!} \re^{-|z|^2} |z|^{2m}  \;,
\end{align}
which is an annulus with its maximum at $|z|^2 = m$.  The mean value of $|z|^2$
is $m+1$ and the logarithmic second derivative at the maximum is $-1/m$, giving
a width of order $\sqrt{m}$.
Both formulas, exact and semiclassical, have the same $z$-dependence,
namely, $|z|^{2m} \re^{-|z|^2}$. They differ only in the coefficient.
Obviously (\ref{hoex}) is correctly normalized and (\ref{ho6}) is not.  The
difference is small, however.  Using Stirling's approximation for $m!$, we can
write (\ref{ho6}) as
\begin{align}
\label{ho6a}
\rho_{\rm Husimi}(z) = \frac{1}{m!} \re^{-|z|^2} |z|^{2m} 
\left(1+\frac{1}{2m+1}\right)^{m+1/2} \; \re^{-1/2} \;.
\end{align}
For large $m$ the quantity to the left of $\re^{-1/2}$ becomes $\re^{+1/2}$.
Thus the semiclassical Husimi becomes exact in the limit of
large quantum numbers, and it is a valid approximation for all phase space
points $z$.

For general Hamiltonians we shall not be able to do such a complete
calculation. As we shall see, the stationary time $t_0$ is given by an
implicit equation that cannot be generally solved. The best thing to
do to get an explicit result will be to expand $t_0$ about
the classical period $nT$. Now we can check this additional
approximation explicitly for the harmonic oscillator. Let
us pretend that \refg{ho3} cannot be solved exactly for $t_0(E,z)$ and,
instead, let us solve it by expanding $t_0$ about $nT$. So we write
$t_0=nT+T_0$ and consider only small values of $T_0$. Taking $\gamma=0$ 
in \refg{ho3} and expanding for small $T_0$, we find
\begin{align}
\label{ho7a}
T_0 \approx \frac{\hbar\omega|z|^2-E}{ i\hbar\omega^2 |z|^2} \;.
\end{align}
Since $T_0$ is independent of $n$ we may go through the same steps as
in \refg{ho4}. The position of the poles does not depend
on $T_0$ and it is not affected by this approximation. The Husimi
distribution, however, does depend on $T_0$ explicitly. For $E=E_m$ we get
\begin{align}
\label{ho7}
T_0 \approx \frac{|z|^2-m-1/2}{ i\omega |z|^2} \;.
\end{align}
$T_0$ is small whenever $|z|^2$ is close to $m+1/2$, which is the classical
orbit with quantized energy $E_m$.  Therefore, for each eigenstate, there is a
phase space region centered on this classical orbit where the approximation is
justified.  This region should encompass most of the distribution when $m$ is sufficiently large.  Expanding \refg{ho5} to second order in $T_0$ and using
(\ref{ho7}) we obtain
\begin{align}
\label{ho8}
 \rho_{\rm Husimi}(z) \approx \displaystyle{\frac{1}{\sqrt{2\pi}|z|} \; 
  \exp{\left[ -\frac{(|z|^2-m-1/2)^2}{2|z|^2}\right]}}  \; .
\end{align}
The maximum of this annulus comes at $|z|^2 = m + 1/8m + \dotsb$, which agrees
with the exact maximum for the first two orders of $1/m$.  The logarithmic
second derivative also agrees for the leading order.  Therefore, the result of
expanding the stationary time about the classical period is to restrict the
region of validity of the Husimi distribution to not too small quantum numbers.


\subsection{The Green's Function for General Hamiltonians}

We now return to our general calculation, \refg{glg141}.  In what follows we
shall omit the subscript $\nu$.  It will be replaced shortly by the multiple
traversals around periodic trajectories.  The stationary exponent condition is
given by $\phi'(t_0)=0$, with $\phi(t)$ given by (\ref{glg142}).  It is
\begin{align}
  \label{phip}
     \phi'(t_0) = \frac{\ri}{\hbar} \left( \left.
     \frac{\partial S}{\partial t}\right|_{t_0} + \left.
     \frac{\partial {\cal I}}{\partial t}\right|_{t_0} + 
     E + \ri \gamma \right) \equiv 0 . 
\end{align}
The third equation (\ref{mb9}) gives
\begin{align}
  \label{dsdt}
           \left.\partial S /\partial t\right|_{t=t_0} =
           -{\cal E}(z,t_0) \; .
\end{align}
This script $\cal E$ is the energy of the classical trajectory, not to be
confused with $E$, which is the energy variable in the  Green's function.
For the first time derivative of ${\cal I}$, defined in \refg{2.mb2}, we have
\begin{align}
\label{didt}
\left.\partial {\cal I} /\partial t\right|_{t=t_0} =
         \left. \frac{1}{2} \frac{\partial^2 {\cal H}}{\partial u \partial v}
      \right|_{t=t_0} + \frac{1}{2} \int_0^{t_0}  
   \left[ \;\; \left. \frac{\partial^3 {\cal H}}{\partial^2 z \partial z^\star}
            \frac{\partial z}{\partial t}\right|_{t=t_0}  + \left.
          \frac{\partial^3 {\cal H}}{\partial z \partial^2 z^\star}
 \frac{\partial z^\star}{\partial t}\right|_{t=t_0} \right] \rd t' \; .
\end{align}
The terms under the integral involve third derivatives of $\cal H$ with
respect to the phase space variables.  We have consistently neglected such
terms up to now and we must do so here again.  Hence we use only the first 
term, to which we give a simpler name 
\begin{align}
\label{epsl}
 \left. \partial {\cal I} /\partial t\right|_{t=t_0}
 \approx \frac{1}{2} \frac{\partial^2 {\cal H}}{\partial z \partial z^\star} 
           =: \epsilon(z,t_0) \; .
\end{align}
The stationary condition is then written
\begin{align}
  \label{spc}
     E + \ri \gamma - {\cal E}(z,t_0) + \epsilon(z,t_0) = 0 \;.
\end{align}
Besides the contributions from the stationary points $t=t_0$, there is also a
contribution to $G$ coming from the vicinity of $t=0$. As we shall see later,
this contribution is not needed in the calculation of the energy levels and the
Husimi distributions, hence we shall ignore it. We have also discarded it in
the previous calculation for the harmonic oscillator.

In order to perform the time integral, we need to expand $\phi(t)$
around $t_0$ to second order. The first derivative at $t_0$ is zero
by definition. The second derivative is
\begin{align}
  \label{phipp}
     \phi''(t_0) = \frac{\ri}{\hbar} \left( \left.
     \frac{\partial^2 S}{\partial t^2}\right|_{t_0} + \left.
     \frac{\partial^2 {\cal I}}{\partial t^2}\right|_{t_0} \right).
\end{align}
For the second derivative of $S$ we introduce the notation
\begin{align}
  \label{glg144am}
\alpha(z,t_0):=\left.\partial^2 S /\partial t^2\right|_{t=t_0} \;\; .
\end{align}
From (\ref{didt}) we see that $\partial^2 {\cal I}/\partial t^2$
involves only third or higher order derivatives of ${\cal H}$ and
therefore we can discard it completely. 

We can now calculate $G(z,E+\ri\gamma)$.  The effect of our expansion is that
we have applied once again the gaussian approximation to the integral in
\refg{glg141}, which is then straightforward:
\begin{align}
  \label{gree}
   G(z,E+\ri\gamma) &= \frac{1}{\ri\hbar} \frac{1}{\sqrt{M_{vv}(t_0)}}
    \displaystyle{ \;
    \re^{ \frac{\ri}{\hbar}\left[ S(z,t_0) + 
          {\cal I}(z,t_0) + \ri\hbar |z|^2 + 
          (E+\ri\gamma)t_0 \right] } 
     \int_0^{\infty} \re^{\frac{\ri \alpha(z,t_0)}{2\hbar} (t-t_0)^2} }
     \rd t \nonumber \\
   & = \frac{1}{\ri\hbar} \sqrt{\frac{2\pi\ri\hbar}{M_{vv}(t_0)\alpha(z,t_0)}}
    \; \exp{ \left\{\frac{\ri}{\hbar}\left[ S(z,t_0) + 
          {\cal I}(z,t_0) + \ri\hbar |z|^2 + 
          (E+\ri\gamma)t_0 \right] \right\}} 
\end{align}
Actually this integral is a sum of gaussians, because there are many solutions
to \refg{spc}, as we have seen with the harmonic oscillator. As we mention
later, $\alpha$ has a positive imaginary part for $t_0\approx nT$; this follows
from \refg{en9}, plus the fact that $\cal S$ is real.  It means that the
integral is strongly convergent for very large times, both positive and
negative.  However, the integral in \refg{glg141} does not extend over the
interval $-\infty<t<+\infty$, but over $0<t<+\infty$: is there a problem at
$t=0$?  Yes, there might very well be a problem, both for $t=0$ and for
$t_0\approx nT$ and small $n$.  There is no problem for large $n$, since $nT$
becomes arbitrarily large.  We intend to use this Green's function only for the
semiclassical calculation of the energy levels and the Husimi distributions.
These are determined by the poles of the Green's function and by their
residues, respectively.  And the poles and residues are determined solely by
the behavior of the integrand of (\ref{glg141}) at very large times.  Hence we
are safe in ignoring possible mistakes at small $t$.  In particular, the
contributions from trajectories with $t\approx 0$ give rise to the Thomas-Fermi
approximation, which we shall not consider here.  Therefore it is all right for
our purpose to calculate the integral as if it extended from $-\infty$ to
$+\infty$.

We shall now transform the prefactor by finding a convenient way of
expressing $\alpha(z,t_0)$. In the many differentiations which follow, it is
important to remember that the independent variables in $S$ are
$u',v'',$ and $t$~. One should do the differentiations first, and only
afterwards may one compute the functions for the stationary orbit by
replacing $t$ by $t_0$, $u'$ by $z$ and $v''$ by $z^*$.  First we
write two different ways of expressing $\partial S /\partial t$
\begin{align}
  \label{en5a}
\partial S /\partial t = -{\cal H}\left[u',v'(u',v'',t)\right] \qquad
\partial S /\partial t = -{\cal H}\left[u''(u',v'',t),v'')\right] \; .
\end{align}
This leads to two different ways of writing $\alpha$
\begin{align}
  \label{en5b}
\alpha(z,t_0) = \left.\frac{\partial^2 S}{\partial t^2}\right|_{t=t_0} 
  &= \left. -\frac{\partial{\cal H}}{\partial v'}
     \frac{\partial v'}{\partial t}\right|_{t=t_0} \\
  \label{en5c}
\alpha(z,t_0) = \left.\frac{\partial^2 S}{\partial t^2}\right|_{t=t_0} 
  &= \left. -\frac{\partial{\cal H}}{\partial u''}
     \frac{\partial u''}{\partial t}\right|_{t=t_0} \; .
\end{align}
In \refg{en5b} we use Hamilton's eqs. (\ref{glg10}) to express $\partial
H/\partial v'$ and the second \refg{mb9} to express $\partial v'/\partial
t$~. In \refg{en5c} we do similar transformations with the other
variables. This gives the two forms
\begin{align}
  \label{en6a}
\alpha(z,t_0) = - \left. \ri\hbar\dot{u'}\frac{\ri}{\hbar}
\frac{\partial^2 S}{\partial u' \partial t}\right|_{t=t_0}
             =   \left. \ri\hbar\dot{v''}\frac{\ri}{\hbar}
\frac{\partial^2 S}{\partial v'' \partial t}\right|_{t=t_0} \; .
\end{align}
Referring now to \refg{en5a}, using Hamilton's eqs. once again, and using the
first and second eqs.(\ref{mb9}) once again, we transform the second
derivatives of $S$ as follows
\begin{align}
  \label{en7a}
\frac{\partial^2 S}{\partial u'\partial t} =
 -\frac{\partial {\cal H}(u'',v'')}{\partial u''}
      \frac{\partial u''}{\partial u'} =
 \ri\hbar\dot{v''}\frac{\ri}{\hbar}\frac{\partial^2 S}{\partial u'\partial v''}
 \\
  \label{en7b}
\frac{\partial^2 S}{\partial v''\partial t} =
 -\frac{\partial {\cal H}(u',v')}{\partial v'}
     \frac{\partial v'}{\partial v''} =
-\ri\hbar\dot{u'}\frac{\ri}{\hbar}\frac{\partial^2 S}{\partial u'\partial v''}
 \; .
\end{align}
After cancellation of the $\hbar$'s, both forms of the equations give the same
result for $\alpha$, namely
\begin{align}
  \label{en8}
           \alpha(z,t_0)
 = -\left. \dot{u'}\dot{v''}\frac{\partial^2 S}{\partial u'\partial v''}
        \right|_{t=t_0} 
 = -\left. \dot{u'}\dot{v''} A_{uv} \right|_{t=t_0}
 = \ri\hbar \dot{u'}\dot{v''}/ M_{vv}(z,t_0) \;.
\end{align}
Amazingly, the denominator under the square root in
\refg{gree} has become simply $\ri\hbar \dot{u'}\dot{v''} $, where the
velocities are computed at the complex trajectory with $t=t_0$. One 
should, however, be careful about the phase of the $M_{vv}$ under the
square-root, as already mentioned in sections 2 and 4. We shall
discuss it in a moment.

In spite of this great simplification it is still hard to find the
poles of $G(z,E)$ by looking at \refg{gree}. This is because the
stationary time $t_0$ is given implicitly by \refg{spc} and it
refers to complex trajectories.  According to our discussion following
\refg{glg142}, large contributions to
$G(z,E+\ri\gamma)$ are expected for $t_0$ close to $nT$, where $T$ is
the period of the real orbit through $z$.  Therefore, as we did for
the harmonic oscillator, we proceed to expand \refg{gree} about
$t_0=nT$, summing over $n$.  We write $t_0=nT+T_0$, substitute it
into \refg{spc}, and expand the terms to find $T_0$:
\begin{align}
 \label{del1}
  E + \ri \gamma - {\cal E}(z,nT) + \epsilon(z,nT) - \left.
\partial  {\cal E}(z,t)/\partial t\right|_{t=nT} T_0 + \left.
\partial \epsilon(z,t)/\partial t\right|_{t=nT} T_0  \approx 0 \;.
\end{align}
Equation (\ref{epsl}) shows again that $\partial \epsilon(z,t)/\partial t$ 
involves only third or higher derivatives of ${\cal H}$ and therefore
we discard it. For real periodic orbits, the functions
${\cal E}$ and $\epsilon$ depend only on $z$ and we have
\begin{align}
  \label{glg144a}
           \left.\partial{\cal E}(z,t)  /\partial t\right|_{t=nT} =
          - \left.\partial^2 S /\partial t^2\right|_{t=nT} = 
           -\alpha(z,nT) =: -\alpha^{(n)}(z) \; .
\end{align}
Equation (\ref{del1}) can then be written
\begin{align}
 \label{del1a}
  \bigl(E + \ri \gamma - {\cal E}(z,nT) + \epsilon(z,nT) \bigr) +
   \alpha^{(n)}(z) T_0 \approx 0 \;.
\end{align}
Our basic assumption is that $T_0 \ll nT$ for each solution $t_0$ labeled by
$n$. Hence the sum of the terms inside the parenthesis above should be small.
Solving \refg{del1a} for $T_0$ gives
\begin{align}
  \label{delt}
T_0 = -\frac{E + \ri \gamma - {\cal E}(z) + \epsilon(z)}
{\alpha^{(n)}(z)} \;.
\end{align}
In the limit $\gamma \rightarrow 0$, the points $z$ satisfying ${\cal
E}(z) - \epsilon(z)=E$ have $t_0(z,E)=nT(z)$. When we expand the
Green's function about $t_0=nT$, we are restricting the validity to
the neighborhood of these points.  Notice that, contrary to the case
of the harmonic oscillator, $T_0$ does depend on $n$ and the sum over
multiple traversals has to be performed after the expansion about $t_0=nT$.

We are now in a position to discuss the pre-factor which, according to
\refg{en8}, is given (up to its phase) by the square root of $\dot{u}'(t_0) \;
\dot{v}''(t_0)$.  Its main contribution comes from periodic orbits and, once
again, we have to expand each of these velocities for $t_0$ close to $nT$.
Using Hamilton's equation for $\dot{u}'(t_0)$ we get
\begin{align}
\dot{u}'(t_0) &= -\frac{i}{\hbar} \left. 
    \frac{\partial{\cal H}}{\partial v'}\right|_{t=t_0} \approx 
    \dot{u}'(nT) -\frac{i}{\hbar} 
    \frac{\partial^2 {\cal H}}{\partial {v'}^2} \left.
    \frac{\partial v'}{\partial t}\right|_{t=nT} T_0 \nonumber \\
 &= \dot{z}-\frac{i}{\hbar} 
    \frac{\partial^2 {\cal H}}{\partial {v'}^2} 
    \left(\left. \frac{i}{\hbar}\frac{\partial^2 S}{\partial t \partial u'}
    \right|_{t=nT} \right) T_0 \nonumber \\
 &= \dot{z}+\frac{i}{\hbar} {\cal H}_{vv} \; \dot{z}^* \; 
    T_0 / M^n_{vv}\nonumber \\
 &= \dot{z}+\frac{1}{\hbar^2 \dot{z}} {\cal H}_{vv} \; T_0 \; \alpha^{(n)}
\end{align}
where we have used eqs.{\hskip 2pt}(\ref{mb9}, \ref{en7a}, \ref{en2b}) and
(\ref{en8}).  Doing a similar calculation for $\dot{v}''(t_0)$ we find
\begin{align}
\label{pre1}
\dot{u}'(t_0) \; \dot{v}''(t_0) \approx |\dot{z}|^2 
\left(1+{\cal O}(\alpha^{(n)} T_0)\right) \;.
\end{align}
Now we want to argue that the terms of order $\alpha^{(n)} T_0$ are small and
that, in fact, the prefactor can be simply calculated at the periodic orbit
itself, so that (\ref{pre1}) is just $|\dot{z}|^2$.  We gave a similar argument
in section 3.3, where the neglected terms were of order $\hbar$ compared to the
terms kept.  Here, however, their order is $\sqrt{\hbar}$, not as small, but
still small!  We shall verify this {\it a posteriori} once we have calculated
the Husimi distribution.  We shall find later that the Husimi attains
significant magnitudes only when $E-{\cal E}(z)+\epsilon(z)$, which is the
same as $\alpha^{(n)} T_0$ by (\ref{delt}), is of order $\sqrt{\hbar}$.

To calculate the phase of the prefactor, or the tangent matrix, for a
real periodic orbit, one must follow it when moving around the orbit. 
The harmonic oscillator provides again a simple illustration of what happens: 
in this case the solution of Hamilton's equations (\ref{glg10}) with initial
conditions $u(0)=u'$, $v(0)=v'$ is simply $u(t')=u' \re^{-i\omega t'}$
and $v(t')=v' \re^{i\omega t'}$.  The tangent matrix is diagonal and
$M_{vv} = \re^{i\omega t}$.  The pre-factor of the time-dependent
propagator is, therefore, $\re^{-i \omega t/2}$ (see \refg{ho1}). When 
this is calculated at a periodic orbit, we see
that after one period, for $t=T=2\pi/\omega$, $M_{vv}$ has rotated by
$2\pi$ and the phase of the pre-factor is $\re^{-i\pi}=-1$. The phase
for $t=2n\pi/\omega$ is just $\re^{-i n\pi}$.

This phase of $-n \pi$ is not particular to the harmonic potential; it is a
consequence of the fact that the motion is periodic in two--dimensional
phase space. To see this, consider a periodic orbit of energy ${\cal E}$ and
period $T$ in a generic system. The tangent matrix propagates small
displacements about this orbit. Any such small displacement is going to point
to a nearby periodic orbit with energy ${\cal E}+\delta {\cal E}$ and period
$T+\delta T$.  As we move once around the orbit the displacement vector stays
hooked to these two reference orbits and, therefore, has to rotate once as
well. The total rotation is not exactly $2\pi$, since the period of the nearby
orbit is slightly different from $T$. The angle of rotation is actually $2\pi
+\theta$, where the $\theta$ piece follows from \refg{glg148} for $M_{vv}$.
When $M_{vv}$ is raised to the $-1/2$ power in the prefactor, this $2\pi$
contributes a phase of $-\pi$ to the propagator, just as in the harmonic
oscillator. Then there is the angle $\theta$, which is a function of $n$ and 
which remains in the prefactor.  Equation (\ref{glg148}) shows that $\theta$
never gets very large; for all $n$'s it is bounded either by 0 and $\pi/2$, or
by $-\pi/2$ and 0, depending on the signs.  However, by virtue of \refg{en8}
the phase $\theta$ disappears completely from the calculation and the only
relevant phase from the square root is $-n\pi$. From now on we shall take this
phase out of the prefactor and include it explicitly in the exponential.
Replacing $\dot{u'}\dot{v''}$ in \refg{en8} by $|\dot{z}|^2$ we get
\begin{align}
  \label{gree1}
   G(z,E+\ri\gamma)  = \frac{\sqrt{2\pi}}{\ri\hbar|\dot{z}|}
    \; \exp{ \left\{\frac{\ri}{\hbar}\left[ S(z,t_0) + 
          {\cal I}(z,t_0) + \ri\hbar |z|^2 + 
          (E+\ri\gamma)t_0 -n\pi\hbar \right] \right\}} \;.
\end{align}

Next we expand $S$ and ${\cal I}$ around $t_0=nT$.  Since the
zero--order orbit is the real periodic one, we may write, using
\refg{glg83}
\begin{align}
  \label{glg144}
           S(z^\star ,z,nT) = n {\cal S}({\cal E}(z)) -
           n {\cal E}(z)\,T - \ri \hbar |z|^2 \; .
\end{align}
Here ${\cal E}(z)$ is the energy of the real periodic orbit going through $z$.
It should not be confused with $E$, the quantal energy variable.  And
${\cal S}({\cal E}(z))$ is the ``reduced action''
\begin{align}
  \label{glg145}
           {\cal S}({\cal E}(z))
  =        \frac{\ri \hbar}{2} \int_0^T (\dot{u}v - \dot{v}u) {\rm d}t'
\end{align}
which, for the real periodic orbit, is a function only of $\cal E$, or of the
period $T$, themselves functions of $z$.  The third \refg{mb9} gives
\begin{align}
  \label{glg144b}
           \left.\partial S /\partial t\right|_{t=nT} =
           \left.\partial S /\partial t\right|_{t=T} = 
           -{\cal E}(z) \; .
\end{align}
Either ${\cal E}$ or $T$ can be used as the independent variable for the
periodic trajectory. The second derivative of $S$ becomes
(see \refgs{glg144am} and (\ref{en8}))
\begin{align}
  \label{glg144c}
\left.\partial^2 S /\partial t^2\right|_{t=nT} &= \alpha^{(n)}(z) =
\ri\hbar |\dot{z}|^2/ M_{vv}^{(n)} \;.
\end{align}
Using \refg{glg148} and $\rd{\cal S}/\rd {\cal E}=T$, we find
\begin{align}
  \label{en9}
           \frac{1}{\alpha^{(n)}} 
  =        \frac{M^{(n)}_{vv}}{\ri \hbar |\dot{z}|^2} 
  =        \frac{1}{\ri\hbar |\dot{z}|^2} - \frac{n}{ |\dot{z}|^2}
           \frac{\rd T}{\rd {\cal E}}\dot{u'}\dot{v''}
  =        \frac{1}{\ri\hbar |\dot{z}|^2} 
           - n\frac{\rd^2 {\cal S}}{\rd {\cal E}^2} \; .
\end{align}
From the definition (\ref{2.mb2}), we have 
${\cal I}(z^\star ,z,nT) = n {\cal I}(z^\star ,z,T)$.  Once again, for 
the periodic orbit, $\cal I$ depends only on the classical energy or on 
the period and we can call it ${\cal I}({\cal E}(z))$.  For the first time 
derivative we have by (\ref{epsl})
\begin{align}
\label{mam1m}
\left.\partial {\cal I} /\partial t\right|_{t=nT}
 \approx \frac{1}{2} \frac{\partial^2 {\cal H}}{\partial z \partial z^\star} 
           = \epsilon(z) \; .
\end{align}
Calling $\psi(E,z)$ the exponent in \refg{gree1} and using 
\refgs{glg144}-(\ref{mam1m}) we get, up to second order in $T_0$, 
\begin{align}
  \label{glg139}
           \psi(E,z)
  \approx &  \frac{\ri}{\hbar}n \left[ {\cal S}\bigl( {\cal E}(z)\bigr) + 
                         {\cal I}\bigl( {\cal E}(z)\bigr) - \pi\hbar
 + \frac{d{\cal S}}{d {\cal E}} 
           \bigl( E-{\cal E}(z)+\ri\gamma\bigr) \right] \nonumber\\
  & +\frac{\ri}{\hbar} \left[\bigl( E-{\cal E}(z)+\epsilon(z)+\ri\gamma\bigr) 
   T_0 + \frac{\alpha^{(n)}}{2} T_0^2 \right]  \; .
\end{align}
Next we add and subtract $\epsilon \; \rd {\cal S}/\rd {\cal E}$
inside the brackets on the first line and use \refg{delt} for $T_0$ in
the second line. The term in $T_0^2$ becomes 
minus one half of the term in $T_0$. When we add these two terms we get
\begin{align}
  \label{glg139a}
 \psi(E,z) 
  =& \frac{\ri}{\hbar}n \left[ {\cal S}\bigl( {\cal E}(z)\bigr) + 
                         {\cal I}\bigl( {\cal E}(z)\bigr) - \pi\hbar
 + \frac{d{\cal S}}{d {\cal E}} 
           \bigl( E-{\cal E}(z)+\epsilon(z)+\ri\gamma\bigr) 
       -\epsilon(z) \frac{d{\cal S}}{d {\cal E}}  \right] \nonumber\\
 &   -  \frac{\ri}{2\hbar \alpha^{(n)}}
     \bigl( E-{\cal E}(z)+\epsilon(z)+\ri\gamma\bigr) ^2 \; .
\end{align}
Finally we use \refg{en9} for $\alpha^{(n)}$:
\begin{align}
  \label{glg139b}
\psi(E,z) =& \frac{\ri}{\hbar}n \left[ {\cal S}\bigl( {\cal E}(z)\bigr) + 
                         {\cal I}\bigl( {\cal E}(z)\bigr) - \pi\hbar
 + \frac{d{\cal S}}{d {\cal E}} 
           \bigl( E-{\cal E}(z)+\epsilon(z)+\ri\gamma\bigr) 
       -\epsilon(z) \frac{d{\cal S}}{d {\cal E}}  \right. \nonumber\\
 & \left. + \frac{1}{2} \frac{d^2{\cal S}}{d {\cal E}^2} 
 \bigl( E-{\cal E}(z)+\epsilon(z)+\ri\gamma\bigr)^2 \right] 
-  \frac{1}{2\hbar^2|\dot{z}|^2}
     \bigl( E-{\cal E}(z)+\epsilon(z)+\ri\gamma\bigr) ^2 \; .
\end{align}

The first, fourth and sixth terms in (\ref{glg139b}) are
the Taylor expansion of the function ${\cal S}(E+i\gamma +\epsilon(z))$ to
second order around ${\cal E}$.  This is an acceptable approximation as long as
$E-{\cal E}(z)+\epsilon(z)+\ri\gamma$, i.e. $T_0 \alpha^{(n)}$, is small.
Since $\epsilon$ is of order $\hbar$ we can further write ${\cal S}(E+i\gamma
+\epsilon(z)) \approx {\cal S}(E+i\gamma)+ T(E+i\gamma)\epsilon(z)$, neglecting
terms of order $\hbar^2$.  Then \refg{glg139b} can be written
\begin{align}
  \label{glg140a}
          \psi(E,z) & \approx \frac{\ri}{\hbar}n \bigl\{ 
          {\cal S}\bigl( E+\ri \gamma\bigr) 
        + {\cal I}\bigl({\cal E}) + \epsilon(z)[T(E+i\gamma)-T({\cal E})]-
         \pi\hbar \bigr\} \nonumber \\
          & - \frac{1}{2 \hbar^2 |
          \dot{z}|^2}\bigl( E-{\cal E}(z)+\epsilon(z) +\ri\gamma\bigr)^2 \;.
\end{align}
Since ${\cal I}$ and $\epsilon$ are themselves of order $\hbar$, we
may also replace ${\cal E}$ by $E+\ri \gamma$ in the argument of both
${\cal I}({\cal E})$ and $T({\cal E})$, the error in doing so being of
order $T_0 \alpha^{(n)} \hbar$. This simplifies $\psi(E,z)$ even more
and we get just
\begin{align}
  \label{glg140}
          \psi(E,z)  \approx \frac{\ri}{\hbar}n \bigl\{ 
          {\cal S}\bigl( E+\ri \gamma \bigr) 
        + {\cal I}\bigl(E+\ri \gamma \bigr) - \pi\hbar \bigr\} 
          - \frac{1}{2 \hbar^2 |\dot{z}|^2}
          \bigl( E-{\cal E}(z)+\epsilon(z) +\ri\gamma\bigr)^2 \;.
\end{align}

Inserting this as the exponent in \refg{gree1} and summing over $n$ we obtain 
\begin{align}
  \label{en10}
           G(z,E+\ri\gamma)
  =       -\frac{\ri}{\hbar} \frac{\sqrt{2 \pi}}{|\dot{z}|}
           \sum_{n=1}^{\infty} \exp & \left\{ \frac{\ri}{\hbar}n \left[  
          {\cal S}\bigl( E+\ri \gamma\bigr) 
    + {\cal I}\bigl(E+\ri \gamma\bigr) - \pi\hbar \right] \right. \nonumber \\
    &     \left. - \; \frac{\bigl( E-{\cal E}(z)+\epsilon(z)+\ri\gamma\bigr)^2}
            {2 \hbar^2 |\dot{z}|^2}   \right\} \; .
\end{align}
We may now perform the sum over all multiple traversals.  The result is
\begin{align}
  \label{en11}
        G(z,E+\ri\gamma)
  =     -\frac{\ri}{\hbar} \frac{\sqrt{2 \pi}}{|\dot{z}|} \;
        \frac{\re^{\ri({\cal S}+{\cal I}-\pi\hbar)/\hbar}}
        {1-\re^{\ri({\cal S}+{\cal I}-\pi\hbar)/\hbar}}
        \exp\left\{-\frac{\bigl( E-{\cal E}(z)+\epsilon(z)+\ri\gamma\bigr)^2}
        {2 \hbar^2 |\dot{z}|^2}\right\}
\end{align}
where $\cal S$ and $\cal I$ are taken at $E+\ri\gamma$.


\subsection{Semiclassical energy levels and Husimi functions}

Since the poles of the exact $G$ are the stationary state energies, and their
residues are the corresponding Husimi distributions, we are now in a position
to calculate semiclassical approximations to these quantities. We let
$\gamma\rightarrow 0$. If $\gamma\ne 0$ the poles are displaced by $i\gamma$.
Poles occur whenever $({\cal S}+{\cal I}-\pi\hbar)/\hbar = 2m\pi$. 
This is the quantization rule, which can be rewritten 
\begin{align}
  \label{en12}
           ({\cal S}+{\cal I})(E_m) = (m+1/2) h \; .
\end{align}
This is what replaces the usual WKB formula, which can be obtained from the
coordinate Green's function.  One should not forget that $\cal S$ here is not
the same as the $\cal S$ in the WKB formula, and that the latter does not
contain any $\cal I$, of course.  One should also realize that the $+1/2$ is
obtained in very different ways in the two cases.  In the usual WKB, the
semiclassical approximation diverges at the turning points, and one must use
some delicate arguments around this problem to derive a connection formula.
Here, on the other hand, there is no divergence and we derived the $+1/2$ with
a simple continuity argument.

The residue of
\begin{align}
       -\frac{\ri}{\hbar}\; \frac{\re^{\ri({\cal S}+{\cal I}-\pi\hbar)/\hbar}}
           {1 -\re^{\ri({\cal S}+{\cal I}-\pi\hbar)/\hbar}}
\end{align}
at $E=E_m$ is 
\begin{align}
  \label{enresi}
\frac{1}{\rd ({\cal S}+{\cal I})/\rd {\cal E}} =
\frac{1}{T(E_m)+ \left. (\rd {\cal I}/\rd {\cal E})\right|_{E_m}}  \; .
\end{align}
Hence the residue of $G$, which is the Husimi distribution for the level
$E_m$~, is 
\begin{align}
\label{hussemi}
           \rho_{\rm Husimi}(z) 
  =        \frac{\sqrt{2 \pi}}{|\dot{z}| \; \left[T(E_m)+ \left.
                             (\rd {\cal I}/\rd {\cal E})\right|_{E_m}\right]} 
           \exp{\left\{\frac{-\bigl( E_m-{\cal E}(z)+
           \epsilon (z)\bigr)^2}{2 \hbar^2 |\dot{z}|^2}\right\}} \; .
\end{align}
It is centered close to the classical trajectory with the quantized energy.
Both its amplitude and its width are modulated by the phase space velocity
$|\dot{z}|$.  The width is given by the gaussian factor.  In terms of either
variable $\cal E$ or $z$, it is of order $\sqrt{\hbar}$.  This
follows from the fact that $\hbar|z|^2$ is of order unity (see \refgs{glg49}
and (\ref{glg47})).  Therefore, our decision not to expand the pre-factor (see
\refg{pre1}), and the approximation after \refg{glg140a}, amount to discarding
corrections proportional to $\sqrt{\hbar}$ in the Green's function. These are
not as small as the other corrections to the semiclassical approximation, which
were of order $\hbar$ (see section 3.3), but they are still small and vanish in
the semiclassical limit.

For the second type of path integral discussed in section 3, the quantization
rule can be obtained by changing ${\cal I}$ to $-{\cal I}$ in the time
dependent propagator and carrying this change all the way through the Green's
function treatment.  The quantization rule and Husimi functions become then
\begin{align}
  \label{mar11}
           ({\cal S}-{\cal I})(E_m) = (m+1/2) h
\end{align}
and 
\begin{align}
\label{mar12}
     \rho_{\rm Husimi}^{(2)}(z) 
  =   \frac{\sqrt{2 \pi}}{|\dot{z}| \; \left[T(E_m)- \left.
        (\rd {\cal I}/\rd {\cal E})\right|_{E_m}\right]} 
         \exp{\left\{\frac{-\bigl( E_m-{\cal E}(z)-\epsilon(z)\bigr)^2}
         {2 \hbar^2 |\dot{z}|^2}\right\}}
\end{align}
where all quantities are computed with $H_2$, instead of $H_1\equiv{\cal H}$.
 
According to our discussion in section 3, one could also use the Weyl
hamiltonian and drop ${\cal I}$. In this case there would be no $\epsilon(z)$
coming from $\partial {\cal I} /\partial t$.  The quantization rule in this
case becomes
\begin{align}
  \label{mar13}
           {\cal S}(E_m) = (m+1/2) h \;,
\end{align}
which is exactly the WKB rule, and the Husimi functions are
\begin{align}
\label{mar14}
           \rho_{\rm Husimi}^{\rm Weyl}(z) 
  =        \frac{\sqrt{2 \pi}}{|\dot{z}| \; T(E_m)} 
           \exp{\left\{\frac{-\bigl( E_m-E(z)
           \bigr)^2}{2 \hbar^2 |\dot{z}|^2}\right\}} \; .
\end{align}
This expression is similar to a formula suggested previously by
Kurchan {\it et al.} (see equations (5.21) and (5.22) in
\cite{Kurc89}; see also \cite{Carv92} for a numerical application).

The semiclassical energy levels will be slightly different for the three
prescriptions, eqs. (\ref{en12}), (\ref{mar11}), and (\ref{mar13}).  In fact it
can be shown, using arguments similar to those in appendix C, that they all
coincide up to first order in $\hbar$. This is in accordance with the fact that
corrections to the WKB energies are of order $\hbar^2$ \cite{Voro77}. Our
quantization rule (\ref{en12}) does include some corrections of order
$\hbar^2$, but not all of them. These corrections might improve the calculation
of the energy levels with respect to WKB, especially for low lying energy
levels. Numerical work is currently being done to explore the possibilities.
In this we still have one big freedom, the choice of $b$, the width of the
coherent state, {\it for each energy level}.  For the ground state, the
variational principle tells us that ${\cal H}$ is always larger than or equal
to the true energy. Therefore we must choose $b$ so as to minimize the energy.
For the other states, however, no such direct rule exists and other
prescriptions leading to an optimal $b$ have to be devised.

For the harmonic oscillator, all three formulae give the exact
result $E_m=(m+1/2)\hbar\omega$ and the same Husimi distributions, independent
of what $b$ is chosen for the coherent state width. For the case of
$b=\sqrt{\hbar/\omega}$ all three semiclassical formulae give
\begin{align}
\label{husos}
 \rho_m (z) = \displaystyle{\frac{1}{\sqrt{2\pi}|z|} \; 
  \exp{\left[ -\frac{(|z|^2-m-1/2)^2}{2|z|^2}\right]}} 
\end{align}
which is the result we already obtained in subsection 6.2.
 

\section{Conclusion}

One inescapable conclusion is that there is more than one
semiclassical approximation to quantum mechanics in phase space, even
when one restricts oneself to the coherent state representation.  It
has often been said that all semiclassical approximations are
identical in the end, that they always consist in expanding some
exponent to second order in the vicinity of the classical trajectory
and then doing a collection of gaussian integrals, that all such
expansions should be the same except for the choice of independent
variables, and therefore that the gaussian integrals should always
produce the same result.  This argument is wrong, the basic reason
being that the classical trajectory about which the expansion takes
place differs from method to method, because the classical hamiltonian
differs from method to method.  The confusion in the literature is in
part due to this, but not entirely.  People have ignored terms that
looked small or unfamiliar, even when these were clearly part of the
approximation they were using.  They have done complex integrals as if
these were real.  They have ignored phases.  They have been
inconsistent with the hamiltonian they used, often changing it in
midstream.

Although the number of different, equally valid approximations is
actually infinite, it is convenient to focus on only three of them,
which have been discussed at some length in section 3.  They
correspond to three different choices for the classical hamiltonian
associated with a given quantal hamiltonian.  They are the Weyl
hamiltonian $H_W$, the smoothed hamiltonian $H_1 \equiv {\cal H}$, and
the antismoothed hamiltonian $H_2$.  It is essential to realize that,
in all semiclassical approximations, the smoothed hamiltonian is
always associated with an exponential term containing a special
correction to the action which we have called $\cal I$.  The
temptation to omit this unfamiliar term is great, but it should be
resisted, as without it one cannot get a correct quantization rule,
for instance.  Similarly, the antismoothed hamiltonian is always
associated with $-\cal I$.  The Weyl hamiltonian does not come with
such a correction term, which in a way makes it the simplest of the
three.  On the other hand, there is no simple approximation involving
coherent states which yields the Weyl hamiltonian.  It occurs in some
of the other methods, for instance the Wigner-Weyl method mentioned in
section 1, and also the Heller method of section 4.3.  The advantage
of the smoothed hamiltonian is that, since it is smoothed,
approximations based on power series expansions are especially good
for it.  This is the hamiltonian that we have used in most of our
work.  The antismoothed hamiltonian, lacking both of the
characteristics that make the other two desirable, does not seem to
have been used in practice by anyone.

The first test cases for a semiclassical approximation should be the
free particle and the harmonic oscillator.  All three of the above
approximations become exact then.  Once again, this is so only when
the $\cal I$ term is appropriately included.  Without $\cal I$, both
$H_1$ and $H_2$ are wrong; only $H_W$ remains.  Beyond these two
systems, the question of which of the three approximations gives
better results has no single answer; it depends very much on the
specific quantity calculated.  We are planning to address this point
in another paper.  It is true that the three versions differ from each
other only by terms of order $\hbar$, but a difference of this order
can be essential in the quantization rule, and it may also be
significant for the long time behavior.

We shall now summarize our other results.  They can be grouped in
three categories: coherent state propagator, initial value
representation (IVR), and energy Green's function.  We gave in section
2 a very complete derivation of the semiclassical coherent state
propagator in one dimension, smoothed hamiltonian version.  This is
\refg{glg119} or \refg{en4}.  Then in section 3 we compared it with
several other semiclassical approximations.  In particular, we
obtained an explicit expression for the antismoothed hamiltonian
version.  We believe both of these results to be new.

In an attempt to provide a rigorous derivation of the well known
Herman-Kluk formula, we derived in section 4 our own new IVR.  Like
HK, we set out to calculate semiclassically the mixed propagator
$\langle x | K(t) | z' \rangle$ in terms of real trajectories starting
from $q(0)=q'$, $p(0)=p'$.  Our result, \refg{3.24}, is very different
from HK's.  Also in section 4, we considered Heller's old idea for an
IVR, which is not based on the coherent state formalism, and we
derived an explicit expression for it, valid for any smooth
hamiltonian function of $q$ and $p$.  We found that Heller's IVR and
ours are very similar, the difference being that Heller's has the Weyl
hamiltonian and no $\cal I$ term.  We returned to the Herman-Kluk IVR
in section 5 and pointed out mistakes made in its derivation.  The
nonconservation of the HK norm is a fatal flaw, we think.  The HK
expression does not follow from the semiclassical limit of the
coherent state propagator, but it can be derived from an ansatz where
one assumes a fixed width for the propagated wave--packet and one
demands that the resulting propagator recover that of Van Vleck when
transformed to the representation $\langle x|K(t)|x'\rangle$. We show,
however, that the applicability of such a formula is very limited and
should fail for long times and/or for chaotic systems.

Finally, in section 6 we derived the corresponding semiclassical
approximations for the energy Green's function in the coherent state
representation.  For each of the approximations, we used this Green's
function to find a quantization rule, which yields the energy levels
in terms of classical quantities.  We also used it to get a
semiclassical approximation to the Husimi distribution for each level.

There are at least two ways that one might go farther in this field.
One needs to carry out numerical comparisons between the various
semiclassical approximations and determine the conditions under which
one or the other might be preferable.  We have made a small start in
this direction, but we leave the results for a future publication.
One needs also to extend this work to two dimensions, which will allow
applications to more interesting problems, in particular problems with
partially chaotic classical mechanics.  There too we have made a
significant start; it promises to be a rather complicated field and we
shall say no more about it here.
\bigskip

\noindent ACKNOWLEDGMENTS 

\noindent MB wishes to thank Marcos Saraceno for an illuminating 
discussion. The work at MIT was partially supported by the
U.S. Department of Energy (DOE) under Contract
No. DE-FC02-94ER40818. MAMA acknowledges financial support from the
Brazilian agencies FAPESP and CNPq. FK, HJK and BS acknowledge the
support granted by the Deutsche Forschungsgemeinschaft (SPP 470
"Zeitabhängige Phänomene und Methoden in Quantensystemen der Physik
und Chemie").
 
\begin{appendix}

\section{Calculating the Prefactor by the Determinantal Method}

In this appendix we present an alternative way of performing the multiple
gaussian integral which occurs in the calculation of the semiclassical
propagator.  In subsection 2.3, we started by doing the integrals one by one,
and then we derived a recursion relation between successive integrals.  Here we
start with the determinant of the quadratic form, and we derive recursion
relations between determinants of successively higher orders.  This method is
very general and can be applied for any number of dimensions.  We use it here
again for an integral in subsection 3.1 concerning the semiclassical
approximation of a different kind of coherent state propagator.

We start with \refg{glg6}, where we had to perform a $2(N-1)$--dimensional
gaussian integral of the form
\begin{align}
  \label{glg98}
      \int \exp\left\{-\frac{x^TMx}{2}\right\}\,{\rm d}^L x 
   =  \frac{(2\pi)^{L/2}}{\sqrt{{\rm det} M}}
\end{align}
where $M$ is a complex, positive definite $L\times L$ matrix.  All the work
comes in calculating the ``prefactor'', or the square root of the determinant.
We rewrite \refg{glg6} as
\begin{align}
  \label{glg99}
  K_1(z'',t;z',0) = {\rm e}^{f(z^\star,z)} 
         \mbox{\large $\displaystyle \int$} \left\{ \prod_{j=1}^{N-1} 
         \frac{{\rm d}\eta_j^\star {\rm d}\eta_j}{2\pi {\rm i}}\right\} 
         {\rm e}^\zeta \; .
\end{align}
The boundary conditions (\ref{glg41}) ensure that $j$ varies only from 1 to
$N-1$.  We already wrote $\zeta$ in \refg{5mb13}.  Here we define the
abbreviations
\begin{align}
  \label{glg101}
   A_j=\frac{{\rm i}}{\hbar} \frac{\partial^2 H_1(z_{j+1}^{\star},z_j)}
         {\partial z_j^2} \qquad
   B_j=\frac{{\rm i}}{\hbar} \frac{\partial^2 H_1(z_j^{\star},z_{j-1})}
         {\partial z_j^{\star2}} \qquad
   C_j=\frac{{\rm i}}{\hbar} \frac{\partial^2 H_1(z_{j+1}^{\star},z_j)}
         {\partial z_{j+1}^\star \partial z_j}
\end{align}
and we write
\begin{align}
  \label{glg100}
  -2\zeta = 2 \sum_{j=1}^{N-1} &  \eta_j \eta_j^\star  + \sum_{j=1}^{N-1} 
      \tau A_j  \eta_j^2 + 2 \sum_{j=1}^{N-2} ( \tau \, C_j-1) 
      \eta_{j+1}^\star \eta_j + \sum_{j=1}^{N-1} \tau B_j \eta_j^{\star2} \; .
\end{align}
The subscript $1$ in eqs.(\ref{glg99}) and (\ref{glg101}) is the notation of
section 3 and means that the first form of path integral is being used. We
shall calculate $K_2(z'',t;z',0)$ as well later in this appendix. We use $\eta$
(without subscript!) to denote a double--sized vector containing all the
$\eta_j$'s as well as their complex conjugates in the order $\eta_{N-1},
\eta_{N-1}^\star, \eta_{N-2},
\eta_{N-2}^\star , \dots , \eta_1, \eta_1^\star $~.
Then eq.(\ref{glg100}) can be rewritten
\begin{align}
  \label{glg102}
  \zeta = -\frac{\eta^T G^{(N-1)}\eta}{2}
\end{align}
where
\begin{equation}
  \label{glg103}
  G^{(N-1)} := 
    \begin{pmatrix}
      \tau A_{N-1} & 1 & 0 & 0 & 0 & 0 & \dots \\
      1 & \tau B_{N-1} & \tau C_{N-2}-1 & 0 & 0 & 0 & \dots \\
      0 & \tau C_{N-2}-1 & \tau A_{N-2} & 1 & 0 & 0 & \dots \\
      0 & 0 & 1 & \tau B_{N-2} & \tau C_{N-3}-1 & 0 & \dots \\
      0 & 0 & 0 & \tau C_{N-3}-1 & \tau A_{N-3} & 1 & \dots \\
      0 & 0 & 0 & 0 & 1 & \tau B_{N-3} & \dots \\
      \vdots & \vdots & \vdots & \vdots & \vdots & \vdots & \ddots
    \end{pmatrix}
\end{equation}
is a tridiagonal band matrix whose dimensionality is twice the value of its
superscript.

The integration formula (\ref{glg98}) {\it cannot} be applied directly to
(\ref{glg99}), because in the latter the variables and the paths are complex.
However, this is a problem that we already considered and solved at the
beginning of subsection 2.3.  If we compare carefully \refg{glg44} with
\refg{mb1}, we find in the complex case that, for each pair of variables, we
must introduce an additional minus sign in front of the determinant in the
square root.  It is also clear that the $2\pi$'s will cancel out.  Hence the
correct formula is
\begin{align}
  \label{glg117}
  K_1(z'',t;z',0) &= {\rm e}^{f(z^\star ,z)}\frac{1}{
         \sqrt{(-1)^{N-1}{\rm det}G^{(N-1)}}} \; .
\end{align}

Then the calculation of the determinant proceeds in the following way.  First,
the determinant is expanded with respect to its first column
\begin{align}
  \label{glg105}
  \det G^{(N-1)} = \tau A_{N-1}  
      &\begin{vmatrix}
        \tau B_{N-1}   & \tau C_{N-2}-1 & 0 & 0 & 0 & \dots & \\
        \tau C_{N-2}-1 & \tau A_{N-2}   & 1 & 0 & 0 & \dots \\
        0 & 1 & \tau B_{N-2}   & \tau C_{N-3}-1 & 0 & \dots \\
        0 & 0 & \tau C_{N-3}-1 & \tau A_{N-3}   & 1 & \dots \\
        0 & 0 & 0 & 1 & \tau B_{N-3} & \dots \\
        \vdots & \vdots & \vdots & \vdots & \vdots & \ddots
      \end{vmatrix} \nonumber \\ \nonumber\\
    \qquad -
      &\begin{vmatrix}
        1 & 0 & 0 & 0 & 0 & \dots \\
        \tau C_{N-2}-1 & \tau A_{N-2}& 1 & 0 & 0 &\dots & \\
        0 & 1 & \tau B_{N-2} & \tau C_{N-3}-1 & 0 & \dots \\
        0 & 0 & \tau C_{N-3}-1 & \tau A_{N-3} & 1 & \dots \\
        0 & 0 & 0 & 1 & \tau B_{N-3} & \dots \\
        \vdots & \vdots & \vdots & \vdots & \vdots &\ddots 
      \end{vmatrix}
     \;\; .
\end{align}
The second determinant can easily be expanded with respect to its first
line:  it is simply ${\rm det}\,G^{(N-2)}$.  The first determinant, on the
other hand, is that of a different quadratic form which we call $F^{(N-1)}$.
Thus we have the recursion relation
\begin{align}
  \label{glg106}
   \det G^{(N-1)} = \tau A_{N-1} \det F^{(N-1)} - \det G^{(N-2)} \, .
\end{align}
Now we can expand ${\rm det} F^{(N-1)}$ analogously, which leads to a second
recursion relation
\begin{align}
  \label{glg107}
  \det F^{(N-1)} & = \tau B_{N-1} \det G^{(N-2)} 
         - (\tau C_{N-2} -1)^2 \det F^{(N-2)} \nonumber \\
        &= \tau B_{N-1} \det G^{(N-2)} + 2\tau C_{N-2} 
         \det F^{(N-2)} - \det F^{(N-2)} + O(\tau^2).
\end{align}
Since eventually we shall go to the limit $\tau\rightarrow 0, \;
N\rightarrow\infty$, we may drop the terms of order $\tau^2$.  We can get rid
of many minus signs by making the definitions
\begin{align}
  \label{glg108}
  g_{N-1}=(-)^{N-1} \det G^{(N-1)}, \qquad f_{N-1}=(-)^{N-1} \det F^{(N-1)}.
\end{align}
Eqs.{\hskip 2pt}(\ref{glg106}) and (\ref{glg107}) can now be rewritten as
\begin{align}
  \label{glg109}
  g_{N-1} & =g_{N-2}+\tau A_{N-1} f_{N-1} \nonumber \\
  f_{N-1} & =f_{N-2}-\tau B_{N-1} g_{N-2} - 2\tau C_{N-2} f_{N-2}  \; .
\end{align}
In the continuous time limit, these recursion relations turn into two
first--order differential equations
\begin{align}
  \label{glg110}
  \dot{g} = A f  \qquad\qquad \dot{f}=-Bg-2Cf \; .
\end{align}
By letting $\tau$ go to 0 in the determinants themselves, one sees easily that
the initial conditions are
\begin{align}
  \label{glg111}
  g(0) = 1 \qquad f(0) = 0 \; .
\end{align}
One way to solve the differential equations is to introduce the functions
\begin{align}
  \label{glg112}
  \tilde{g} &= g \exp \left[ \int_0^t C(t') {\rm d} t' \right] \qquad\qquad
  \tilde{f} = f \exp \left[ \int_0^t C(t') {\rm d} t' \right] \; .
\end{align}
Eqs.{\hskip 2pt}(\ref{glg110}) become now
\begin{align}
  \label{glg113}
  \dot{\tilde{g}} = A \tilde{f} + C \tilde{g} 
  \qquad\qquad
  \dot{\tilde{f}} = -B \tilde{g} - C \tilde{f} 
\end{align}
with initial conditions $\tilde{g}=1$ and $\tilde{f}=0$. Replacing the
abbreviations $A,B,C$ by their definitions in \refg{glg101}, we see that
(\ref{glg113}) is formally identical to the differential equations
(\ref{glg30}) if we let $\tilde{g}$ be $\delta v$ and $\tilde{f}$ be $\delta
u$ .  The initial values are $\delta u(0) \equiv \delta u' = 0$ and 
$\delta v(0) \equiv \delta v' = 1$ .

We need to know $g(t)$, which is the quantity under the square root in 
(\ref{glg117}); we do not need $f(t)$. Since $\tilde{g}(t)=\delta v''$, we
can use eq.{\hskip 2pt}(\ref{mb11}) with $\delta v'=1$ to write
\begin{align}
  \label{glg114}
  g(t) = \delta v''\exp \left( - \int_0^t C(t') {\rm d} t' \right) 
         = \left( \frac{{\rm i}}{\hbar}\frac{\partial^2 S}
         {\partial u' \partial v''}\right)^{-1} \exp \left( - \int_0^t C(t') 
         {\rm d} t' \right).
\end{align}
Substituting back into \refg{glg117} and replacing $C$ by its definition gives
\begin{align}
  \label{glg118}
  K_1(z'',t;z',0) = \re^{f(z^\star ,z)}
        &\sqrt{\frac{\ri}{\hbar}\frac{\partial^2 S}{\partial 
         u' \partial v''}} \exp\left\{ \frac{\ri}
         {2\hbar}\mbox{\large $\displaystyle \int$}^t_{\!\!\!\!0} \rd
         t' \frac{\partial^2H_1}{\partial u\partial v} \right\}.
\end{align}
Finally, by calculating $\exp\{f(z^\star ,z)\}$ in the same way as in section
2 (see \refg{glg15} and subsection 2.5), we obtain a result identical to
\refg{glg119}:
\begin{align}
  \label{glg116}
  K_1(z'',t;z',0) = 
        &\sqrt{\frac{\ri}{\hbar}\frac{\partial^2 S}{\partial 
         u' \partial v''}} \exp\left\{ \frac{\rm i}
         {2\hbar}\mbox{\large $\displaystyle \int$}^t_{\!\!\!\!0} \rd
         t' \frac{\partial^2H_1}{\partial u\partial v} \right\} \nonumber\\
        &\exp \left\{ \frac{\ri}{\hbar}S(v'',u',t) - \frac{1}{2} 
         \bigl( |v''|^2 + |u'|^2\bigr) \right\} .
\end{align}

Let us now repeat this calculation for Klauder and Skagerstam's ``second form
of the path integral'' discussed in subsection 3.1.  We follow very closely
what we just did for $K_1$~, with small but important differences.  According
to eq.(\ref{5mb12}), the quadratic form for $K_2$ is
\begin{align}
  \label{apx1}
-2 \zeta =  2 \sum_{j=1}^{N-1}\eta_j^\star\eta_j 
-2 \sum_{j=1}^{N-2}\eta_{j+1}^\star\eta_j +
\sum_{j=1}^{N-1} \tau A_j \eta_j^2 + 
\sum_{j=1}^{N-1} \tau B_j \eta_j^{\star2} +
2\sum_{j=1}^{N-1} \tau C_j \eta_j^{\star} \eta_j 
\end{align}
with the (different) definitions
\begin{align}
  \label{apx2}
   A_j=\frac{{\rm i}}{\hbar} \frac{\partial^2 H_2(z_j^{\star},z_j)}
         {\partial z_j^2},\qquad
   B_j=\frac{{\rm i}}{\hbar} \frac{\partial^2 H_2(z_j^{\star},z_j)}
         {\partial z_j^{\star2}},\qquad
   C_j=\frac{{\rm i}}{\hbar} \frac{\partial^2 H_2(z_j^{\star},z_j)}
         {\partial z_j^\star \partial z_j} \; .
\end{align}
The matrix $G^{N-1}$ becomes
\begin{equation}
  \label{apx3}
  G^{(N-1)} := 
    \begin{pmatrix}
      \tau A_{N-1} & 1 + \tau C_{N-1}& 0 & 0 & 0 & 0 & \dots \\
      1 + \tau C_{N-1}& \tau B_{N-1} & -1 & 0 & 0 & 0 & \dots \\
      0 & -1 & \tau A_{N-2} & 1+ \tau C_{N-2} & 0 & 0 & \dots \\
      0 & 0 & 1+ \tau C_{N-2} & \tau B_{N-2} & -1 & 0 & \dots \\
      0 & 0 & 0 & -1 & \tau A_{N-3} & 1 + \tau C_{N-3}& \dots \\
      0 & 0 & 0 & 0 & 1 + \tau C_{N-3}& \tau B_{N-3} & \dots \\
      \vdots & \vdots & \vdots & \vdots & \vdots & \vdots & \ddots
    \end{pmatrix}
\end{equation}
Calling $F^{N-1}$ the matrix obtained by removing the first column and the first
row of $G^{N-1}$, and expanding both $G^{N-1}$ and $F^{N-1}$ with respect
to their first lines, we get
\begin{align}
  \label{apx4}
   \det G^{(N-1)} &= \tau A_{N-1} \det F^{(N-1)} - 
    (1 + \tau C_{N-1})^2\det G^{(N-2)} \nonumber \\
\det F^{(N-1)}  &= \tau B_{N-1} \det G^{(N-2)} - \det F^{(N-2)} \; .
\end{align}
Once again we get rid of signs with the help of definitions (\ref{glg108}) and
we take the limit $\tau \rightarrow 0, \; N \rightarrow \infty$ to get
\begin{align}
  \label{apx5}
  \dot{g} = A f+2Cg  \qquad \dot{f}=-Bg \; .
\end{align}
Finally we define
\begin{align}
  \label{apx6}
  \tilde{g} &= g \exp \left[ -\int_0^t C(t') {\rm d} t' \right] \qquad\qquad
  \tilde{f} = f \exp \left[ -\int_0^t C(t') {\rm d} t' \right].
\end{align}
Note that, compared to the analogous definitions (\ref{glg112}), the present
ones have a minus sign!  We obtain again
\begin{align}
  \label{apx7}
  \dot{\tilde{g}} = A \tilde{f} + C \tilde{g} 
  \qquad\qquad
  \dot{\tilde{f}} = -B \tilde{g} - C \tilde{f} 
\end{align}
which are the same equations as (\ref{glg113}). Therefore the propagator $K_2$
is given again by (\ref{glg118}) except that, this time, there is a minus sign
in front of the $\cal I$ term!

\section{Proof of \refg{sec27ca}}

We start by rewriting \refg{sec27a} as

\begin{align}
\label{sec27d}
A = \int_{-\infty}^{+\infty} e^{\frac{i}{\hbar}\phi(x)} \; \rd x 
\end{align}
where $\phi(x):=f(x)-i\hbar \log{g(x)}$. To calculate $A$ beyond the
stationary phase approximation (SPA) we expand $\phi$ about its
stationary point $X$ to fourth order:
\begin{align}
\label{sec27e}
\phi(x) = \phi(X)+\frac{1}{2!}\phi^{(2)}\delta x^2 + 
\frac{1}{3!}\phi^{(3)}\delta x^3 + 
\frac{1}{4!}\phi^{(4)}\delta x^4 + ...
\end{align}
where $\delta x = x-X$ and $\phi^{(n)}= \rd^n \phi/\rd x^n (X)$. Next
we change the integration variable to
\begin{align}
\label{sec27f}
y=\sqrt{\frac{|\phi^{(2)}|}{2\hbar}}\; e^{i\alpha/2-i\pi/4} \; \delta x \;,
\end{align}
where $\alpha$ is the phase of $\phi^{(2)}$. Substituting
(\ref{sec27e}) and (\ref{sec27f}) into (\ref{sec27d}) we get
\begin{align}
\label{sec27g}
A & \approx \sqrt{\frac{2\hbar}{|\phi^{(2)}|}} e^{i\pi/4-i\alpha/2} 
e^{\frac{i}{\hbar}\phi(X)} \int_{-\infty}^{+\infty} \displaystyle{ e^{ -y^2 + 
i\frac{\phi^{(3)}}{6 \hbar}\left(\frac{2\hbar}{|\phi^{(2)}|}\right)^{3/2} 
e^{3i\pi/4-3i\alpha/2} y^3 + 
\frac{i\hbar \phi^{(4)}}{6|\phi^{(2)}|^2 } e^{i\pi-2i\alpha} y^4
} }\rd y \nonumber \\
 & \approx  \sqrt{\frac{2\hbar}{|\phi^{(2)}|}} e^{i\pi/4-i\alpha/2} 
e^{\frac{i}{\hbar}\phi(X)} \int_{-\infty}^{+\infty} e^{-y^2} \left[
1 + \frac{i\hbar \phi^{(4)}}{6|\phi^{(2)}|^2 } e^{i\pi-2i\alpha} y^4 -
\frac{\hbar (\phi^{(3)})^2}{9 |\phi^{(2)}|^3} 
e^{3i\pi/2-3i\alpha} y^6 \right]
\rd y \nonumber \\
 &= \sqrt{\frac{2\pi\hbar}{|\phi^{(2)}|}} e^{i\pi/4-i\alpha/2} 
e^{\frac{i}{\hbar}\phi(X)} \left[1 - 
\frac{i\hbar \phi^{(4)}}{8|\phi^{(2)}|^2 } e^{-2i\alpha} + 
\frac{5i \hbar (\phi^{(3)})^2}{24 |\phi^{(2)}|^3} e^{-3i\alpha}
\right] \nonumber \\
 & \approx \sqrt{\frac{2\pi\hbar}{|\phi^{(2)}|}} e^{i\pi/4-i\alpha/2} 
\exp{\left\{\frac{i}{\hbar}\left[ \phi(X) + \frac{\hbar^2}{24|\phi^{(2)}|^3} 
\left(5 e^{-3i\alpha} (\phi^{(3)})^2 -
3 e^{-2i\alpha} |\phi^{(2)}| \phi^{(4)} \right) \right]\right\}} \; .
\end{align}
We now proceed to expand the various terms above in powers of $\hbar$
so as to compare this result with SPA. The stationary point $X$ of $\phi$   
can be written in terms of the stationary point $x_0$ of $f$ as
\begin{align}
\label{sec27h}
X=x_0 + i \hbar \frac{g^{(1)}}{f^{(2)} g^{(0)}} + {\cal O}(\hbar^2)
=: x_0 + \ri \hbar x_1 + {\cal O}(\hbar^2) \; .
\end{align}
Also
\begin{align}
\label{sec27i}
\phi(X) &= f(x_0)-i\hbar \log{g(x_0)} + \hbar^2 \left[
 \frac{x_1 g^{(1)}}{g^{(0)}} -\frac{f^{(2)} x_1^2}{2} \right] + 
{\cal O}(\hbar^3) \; , \\
\phi^{(2)} &= f^{(2)}+ \ri \hbar f^{(3)}x_1-i\hbar 
\frac{g^{(2)}g^{(0)}-{g^{(1)}}^2}{{g^{(0)}}^2} + {\cal O}(\hbar^2)
\end{align}
where $g^{(n)}= \rd^n g/\rd x^n (x_0)$ and $f^{(n)}= \rd^n f/\rd x^n (x_0)$.
From the last expression we find
\begin{align}
\label{sec27ia}
|\phi^{(2)}|^{-1/2} &= |f^{(2)}|^{-1/2} \; (1+ {\cal O}(\hbar^2)) 
\end{align}
and
\begin{align}
\label{sec27ii}
\alpha=\frac{\pi}{2}(s-1) + \frac{\hbar f^{(3)}x_1}{f^{(2)}} 
-\hbar\frac{g^{(2)}g^{(0)}-{g^{(1)}}^2}{{g^{(0)}}^2 f^{(2)}} 
+ {\cal O}(\hbar^2) \;.
\end{align}
Substituting these equations into (\ref{sec27g}) we obtain, after
some simplifications, the desired result, \refg{sec27ca}, with
\begin{align}
\label{sec27k}
R(x_0) =  \frac{f^{(2)} g^{(2)} -f^{(3)} g^{(1)}}{2{f^{(2)}}^2 g^{(0)}}  +
 \frac{5  (f^{(3)})^2 - 3 f^{(2)} f^{(4)}}{24(f^{(2)})^3}  \;.
\end{align}
%

\section{Cancelation of first order terms in $S+{\cal I}$} 

Our goal here is to show that the quantity $(S+{\cal I})$ appearing in
the phase of the propagator, \refg{en4}, can be written as $S_W +
{\cal O}(\hbar^2)$, where $S_W$ is the action computed with the
classical Hamiltonian $H_C$, which we assume to be the same as the
Weyl symbol $H_W$. In order to do that we must relate ${\cal H}$ to
$H_W$ and write the classical trajectories governed by ${\cal H}$ in
terms of those governed by $H_W$.

The relation between $H_W$ and ${\cal H}\equiv H_1$ is given by the
second of \refgs{5mb24}. Expanding the exponential operator we obtain
\begin{align}
  \label{sec331}
H_W(u,v) = {\cal H}(u,v)-\frac{1}{2}\frac{\partial^2 {\cal H}}
{\partial u \partial v}(u,v) + ... \;.
\end{align}
The second term on the right is what we have called $\epsilon(u,v)$ in
section 6 (see \refg{epsl}).  We assume that $H_W$ does not depend
explicitly on $\hbar$, and that ${\cal H}$ may be written as a power
series in $\hbar$, starting at $\hbar^0$.  The same holds for
$\epsilon(u,v)$, with the series starting at $\hbar^1$. Since the next 
term in the expansion (\ref{sec331}) is of order $\hbar^2$, 
 \refg{sec331} can be rewritten as
\begin{align}
  \label{sec333}
H_W(u,v) = {\cal H}(u,v)- \epsilon(u,v) + {\cal O}(\hbar^2) \;.
\end{align}
The inverse equation, where ${\cal H}$ is written in terms of $H_W$, is
\begin{align}
  \label{sec333a}
{\cal H}(u,v) = H_W(u,v) + \epsilon_W(u,v) + {\cal O}(\hbar^2)
\end{align}
where $\epsilon_W(u,v)$ is defined by \refg{epsl} with $H_W$
replacing ${\cal H}$.

Let $u=u(v'',u',t')$ and $v=v(v'',u',t')$ be the solutions of Hamilton's
equations (\ref{glg10}) with boundary conditions $u(v'',u',0)=u'$, 
$v(v'',u',t)=v''$. Let also  $u_0=u_0(v'',u',t')$ and 
$v_0=v_0(v'',u',t')$ be the solutions of the same Hamilton's 
equations but with ${\cal H}$ replaced by the classical Hamiltonian $H_W$ 
and the same boundary conditions, $u_0(v'',u',0)=u'$, 
$v_0(v'',u',t)=v''$. Since the difference between the functions
${\cal H}$ and $H_W$ is of order $\hbar$ we write
\begin{align}
  \label{sec334}
u(v'',u',t') &= u_0(v'',u',t')+\hbar u_1(v'',u',t')
+{\cal O}(\hbar^2)  \nonumber \\
v(v'',u',t') &= v_0(v'',u',t')+\hbar v_1(v'',u',t')+{\cal O}(\hbar^2)\;.
\end{align}
Notice that $t$ is the total propagation time, whereas $t'$ is
used for intermediate instants, $0 < t' < t$. Due to the 
boundary conditions we have
\begin{align}
  \label{sec335}
u_1(v'',u',0)&=0 \nonumber \\
v_1(v'',u',t)&=0 \; .
\end{align}
For the sake of clarity we rewrite formula (\ref{glg83}) for the action:
\begin{align}
  \label{sec335a}
  S(v'',u',t) : 
         &=\int\limits_0^t \rd t' \left[\frac{\ri \hbar}{2}
          (\dot{u}v-\dot{v}u) - {\cal H}(u,v,t') \right]
          - \frac{\ri \hbar}{2} ( u''v'' + u'v' )\; .
\end{align}
We shall now expand the various terms in $S(v'',u',t)$ in powers of
$\hbar$. We start with ${\cal H}(u,v)$ calculated along its own
trajectory:
\begin{align}
  \label{sec336}
{\cal H}(u,v) &= H_W(u_0+\hbar u_1,v_0+\hbar v_1)+ \epsilon_W(u_0,v_0) 
 + {\cal O}(\hbar^2) \nonumber \\
 &= H_W(u_0,v_0)+ \frac{\partial H_W}{\partial u}(u_0,v_0) \hbar u_1 +
\frac{\partial H_W}{\partial v}(u_0,v_0) \hbar v_1 + \epsilon(u_0,v_0) 
 + {\cal O}(\hbar^2) \nonumber \\
 &= H_W(u_0,v_0)-\ri\hbar^2 \dot{v_0} u_1 +\ri\hbar^2 \dot{u_0} v_1 
+ \epsilon(u_0,v_0) + {\cal O}(\hbar^2) 
\end{align}
where the dot means $\rd/\rd t'$. Using \refgs{sec334} we also 
find 
\begin{align}
  \label{sec337}
\frac{\ri \hbar}{2} (\dot{u}v-\dot{v}u) &= 
\frac{\ri \hbar}{2}(\dot{u}_0 v_0-\dot{v}_0 u_0) + \frac{\ri \hbar^2}{2}
(\dot{u}_1 v_0+ \dot{u}_0 v_1 -\dot{v}_0 u_1-\dot{v}_1 u_0) +
{\cal O}(\hbar^2) \nonumber \\
 &=  \frac{\ri \hbar}{2} (\dot{u}_0 v_0-\dot{v}_0 u_0) 
- \ri \hbar^2(\dot{v}_0 u_1-\dot{u}_0 v_1) 
+\frac{\ri \hbar^2}{2}\frac{\rd}{\rd t'} (v_0 u_1-v_1 u_0)  + 
{\cal O}(\hbar^2) \; .
\end{align}
Finally, using \refgs{sec335}, we have
\begin{align}
  \label{sec338}
\frac{\ri \hbar}{2}(u''v''+u'v') &= 
\frac{\ri \hbar}{2}(u_0''+\hbar u_1'')v'' + 
\frac{\ri \hbar}{2} u'(v_0'+\hbar v_1') + {\cal O}(\hbar^2) \nonumber \\
&= \frac{\ri \hbar}{2}(u_0''v''+u'v_0') + 
\frac{\ri \hbar^2}{2} (u_1'' v''+ u'v_1') + {\cal O}(\hbar^2)
\end{align}
Substituting \refgs{sec336}--(\ref{sec338}) in \refg{sec335a} we
see that the second and third terms in \refg{sec336} cancel
the terms in the second parenthesis of (\ref{sec337}). We get 
\begin{align}
  \label{sec339}
S(v'',u',t)=S_W(v'',u',t) &+ \int_0^t \rd t' \left[
\frac{\ri\hbar^2}{2} \frac{\rd}{\rd t'} (v_0 u_1-v_1 u_0) 
-\epsilon(u,v) \right] \nonumber \\
& -  \frac{\ri \hbar^2}{2}(u_1'' v''+ u'v_1') + {\cal O}(\hbar^2) \;.
\end{align}
Using \refg{sec335} once again, we cancel the first two terms
on the second line against those under the total derivative 
sign. Recalling the definition  (\ref{2.mb2}) of ${\cal I}$, we 
finally get
\begin{align}
  \label{sec3310}
S(v'',u',t) + {\cal I}(v'',u',t) =S_W(v'',u',t) + {\cal O}(\hbar^2) \;.
\end{align}

\end{appendix}

\bibliographystyle{alphamt}
\bibliography{abbrev,dipldiss,paper60,paper70,paper80,paper90,publko,rest}

\end{document}